\documentclass[10pt,nofootinbib,aps,prd,groupedaddress,preprintnumbers,showpacs,showkeys,floatfix,amssymb,amsfonts,superscriptaddress]{revtex4-2}
\usepackage{graphicx}
\usepackage{amsmath}
\usepackage[dvipsnames]{xcolor}
\usepackage{hyperref}
\usepackage{cleveref}
\usepackage{booktabs}
\usepackage{physics}
\usepackage{dsfont}
\usepackage{tcolorbox}
\usepackage{enumitem}
\usepackage{graphicx}
\usepackage{wrapfig}
\usepackage[export]{adjustbox}
\usepackage{setspace}
\usepackage{bbm}
\usepackage{bm}
\usepackage{orcidlink}
\usepackage{soul}
\usepackage{array}

\newif\ifred
\redfalse % \redtrue

\ifred
\newcommand{\del}[1]{{\setstcolor{red}\st{#1}}}

\newcommand{\delB}[1]{{\setstcolor{red}\st{#1}}}
\else
\newcommand{\del}[1]{}

\newcommand{\delB}[1]{}

\fi

\makeatletter
\AddToHook{cmd/appendix/before}{\def\cref@section@alias{appendix}}
\makeatother
\crefformat{section}{Sec.~#2#1#3}
\crefmultiformat{section}{Secs.~#2#1#3}
    { and~#2#1#3}{, #2#1#3}{ and~#2#1#3}
\crefrangeformat{section}{Secs.~#3#1#4\,--\,#5#2#6}
\crefformat{subsection}{Sec.~#2#1#3}
\crefmultiformat{subsection}{Secs.~#2#1#3}
    { and~#2#1#3}{, #2#1#3}{ and~#2#1#3}
\crefrangeformat{subsection}{Secs.~#3#1#4\,--\,#5#2#6}
\crefformat{subsubsection}{Sec.~#2#1#3}
\crefmultiformat{subsubsection}{Secs.~#2#1#3}
    { and~#2#1#3}{, #2#1#3}{ and~#2#1#3}
\crefrangeformat{subsubsection}{Secs.~#3#1#4\,--\,#5#2#6}
\crefformat{figure}{Fig.~#2#1#3}
\crefmultiformat{figure}{Figs.~#2#1#3}
    { and~#2#1#3}{, #2#1#3}{ and~#2#1#3}
\crefrangeformat{figure}{Figs.~#3#1#4\,--\,#5#2#6}
\crefformat{table}{Table~#2#1#3}
\crefmultiformat{table}{Tables~#2#1#3}
    { and~#2#1#3}{, #2#1#3}{ and~#2#1#3}
\crefrangeformat{table}{Tables~#3#1#4\,--\,#5#2#6}
\crefformat{equation}{Eq.~(#2#1#3)}
\crefmultiformat{equation}{Eqs.~(#2#1#3)}{ and~(#2#1#3)}{, (#2#1#3)}{ and~(#2#1#3)}
\crefrangeformat{equation}{Eqs.~(#3#1#4)--(#5#2#6)}
\crefformat{appendix}{Appendix~#2#1#3}
\crefmultiformat{appendix}{Apps.~#2#1#3}
    { and~#2#1#3}{, #2#1#3}{ and~#2#1#3}
\crefrangeformat{appendix}{Apps.~#3#1#4\,--\,#5#2#6}

\newcommand{\smallcap}[1]{{\mathchoice
  {\scalebox{1.1}{$\scriptstyle \mathsf{#1}$}}
  {\scalebox{1.1}{$\scriptstyle \mathsf{#1}$}}
  {\scalebox{1.1}{$\scriptscriptstyle \mathsf{#1}$}}
  {\scalebox{1.0}{$\scriptscriptstyle \mathsf{#1}$}}
}}

\newcommand{\EV}{{\rm ev}}
\newcommand{\W}{\smallcap{W}}
\newcommand{\R}{\smallcap{R}}
\newcommand{\TM}{\smallcap{TM}}
\newcommand{\IR}{\smallcap{IR}}
\newcommand{\MG}{\smallcap{MG}}
\newcommand{\UV}{\smallcap{UV}}
\newcommand{\LMA}{\smallcap{LMA}}

\newcommand{\gfi}{\gamma_{\scalebox{0.9}{$\scriptstyle 5$}}}
\newcommand{\fm}{\mathrm{fm}}
\newcommand{\mev}{\mathrm{MeV}}
\newcommand{\gev}{\mathrm{GeV}}
\newcommand{\MSb}{\overline{\text{MS}}}
\newcommand{\proj}[2]{T_{\vb*{#1}}^{#2}\,}
\newcommand{\projc}[2]{T_{\vb*{#1},c}^{#2}\,}
\newcommand{\mybox}[1]{
\begin{tcolorbox}[colback=Blue!5!white, colframe=Blue!75!black, valign=center, leftrule=6pt, top=0pt]
#1 
\end{tcolorbox}
}

\newcommand{\gmu}{\gamma^{\mu}}

\newcommand{\cgfi}{\left[\mathcal{C}\gfi\right]}
\newcommand{\CyI}{Computation-based Science and Technology Research Center, The Cyprus Institute, 20 Kavafi Str., Nicosia 2121, Cyprus}
\newcommand{\UCY}{Department of Physics, University of Cyprus, P.O. Box 20537, 1678 Nicosia, Cyprus}
\newcommand{\ToV}{Dipartimento di Fisica \& INFN, Universit\`a di Roma ``Tor Vergata'', Via della Ricerca Scientifica 1, I-00133 Rome, Italy}
\newcommand{\RmT}{INFN, Sezione di Roma Tre, Via della Vasca Navale 84, I-00146 Rome, Italy}

\begin{document}

\title{Performance of Low Mode Averaging on Twisted-Mass Fermion Ensembles at the physical pion mass point}

\author{C.~Alexandrou\,\orcidlink{0000-0001-9136-3621}}
\affiliation{\CyI}
\affiliation{\UCY}
\author{S.~Bacchio\,\orcidlink{0000-0002-5532-450X}}
\affiliation{\CyI}
\author{A.~Evangelista\,\orcidlink{0000-0002-3320-3176}}
\affiliation{\UCY}
\author{R.~Frezzotti\,\orcidlink{0000-0001-5746-0065}}
\author{F. Margari\,\orcidlink{0000-0003-2155-7679}} 
\affiliation{\ToV}
\author{F.~Sanfilippo\,\orcidlink{0000-0002-1333-745X}}
\affiliation{\RmT}
\author{C.~Schneider}
\affiliation{\UCY}
\date{June 26, 2026}

\begin{abstract}
We study the performance of low-mode averaging (LMA) on twisted-mass fermion ensembles at near-physical quark masses, assessing both its theoretical framework and practical cost-effectiveness in modern lattice QCD.
In particular, we present a numerical study of light-quark meson and baryon observables. For mesons, we analyse two-point functions, including the vector–vector correlator relevant for the hadronic vacuum polarisation contribution to the muon anomalous magnetic moment, comparing two implementations of LMA: an exact approach based on explicit low modes and an approximate, high-statistics variant using multigrid techniques. For baryons, we restrict to the exact approach and study both two- and three-point functions, quantifying the resulting noise and cost reductions at large Euclidean times.
In addition, we compute the eigenvalue density of the massless Wilson operator and determine the renormalised chiral condensate via the Banks–Casher relation, obtaining $\sqrt[3]{\Sigma_{\R}}=269.6(4.6)~\mev$ for $N_f{=}2{+}1{+}1$ isospin-symmetric QCD at a scale $2~\gev$ in the $\overline{\mathrm{MS}}$ scheme, with an uncertainty dominated by the chiral extrapolation. Additionally, from the pion-mass dependence of $\Sigma_{\R}$, we extract the scale-independent low-energy constant $\bar{h}_1=5.2(1.2)$.
\end{abstract}

\maketitle

%%%%%%%%%%%%%%%%%%%%%%%%%%%%%%%%%%%%%%%%%%%%%%%%%%
\setlength{\parskip}{6pt}
\setlength{\parindent}{0pt}
\addtolength{\jot}{4pt} 
%%%%%%%%%%%%%%%%%%%%%%%%%%%%%%%%%%%%%%%%%%%%%%%%%
\section{Introduction}

As computational techniques, the inclusion of more complex observables, and higher precision in lattice QCD continue to advance, reducing statistical uncertainties has become essential for meaningful comparisons with experimental data. Computing correlation functions at large Euclidean time is central to these efforts, both for isolating hadronic ground states, such as in nucleon matrix elements~\cite{Walker-Loud:2019cif,Bali:2023sdi,Djukanovic:2024krw,Alexandrou:2024ozj,FlavourLatticeAveragingGroupFLAG:2024oxs,Jang:2023zts,Djukanovic:2022wru,Alexandrou:2023qbg,Djukanovic:2023beb,Tsuji:2023llh,Alexandrou:2025vto}, and for applications of spectral reconstruction methods~\cite{ExtendedTwistedMassCollaborationETMC:2022sta,Evangelista:2023fmt,ExtendedTwistedMass:2025tpc,ExtendedTwistedMass:2024myu,DeSantis:2025qbb}, including those used to determine the long-distance hadronic vacuum polarization contribution to the muon anomalous magnetic moment~\cite{Borsanyi:2020mff,Djukanovic:2024cmq,Boccaletti:2024guq,FermilabLatticeHPQCD:2024ppc,Moningi:2025dlf,ETMC:2026mpp,ExtendedTwistedMass:2026nbv}. In both settings, state-of-the-art simulations at physical quark masses demand reliable control of correlators at source–sink time separations exceeding 2~fm or  even 3~fm. The exponential degradation of the signal-to-noise ratio at such Euclidean time separations, in particular for baryon observables, renders the long-distance regime one of the most critical and computationally demanding aspects of modern lattice calculations. A key strategy to address this challenge is low-mode averaging (LMA), which targets the infrared (IR) sector of the Dirac spectrum governing long-distance physics and enhances its precision by explicitly isolating its contribution.

Low-mode averaging has a long history in lattice QCD. Its first applications in the early 2000s~\cite{Neff:2001zr,Giusti:2004yp,DeGrand:2004qw,DeGrand:2005vb,Giusti:2005sx} identified it as a promising variance-reduction technique, although its impact was then limited by the fact that lattice QCD simulations used relatively small lattice volumes and unphysical quark masses. Subsequently, LMA was combined with the truncated solver method (TSM)~\cite{Bali:2009hu} to form all-mode averaging (AMA)~\cite{Blum:2012uh,Shintani:2014vja}, thanks to their complementarity: LMA improves the IR sector of the Dirac spectrum, while TSM efficiently suppresses fluctuations originating from ultraviolet (UV) modes. Owing to its conceptual simplicity and ease of implementation,  TSM without LMA quickly became widely adopted. It is important to note, however, that TSM alone predominantly improves short-distance observables and has little impact at large Euclidean time separations, as its action is largely confined to the UV sector~\cite{FermilabLatticeHPQCD:2024ppc}.

The landscape shifted markedly with the advent of simulations performed near the physical pion mass, hereafter referred to as the \textit{physical point}. In this regime, as the light quark mass is reduced, the condition number of the Dirac operator grows rapidly, rendering inversions with standard Krylov solvers prohibitively expensive. This development revived interest in LMA and, more broadly, in deflation techniques~\cite{Luscher:2007se,xQCD:2010pnl}. These methods also exploit low-lying eigenmodes to accelerate solver convergence~\cite{Luscher:2007es}; however, their substantial setup costs—due to the computation of exact eigenvectors, which scales poorly with the volume—limit their applicability. A major turning point was the successful deployment of multigrid solvers for Wilson-type fermions~\cite{Brannick:2007ue,Babich:2010qb,Frommer:2013fsa,Frommer:2013kla}, including variants for twisted mass~\cite{Alexandrou:2016izb,Alexandrou:2018wiv}, overlap~\cite{Brannick:2014vda}, domain-walls~\cite{Brower:2020xmc}, and staggered~\cite{Brower:2018ymy} fermions. Multigrid algorithms feature substantially cheaper setup costs and can reduce inversion times by up to two orders of magnitude compared to standard Krylov methods. This breakthrough proved transformative for simulations at physical quark masses~\cite{Bacchio:2017pcp,Finkenrath:2017cau,Alexandrou:2018egz}, effectively displacing exact-deflation solvers and TSM techniques. With Dirac operator inversions no longer the dominant component of the overall computational cost, the benefits of loose stopping criteria within TSM largely vanished. Moreover, TSM does not integrate well with the MG setup, as the solve time depends highly nonlinearly on the target residual.

The successful deployment of multigrid algorithms brings back the focus on LMA  and motivates  the present work. In the current computational regime, characterised by light quark masses, affordable inversions, and a growing demand for precision at long Euclidean time separations, LMA alone re-emerges as a particularly well-suited noise-reduction technique. By directly targeting the IR modes responsible for long-distance physics, it is especially relevant for modern high-precision lattice QCD calculations. In this work, we consider two alternative realisations of LMA: The first approach is the original formulation based on exact all-to-all treatments obtained by explicitly computing a subset of low-lying eigenvectors of the Dirac operator; this approach allows the IR contribution to correlation functions to be evaluated exactly on each gauge configuration, fully saturating the associated statistical fluctuations. We refer to this method as exact LMA (exLMA). The second approach, explored more recently, employs approximate but high-statistics implementations based on multigrid techniques~\cite{Frommer:2021tqd,Gruber:2024cos,Frommer:2025wew}, which we denote as multigrid LMA (mgLMA). The aim of this paper is to present and discuss the optimisation and performance of these low-mode averaging strategies using modern physical-point twisted mass ensembles, drawing on our practical experience.

The remainder of this work is organised as follows. In \cref{sec:LMA}, we introduce the LMA procedure and discuss a key feature, namely, in commonly used correlation-function setups, LMA improves only the IR contribution, leaving the UV sector and mixed IR--UV terms unaffected. This has important implications for both performance and optimisation. In \cref{sec:scaling}, we introduce the twisted-mass operator and its properties, and discuss the expected scaling of LMA-based approaches, identifying the physical volume and the quark mass as the dominant dependences. Combining these observations with a conceptual understanding of LMA, we argue that LMA is effective primarily on observables involving only light-quark propagators. Numerical results follow: In \cref{sec:mesons}, we present results for exLMA applied to quark-bilinear correlation functions on several physical-point ensembles, in \cref{sec:mesons_mg}, we present the corresponding results obtained with mgLMA, which should become advantageous for  large physical volumes, and in \cref{sec:baryon_results}, we present results for baryon correlation functions using exLMA, quantifying the improvements achieved in both two- and three-point functions  and highlighting the observable-dependent nature of the overall gain. In \cref{sec:conclusion}, we summarise and conclude.
\section{LMA improvement of IR part in common applications}\label{sec:LMA}

Let $P_{\IR}$ be a projector onto an infrared (IR) subspace of the Dirac operator $D$, and define the complementary ultraviolet (UV) projector as $P_{\UV}\equiv \mathds{1} - P_{\IR}$. This induces the decomposition of the all-to-all propagator $S$ as
\begin{equation} 
   S^{\IR}\equiv D^{-1} P_{\IR}\,, \qquad S^{\UV}\equiv D^{-1} P_{\UV} \qquad{\Longrightarrow}\qquad S \equiv D^{-1} =  S^{\IR} + S^{\UV}\,.
\end{equation}
The role of $P_{\IR}$ is to enable an exact evaluation of the IR contribution $S^{\IR}$, while the remaining UV component $S^{\UV}$ is estimated stochastically. To this end, we introduce a set of stochastic sources $\eta$ with support on a subset of lattice spacetime sites $\{x\}_\eta$ and internal degrees of freedom $\{\alpha\}_\eta$, such that any source $\eta$ satisfies
\begin{equation}\label{eq:stoch_sources_loc}
    \eta(x,\alpha)\, \eta(x,\alpha)^\star = \begin{cases} 
    1 & \text{if}\quad x\in\{x\}_\eta\qq{and}\alpha\in\{\alpha\}_\eta\\
    0 & \text{if}\quad x\not\in\{x\}_\eta\qq{~or~}\alpha\not\in\{\alpha\}_\eta
    \end{cases}\,.
\end{equation}
These sources then also satisfy
\begin{equation}
H_\eta \equiv \eta\eta^\dagger\qquad\text{and}\qquad\lim_{N_\eta\to\infty}\frac{1}{N_\eta}\sum_{\eta} H_\eta = \mathds{1}_\eta\qquad\text{and}\qquad
H_\eta H_\eta = V_\eta H_\eta
\,,
\label{eq:stoch_sources}
\end{equation}
where $V_\eta=\eta^\dagger\eta={\rm const.}$ is the volume of the source support, and $\mathds{1}_\eta$ denotes the resolution of the identity within that support. In practice, we consider two types of sources: time- and spin-diluted $\mathbb{Z}_n$-noise stochastic sources, which are nonzero on the spatial components and have $V_\eta=(L/a)^3N_c$ nonzero entries, with $N_c=3$ the number of colours; and point sources, for which there is no stochasticity, $V_\eta=1$, and the identity is recovered directly for a single spacetime coordinate $x$ and a single spin-colour component.

An approximate all-to-all propagator is then defined as
\begin{align}
S_{\eta}^{\LMA} \equiv S^{\IR}\, \mathds{1}_\eta + S_{\eta}^{\UV} 
\qquad\text{with}\qquad 
S_{\eta}^{\UV}\equiv S^{\UV}  H_\eta\qq{and} \lim_{N_\eta\to\infty}\frac{1}{N_\eta}\sum_{\eta} S_\eta^{\LMA} = S\, \mathds{1}_\eta,
\label{eq:prop_defl}
\end{align}
which corresponds to the standard exact low-mode averaging (exLMA) construction~\cite{Giusti:2004yp}. In this approach, the IR contribution is known exactly all-to-all, whereas the UV part is entirely determined by the stochastic estimator, to be 
compared with the standard stochastic approach, where the full propagator is computed stochastically on the same set of sources
\begin{equation}
    S_{\eta} \equiv SH_{\eta}= S_{\eta}^{\IR} + S_{\eta}^{\UV}\qq{with}  
S_{\eta}^{\IR} \equiv  S^{\IR}H_{\eta}\,.
\label{eq:prop_stoc}
\end{equation}
Using the definitions above, an alternative but numerically identical approach to compute the LMA propagator can be defined as
\begin{equation}
S_{\eta}^{\LMA} = S_\eta  + (S^{\IR}\,\mathds{1}_\eta- S_{\eta}^{\IR})\,.
\label{eq:impr_prop}
\end{equation}
Both here and in~\cref{eq:prop_defl}, $S^{\IR}$ is computed exactly; however, in~\cref{eq:prop_defl} the stochastic source is first projected onto the UV subspace and then propagated, whereas in~\cref{eq:impr_prop} the full stochastic propagator is computed and the stochastic IR contribution is subsequently subtracted and replaced by its non-noisy counterpart on the support of $\eta$, $S^{\IR}\, \mathds{1}_\eta$.

The procedural distinction between these approaches becomes more pronounced at the level of correlation functions. Consider a generic correlation function $C$, obtained by contracting and reducing several propagators. 
We define five realisations of $C$ using the propagators introduced above:
\begin{description}[font=\normalfont\itshape]
    \item[$C_\eta(t)$\,] The reference correlation function obtained by contracting standard stochastic propagators $S_\eta$.
    \item[$C_\eta^{\IR}(t)$ {\rm and} $C_\eta^{\UV}(t)$\,] The same correlation function, whereas the stochastic propagators are restricted to their IR or UV parts, $S_\eta^{\IR}$ or $S_\eta^{\UV}$, with contractions performed identically to $C_\eta(t)$.
    \item[$C^{\IR}(t)$\,] A conceptually and computationally distinct quantity. Here, the exact IR propagator $S^{\IR}$ is employed to construct the full all-to-all correlation function, eliminating stochastic noise and any restriction to a lattice subset, such that the variance is solely due to gauge fluctuations.
    \item[$C_{\eta}^{\LMA}(t)$\,] The correlation function constructed from $S_{\eta}^{\LMA}$ as defined in~\cref{eq:prop_defl}. Upon contraction, it generates $2^n$ terms, where $n$ is the number of quark propagators involved in the contraction, two of which correspond to the decomposition of $S_{\eta}^{\LMA}$ into two pieces, $S^{\IR}$ and $S_{\eta}^{\UV}$. Therefore, among the resulting terms in the correlator, only two correspond to the purely IR and purely UV contributions, $C^{\IR}(t)$ and $C_\eta^{\UV}(t)$, defined above; the remaining $2^n{-}2$ terms mix IR and UV components and need to be computed individually.
\end{description}
A central statement of this work is that, although $C_{\eta}^{\LMA}(t)$ contains several mixed IR--UV contributions in which exact and stochastic components appear simultaneously, in commonly used computational setups LMA improves only the purely IR contribution $C^{\IR}(t)$. All remaining terms continue to be limited by stochastic sources, even when the IR contribution is treated exactly in every term in which it appears.
Specifically, we show that the following equality holds.
\mybox{
\vspace{-5pt}
\begin{equation}
C^{\LMA}_\eta(t) =  C_{\eta}(t) - C_{\eta}^{\IR}(t) + C^{\IR}(t) 
\qq{in commonly used  computational setups.}
\label{eq:corr_defl}
\end{equation}
\vspace{-15pt}
}

The proof is presented in \cref{app:proof}. For disconnected loops, the relation in \cref{eq:corr_defl} follows directly from~\cref{eq:impr_prop}, and we will demonstrate it explicitly for spin-diluted quark-bilinear (meson) two-point functions and point-source baryon two- and three-point functions. In these cases, the specific properties of the stochastic sources employed are used to establish the relation exactly. 

\subsection{Discussion on~\texorpdfstring{\cref{eq:corr_defl}}{Eq.~(7)}}
\label{sec:LMA_remarks}

A direct contraction of propagators built from $S_{\eta}^{\LMA}$ as in~\cref{eq:prop_defl} produces purely IR and purely UV contributions, along with several mixed IR–UV terms. In practice, this typically requires dedicated contraction kernels and considerable implementation effort. By contrast, the right-hand side (r.h.s) of~\cref{eq:corr_defl} allows for a much simpler procedure: compute the full stochastic correlator, subtract its stochastic IR component, and replace it with an improved—ideally exact—estimate. Moreover, even in lattice setups where the two sides of \cref{eq:corr_defl} are not identical, the construction in the r.h.s. still provides a valid and more efficient realisation of LMA, with double counting of the IR contribution properly removed.

We note that both approaches, left- and right-hand sides of~\cref{eq:corr_defl}, have been employed in the literature. In fact, the formulation based on the full $S_{\eta}^{\LMA}$ dates back to the original LMA works~\cite{Giusti:2004yp}, while the second strategy was introduced in Ref.~\cite{xQCD:2010pnl} and in the context of combining LMA with AMA~\cite{Blum:2012uh}, as it naturally enables independent improvement of the IR contribution via LMA and the UV contribution via TSM. To our knowledge, however, it has not been emphasized that, in certain commonly used computational setups, the two procedures are in fact exactly identical\footnote{Actually, in parts of the literature, the full correlation function $C_{\eta}^{\LMA}$ is sometimes regarded as superior because the mixed terms are assumed to yield additional improvement since their IR part is computed exactly. This may indeed happen in setups where \cref{eq:corr_defl} does not hold, e.g. as discussed in Ref.~\cite{Giusti:2005sx} for baryons computed with stochastic sources. For the cases considered here, however, we demonstrate that the two constructions are identical.
}. Recognising this equivalence or, more generally, adopting the second strategy in place of the first, has several important consequences:

\begin{description}[font=\itshape]
\item[Noise reduction] When \cref{eq:corr_defl} holds, it makes explicit that LMA improves only the purely IR contribution $C^{\IR}(t)$. All UV contributions, as well as any IR--UV mixing, are governed by stochastic noise. This observation is crucial for identifying the class of observables, Euclidean-time regions, and quark-mass regimes in which LMA can deliver significant variance reduction and meaningful performance gains. In particular, noise reduction is expected only where $C^{\IR}(t)$ dominates the correlator. In general, this is not straightforward to identify, as the extent of this region depends sensitively on the number of deflated modes, as we will demonstrate with our numerical results.

\item[Reduced contraction costs] Evaluating the r.h.s. of \cref{eq:corr_defl} is considerably easier and more efficient than computing $C^{\LMA}_\eta$ through explicitly expanding it in terms of the IR and UV contractions. The correlators $C_{\eta}$ and $C_{\eta}^{\IR}$ are expressed in terms of ordinary stochastic contractions differing only in the contracted propagator, while $C^{\IR}$ is the only genuinely new object. Effectively, the $2^n$ IR-UV contraction terms---with $n$ being the number of propagators involved in the contraction---are replaced by only three, since all interference contributions are avoided.

\item[Post-production LMA] If the stochastic sources $\eta$ are reproducible---i.e., the same exact stochastic source can be generated in independent runs by e.g. storing the point-source coordinates or the random seeds---then stochastic correlators can be improved \textit{a posteriori}. Suppose $C_{\eta}$ has already been computed using the full stochastic propagator, and one subsequently wishes to apply LMA. Then $C_{\eta}^{\IR}$ and $C^{\IR}$ can be evaluated independently, without recomputing the full propagator, but using only the exact IR information. 

\item[Optimised production with multigrid solvers] The separation implied by \cref{eq:corr_defl} naturally suggests an optimised computational strategy, particularly at physical quark masses. The evaluation of $C_{\eta}$ requires Dirac matrix inversions, for which modern multigrid solvers are optimal but exhibit poor strong scaling, and are therefore best run on as few nodes as possible. By contrast, the computation of the IR sector involves a large number of exact eigenvectors; while memory-intensive, it scales efficiently with the number of nodes. The correlators $C_{\eta}^{\IR}$ and $C^{\IR}$ can thus be computed in a separate stage on large node counts, decoupled from the full Dirac operator inversions, provided that the stochastic sources $\eta$ are reproducible. On the other hand, in the direct $C^{\LMA}_\eta$ construction, the UV projection forces inversions and IR modes to be treated simultaneously within a single workflow, often requiring execution on more nodes than optimal due to the increased memory demands.

\item[Solver acceleration trade-off] When standard Krylov solvers are used, treating the IR and UV sectors simultaneously, instead, can provide a benefit: the low modes can be deflated from the solver, thereby accelerating the inversions associated with the remaining UV modes. This advantage has been discussed in Refs.~\cite{Giusti:2004yp,Borsanyi:2020mff,Djukanovic:2024cmq} and is also consistent with the use of deflation techniques in iterative solvers~\cite{Luscher:2007se,xQCD:2010pnl}. For the setup considered here, however, where multigrid solvers are employed for inversions at the physical quark mass, this benefit is largely absent, and we observe no significant improvement in the inversion cost.
\end{description}

In conclusion, if exact deflation is not also required to accelerate the calculation of propagators, then we advocate implementing LMA as defined by the right-hand side of \cref{eq:corr_defl}, or better, its generalisation in~\cref{eq:corr_defl_var} of the conclusions, and in general ensuring that correlation functions are reproducible so that LMA can be applied \textit{a posteriori} when needed.

\section{LMA Dependence on Quark Mass and Physical Volume}\label{sec:scaling}

Another important aspect of LMA concerns its computational cost and, in particular, its scaling with the quark mass and the physical volume. The scaling of the Dirac spectrum with the volume is well understood: in the IR region, the number of modes scales with the physical volume  $V$, rather than with the number of lattice points $V/a^4$~\cite{Banks:1979yr, Leutwyler:1992yt, Luscher:2007se, Giusti:2004yp, Giusti:2008vb}. 
By contrast, the scaling with the quark mass is less well understood and may depend on the discretisation of the Dirac operator used in the analysis. In several works~\cite{Neff:2001zr,Giusti:2004yp,DeGrand:2004qw,Giusti:2005sx}, it has been observed that the computational gain from using LMA decreases as the quark mass increases, although this has typically been reported as an empirical finding rather than derived from theoretical considerations. Here, we show that the twisted-mass fermion discretisation provides a particularly transparent framework to understand this mechanism and yields a simple theoretical explanation. In particular, we demonstrate that maintaining constant the LMA induced noise reduction as the quark mass increases requires proportionally increasing the number of deflated modes.

\subsection{Twisted Mass fermions and eigen-decomposition}\label{sec:deflation}

We first briefly summarise the key properties of Wilson twisted-mass fermions that are relevant for discussing their spectrum and deriving the mass dependence of LMA. In the twisted basis, \textit{i.e.}, the one in which the mass parameter $\mu$ couples to a pseudoscalar density, the \textit{maximally-twisted} Wilson--Dirac operator with a quark mass $\mu$ reads as\footnote{
For simplicity in the following discussion, we set the Wilson parameter to $r=1$; the case $r=-1$ is completely equivalent.
}
\begin{equation}\label{eq:TMDirac}
D(\mu) = D_{\W} + i\gfi\,\mu \qq{with} D_{\W} = \left[\gamma\cdot \Tilde{\nabla} + W_{\mathrm{cr}}\right] \, ,
\end{equation}
where $D_{\W}$ is the massless $\gfi$-Hermitian Wilson--Dirac operator, consisting of the hopping term $\gamma{\cdot} \Tilde{\nabla}$ and a diagonal mass term $W_{\rm cr}$ tuned to its critical value, \textit{i.e.} to vanishing residual mass, with an optional clover term included. The associated Hermitian operator
\begin{equation}
Q_{\W} \equiv D_{\W} \gfi = Q_{\W}^\dagger
\end{equation}
has eigenpairs $\left(\Lambda_j, v_j\right)$ satisfying
\begin{equation}
Q_{\W} v_j = \Lambda_j v_j \qq{with} \Lambda_j \in \mathbb{R}\, .
\end{equation}
It follows that the equivalent
twisted-mass Dirac operator reads as
\begin{equation}
Q(\mu) \equiv D(\mu) \gfi = Q_{\W} + i\mu
\end{equation}
and since the mass term enters multiplied by the identity operator, it shares the same eigenvectors as $Q_\W$, while its eigenvalues are shifted into the complex plane by $\mu$,
\begin{equation}
Q(\mu)\, v_j = \lambda_j(\mu)\, v_j \qq{with} \lambda_j(\mu) \equiv \Lambda_j + i\mu \, .
\end{equation}
These vectors are also eigenmodes of the positive-definite squared operator,
\begin{equation}
Q(\mu)^\dagger Q(\mu)\, v_j = Q(-\mu)Q(\mu)\, v_j
= \left(\Lambda_j^2 + \mu^2\right) v_j
\qquad\Longrightarrow\qquad
|\lambda_j(\mu)| \ge |\mu|{\rm ~~for~any~}j\,.
\end{equation}
The above relation highlights a central property of the twisted-mass operator, namely, the parameter $\mu$ provides an infrared cutoff that prevents the appearance of near-zero modes~\cite{Frezzotti:2000nk}. As a consequence, at non-zero mass, the squared twisted operator is always positive definite, with condition number $\kappa \le \lambda_{\rm max}^2/\mu^2$, and its inverse is well defined.

Another important property of the twisted mass operator is that its eigenvectors are independent of the value of $\mu$, being entirely determined by the massless Hermitian Wilson operator $Q_\W$. 
As a result, the cost of the eigensolver does not depend on the quark mass, and the same set of deflated modes can be employed for any value of the twisted mass parameter $\mu$. This contrasts with, e.g., the Wilson operator and other discretisations, where the eigenvectors must be recomputed for each quark mass. The inverse of the twisted mass Dirac operator then follows directly from the spectral decomposition of $Q_\W$ as
\begin{equation}\label{eq:S_IR}
    Q_\W = D_\W \gfi = \sum_{j=0}^{N_{\rm ev}} \Lambda_j\,v_j v^\dagger_j \qquad\Longrightarrow\qquad S(\mu) =  D^{-1}(\mu) = \sum_{j=0}^{N_{\rm ev}} \frac{1}{\lambda_j(\mu)}\,\gfi v_j v^\dagger_j \qquad\Longrightarrow\qquad S^\IR =  \sum_{j=0}^{N^\IR_{\rm ev}} \frac{1}{\lambda_j}\,\gfi v_j v^\dagger_j\,,
\end{equation}
where the modes are ordered by magnitude $|\Lambda_j|$ and the IR propagator, $S^\IR$, is obtained by restricting the sum to the first $N_{\rm ev}^\IR$ modes.

\subsection{Spectral density}\label{sec:spectrum}

We now examine the spectral densities of the massless Wilson Dirac operator, as well as those of the twisted-mass operator at the light quark mass. It has been shown in Refs.~\cite{Banks:1979yr, Leutwyler:1992yt, Luscher:2007se, Giusti:2004yp, Giusti:2008vb} that, on sufficiently large-volume ensembles—i.e., where the spectrum is dense and finite-volume effects are negligible—the number of eigenvalues of the massless Wilson Dirac operator whose norms lie in the IR interval $|\Lambda_j|\in\left[\Lambda,\Lambda{+}\delta \Lambda\right]$ is proportional to both the interval width $\delta\Lambda$ and the physical volume $V$. Therefore, the spectral density of the eigenvalue norms per unit volume and per unit eigenvalue interval is approximately constant in the IR region. We thus define
\begin{equation}
f_{\W}\left(\Lambda, \delta \Lambda; V \right) \equiv 
\frac{1}{V\,\delta \Lambda} \int_{\Lambda}^{\Lambda + \delta\Lambda}\!\!\!\!\rho_{\!{\textstyle\mathstrut}{\W}}(\Tilde{\Lambda};V) \dd{\Tilde{\Lambda}}
\quad=\quad 
\Bar{\rho}_{\W} + df_{\W}(\Lambda,\delta \Lambda; V)
\quad\simeq\quad \Bar{\rho}_{\W} \,,
\label{eq:banks}
\end{equation}
where $\rho_{\W}(\Lambda)$ denotes the eigenvalue density as a function of the eigenvalue norm $\Lambda$, while $\Bar{\rho}_{\W}$ is its leading constant value related to the chiral condensate $\Sigma$ in the chiral limit~\cite{Banks:1979yr}, see \cref{sec:Banks}, and $df_{\W}$ is a smooth function, providing subleading corrections across physically relevant IR eigenvalue intervals. Similarly, for the twisted mass operator we define 
\begin{equation}\label{eq:tm_density}
f_{\TM}\left(\lambda, \delta \lambda; V \right) \equiv \frac{1}{V\,\delta \lambda}
\int_{\lambda}^{\lambda + \delta\lambda}\!\!\!\!\rho_{\!{\textstyle\mathstrut}{\TM}}(\Tilde{\lambda};V) \dd{\Tilde{\lambda}}\,,
\end{equation}
which is no longer approximately constant, as the eigenvalue density is modified by the twisted mass $\mu$, as discussed below.

\begin{figure}[htb]
\centering
\includegraphics[width=0.49\textwidth]{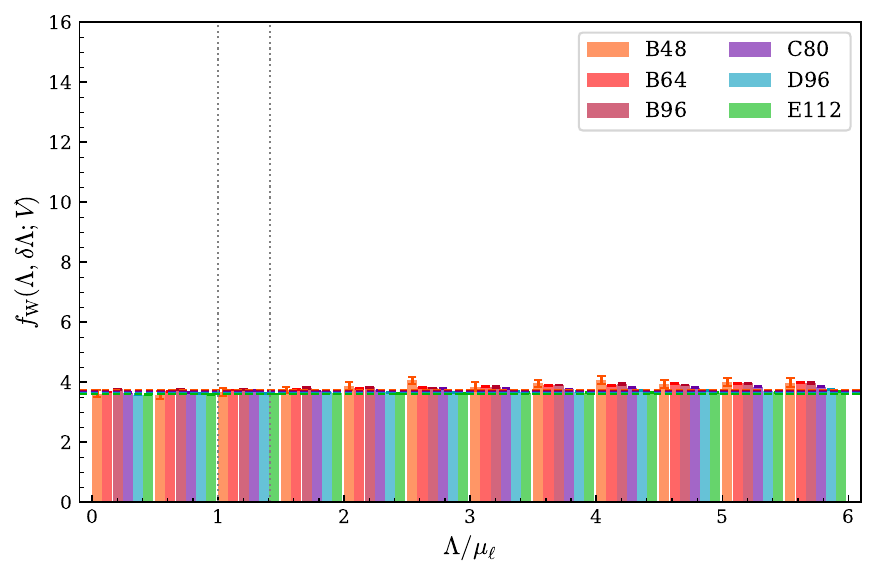}
\includegraphics[width=0.49\textwidth]{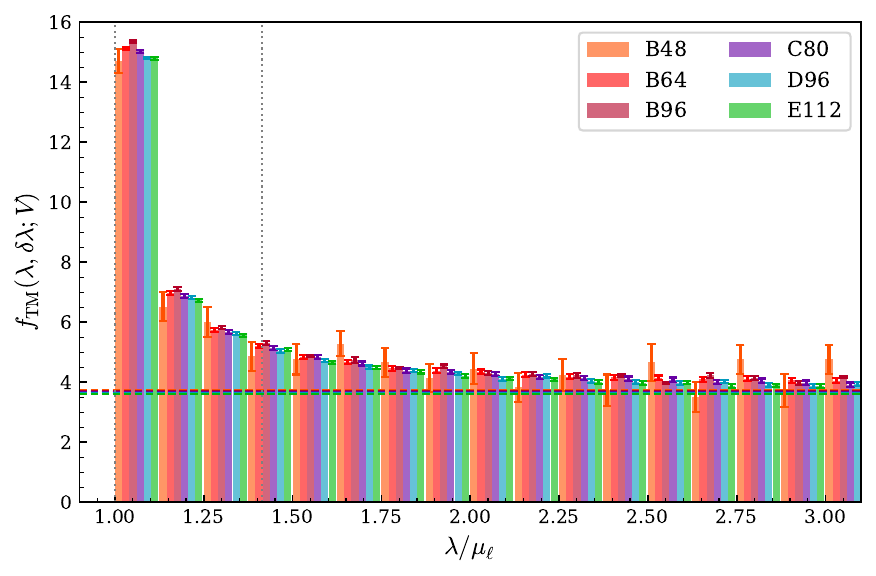}
\caption{The left and right panels show $f_{\W}(\Lambda,\delta\Lambda;V)$ and $f_{\TM}(\lambda,\delta\lambda;V)$, respectively, defined in~\cref{eq:banks} and in~\cref{eq:tm_density}. These quantities represent the normalised density of measured eigenvalues whose norm lies in the interval $[\Lambda,\Lambda+\delta\Lambda]$ or $[\lambda,\lambda+\delta\lambda]$, for the massless Wilson or twisted-mass operator, respectively. The results are shown for the six near-physical-pion-mass ensembles described in \cref{tab:spectrum}. Each ensemble is indicated by a different colour (with red shades corresponding to the same lattice spacing but increasing physical volume), as specified in the legend. The left panel shows the distribution for the massless Wilson case with $\delta\Lambda/\mu_\ell = 0.5$, while the right panel corresponds to the twisted-mass case with $\delta\lambda(\mu_\ell)/\mu_\ell = 0.125$. The horizontal dashed lines indicate the fitted values of $\bar{\rho}_{\W}$ for the respective ensembles (see \cref{tab:spectrum} for details). The vertical grey dashed lines mark the interval $[\mu_\ell,\sqrt{2}\mu_\ell]$ in both panels, which in the case of $f_{\TM}$ in the right panel corresponds to the region in which eigenvalues of $Q_\W$ below $\mu_\ell$ are accumulated, as defined in~\cref{eq:tm_evs_accum}.
}
\label{fig:hist}
\end{figure}
We study the eigenvalue densities of both operators using six near-physical-pion-mass $N_f{=}2+1+1$ ensembles generated by the Extended Twisted Mass Collaboration (ETMC) at four different lattice spacings and three different physical volumes (see Ref.~\cite{ExtendedTwistedMass:2026nbv} for the details about the exact matching of simulations to the FLAG/Edinburg definition of isoQCD). Our results are summarised in~\cref{tab:spectrum} and shown  in~\cref{fig:hist}. In the left panel, we show the dependence of the normalised density as a function of the eigenvalue norm $\Lambda$.
For our ensembles, we determine $\Bar{\rho}_{\W}$ by performing a linear fit of $f_{\W}\left(\Lambda, \delta \Lambda; V \right)$ in the $\Lambda/\mu_\ell$ interval $\left[1.5, 5 \right]$ with $\mu_\ell$ the simulated light quark mass and $\delta \Lambda/\mu_\ell = 0.5$. 
Furthermore, we assess the stability of this result by varying the fit range, including or excluding one bin at either end of the aforementioned interval. The maximum deviation from the result obtained with the reference fit range is then added in quadrature as a systematic uncertainty.
We observe that its values have a mild dependence on the lattice spacing and are approximately constant with respect to the volume.

\begin{figure}[htb]
\centering
\includegraphics[width=0.8\textwidth]{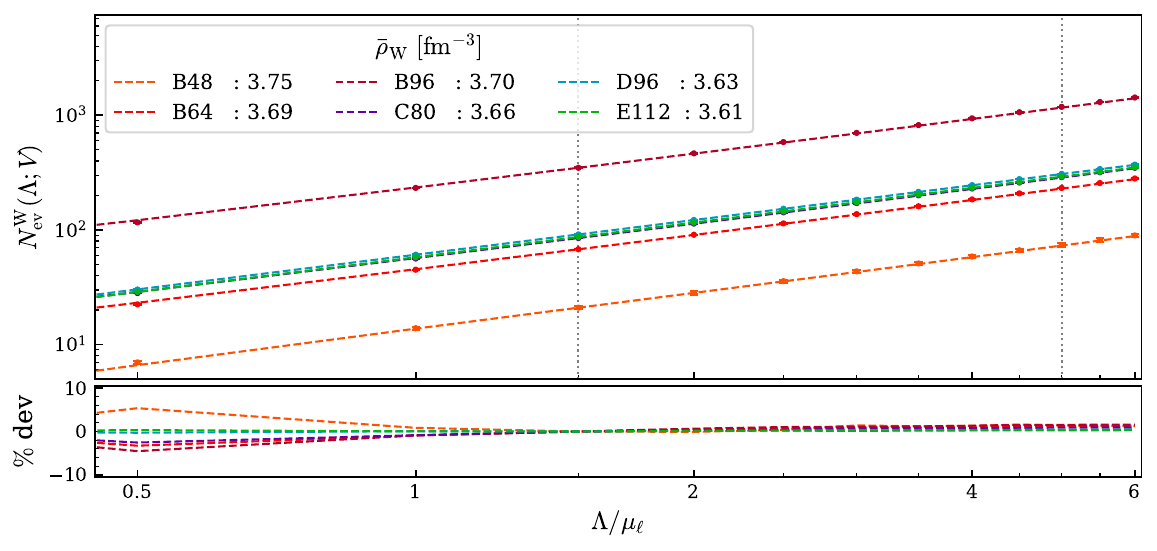}
\caption{
\textit{Top panel}: For each ensemble, the number of eigenvalues $N_{\mathrm{ev}}^{\W}$ in the range $\left[0, \Lambda \right]$ measured according to the left-hand side of \cref{eq:Nwilson}. The grey vertical dotted lines represent the fit interval $\Lambda/\mu\in[1.5,5]$ within which $\Bar{\rho}_{\W}$ is extracted. The dashed lines represent the linear behaviour expected according to the right-hand side  of \cref{eq:Nwilson_1}. \textit{Bottom panel}: the percentage deviation of the measured values from this prediction.
}
\label{fig:lamscale}
\end{figure}
The number of eigenvalues up to a threshold $\Lambda$ is then given by
\begin{equation}
N_{\rm ev}^{\W}(\Lambda; V) \equiv \int_{0}^{\Lambda}\!\!\!\!\rho_{\!{\textstyle\mathstrut}{\W}}(\Tilde{\Lambda};V) \dd{\Tilde{\Lambda}} \quad=\quad f_{\W}(0, \Lambda; V)\,V\,\Lambda \quad\simeq\quad \bar{\rho}_{\W}\,\Lambda\,V\,.
\label{eq:Nwilson}
\end{equation}
where, in the last (approximated) step of~\cref{eq:Nwilson}, we neglected the very mild higher-order dependence in $\Lambda$ of the spectral density. We show this expectation in~\cref{fig:lamscale} for the aforementioned ETMC ensembles. Due to the scarcity of modes in the near-zero region, the r.h.s. of~\cref{eq:Nwilson} does not adequately describe the measured $N_{\rm ev}^{\W}(\Lambda; V)$ at small $\Lambda$. In~\cref{fig:lamscale}, we then depict
\begin{equation}
N_{\rm ev}^{\W}(\Lambda; V) \simeq  N_{\rm ev}^{\W}(\Lambda_{\rm min}; V) + \bar{\rho}_{\W}\,V (\Lambda - \Lambda_{\rm min})\,,
\label{eq:Nwilson_1}
\end{equation}
where $\Lambda_{\rm min}=1.5\mu_{\ell}$ defines the lower bound of the fit region of $f_\W$. The bottom panel shows the relative measured--predicted deviation, which remains within $10\%$ in the low-lying part of the spectrum.
\begin{table}
\centering
\setlength{\tabcolsep}{0.35em}
\renewcommand{\arraystretch}{1.4}
\aboverulesep=0ex % Solution part 1 of 3
\belowrulesep=0ex % Solution part 1 of 3
\small
\begin{tabular}{l|ccccc|ccc|cc}
\toprule  \rule{0pt}{1.1EM}
& $V/a^4$ &  $a~[\fm]$ & $L~[\fm]$ & $\xi_\pi{\cdot} 10^3$ & $a\mu_\ell$ & $\Bar{\rho}_{\W}~[\fm^{-3}]$ & $\sqrt[3]{\frac{\pi}{2}Z_{P}\Bar{\rho}_\W}$~[MeV] & $\sqrt[3]{\Sigma_{\R}}$~[MeV] & $N^\TM_{\rm ev}(\sqrt{2}\mu_{\ell})$  & $\bar{\rho}_{\W}\,\mu_\ell \, V$ \\
\midrule  
B48   & $\phantom{1}48^3 {\times} \phantom{1}96$ & 0.07968(13) & 3.82 & 7.251(05) & 0.00072 & $\sim$\,3.7 & $\sim$\,276  & $\sim$\,267  &  13 &  14 \\
B64   & $\phantom{1}64^3 {\times} 128$           & 0.07968(13) & 5.10 & 7.251(05) & 0.00072 & 3.686(18)     & 275.51(84)   & 266.6(3.2) &  44 &  45 \\
B96   & $\phantom{1}96^3 {\times} 192$           & 0.07968(13) & 7.65 & 7.251(05) & 0.00072 & $\sim$\,3.7 & $\sim$\,276  & $\sim$\,267  & 232 & 229 \\
\midrule
C80   & $\phantom{1}80^3 {\times} 160$           & 0.06801(10) & 5.44 & 6.936(11) & 0.00060 & 3.658(27) & 275.94(79) & 267.4(3.1) &  55 &  56 \\
D96   & $\phantom{1}96^3 {\times} 192$           & 0.05690(13) & 5.46 & 7.344(18) & 0.00054 & 3.632(29) & 277.08(80) & 268.0(3.3) &  62 &  61 \\
E112  & $112^3 {\times} 224$                     & 0.04881(08) & 5.47 & 6.892(17) & 0.00044 & 3.608(22) & 277.03(62) & 268.5(3.1) &  58 &  58 \\
\midrule 
\multicolumn{7}{r}{Continuum limit} & 278.09(95)   &  \multicolumn{1}{c}{269.6(4.6)} & \multicolumn{2}{c}{}\\
\bottomrule
\end{tabular}
\caption{Parameters of the $N_f=2+1+1$ ETMC gauge ensembles used in this study (see Ref.~\cite{ExtendedTwistedMass:2026nbv}). The first six columns list the ensemble short name, lattice size in lattice units, lattice spacing, physical linear extent, the mass-over-decay constant of the pion (see \cref{eq:scalar_density}), and the light quark mass in lattice units. Columns seven and eight report the fitted bare $\Bar{\rho}_{\W}$ and the renormalised quantity $\sqrt[3]{\frac{\pi}{2}Z_P \Bar{\rho}_\W}$, where $Z_P =Z_P^{\MSb}$ is the flavour non-singlet pseudoscalar density renormalisation factor in the $\MSb$ scheme at 2~GeV (renormalisation constants will be presented in a forthcoming dedicated paper \cite{renoEtmc}). Column nine corresponds to the renormalised chiral condensate $\Sigma_{\R}$ (see \cref{eq:BanksCasher}). For ensembles B48 and B96, no uncertainties are quoted because only a small number of configurations were analysed for testing purposes. The final two columns show the measured and predicted, from the right-hand side of \cref{eq:Nev_tm}, number of eigenvalues $N^\TM_{\rm ev}(\sqrt{2}\mu_{\ell};V)$ in the interval $\left[\mu_{\ell}, \sqrt{2}\mu_{\ell}\right]$.}
\label{tab:spectrum}
\end{table}

Given the smoothness of the massless Wilson operator, the density of the spectrum of the twisted operator follows immediately. Wilson eigenvalues with $|\Lambda_j|\le \mu$ are mapped into the region
\begin{equation}\label{eq:tm_evs_accum}
\mu \le \left(|\lambda_j(\mu)| = \sqrt{\Lambda_j^2+\mu^2}\right) \le \sqrt{2}\,\mu \, ,
\end{equation}
producing an accumulation of modes in the interval $[\mu,\sqrt{2}\mu]$. For eigenvalues much larger than $\mu$, the spectrum resumes its approximately linear growth in $|\Lambda|$. This pattern is illustrated in the right panel of~\cref{fig:hist} at the physical light-quark mass for each ensemble given in \cref{tab:spectrum}.
This leads directly to the central observation of this section. As the quark mass $\mu$ increases, all Wilson modes with $\abs{\Lambda_j}\le\mu$ are compressed into a narrow region of the twisted-mass spectrum and therefore contribute uniformly to its infrared part. In~\cref{tab:spectrum}, we report the measured and predicted value for $N^\TM_{\rm ev}(\sqrt{2}\mu;V)$ according to
\begin{equation}
N^\TM_{\rm ev}(\sqrt{2}\mu;V)
\equiv \int_{\mu}^{\sqrt{2}\mu} \!\!\!\!\!\!\rho_{\!{\textstyle\mathstrut}\TM}(\abs{\lambda};V,\mu)\, \dd{\abs{\lambda}}
\quad=\quad \int_{0}^{\mu}\!\!\!\!\rho_{\!{\textstyle\mathstrut}{\W}}(\Lambda;V) \dd{\Lambda}
\quad\equiv\quad N^\W_{\rm ev}(\mu;V)
\quad\simeq\quad \bar{\rho}_{\W}\,\mu \, V\,.
\label{eq:Nev_tm}
\end{equation}
Indeed, given the above considerations, the integrated density of eigenvalues in $[\mu,\sqrt{2}\mu]$ of \cref{eq:Nev_tm} grows linearly with $\mu$.

This relation can be generalized to the entire spectral density, yielding
\begin{equation}
   N^\TM_{\rm ev}(n\mu;V)  \equiv \int_{\mu}^{n\mu} \!\!\!\!\rho_{\!{\textstyle\mathstrut}\TM}(\abs{\lambda};V,\mu)\, \dd{\abs{\lambda}}
\quad=\quad \int_{0}^{\mu\sqrt{n^2-1}}\!\!\!\!\!\!\!\!\rho_{\!{\textstyle\mathstrut}{\W}}(\Lambda;V) \dd{\Lambda} \quad\equiv\quad N^\W_{\rm ev}(\mu\sqrt{n^2-1};V)
\quad\simeq\quad \bar{\rho}_{\W}\sqrt{n^2-1}\,\mu \, V
\end{equation}
showing that, in general, the number of twisted-mass eigenvalues in the IR region scales as $\mu V$, assuming that the relevant IR region corresponds to the part of the spectrum where the normalized Wilson spectral density $\rho_{\W}$ can be well approximated by the constant $\bar{\rho}_{\W}$, as in~\cref{eq:banks}.

\subsection{From spectral density to LMA performance}
\label{sec:LMA_perf}

The results of the previous section, together with those of earlier studies~\cite{Giusti:2008vb,Engel:2014eea}, indicate that the spectral density of the massless Wilson operator can be well approximated by a constant over a wide IR region. In particular, Ref.~\cite{Engel:2014eea} finds that deviations from a constant behaviour are approximately described by a factor $(1+0.001\,\Lambda_R/\mev)$, where $\Lambda_R$ is the renormalised eigenvalue, implying deviations of less than $10\%$ up to $\Lambda_R=100\,\mev$. We denote by $\Lambda^\IR_{\max}$ the upper limit of this region, within which the spectral density $\rho_\W(\Lambda; V)$ is an approximately constant function of the eigenvalue $\Lambda$. In the following, we focus on observables dominated by infrared modes, such as correlation functions at large source--sink separations, with propagators corresponding to light quarks satisfying $\mu\ll\Lambda^\IR_{\max}$.

For such observable $\mathcal{O}$, we introduce a further scale, $\Lambda^\IR_{\mathcal{O}}$, defined as the upper limit of the spectral region to be deflated in order to capture a specified fraction of the observable's infrared contribution (e.g. 90\%). This scale is therefore observable-dependent, and we assume it to satisfy $\Lambda^\IR_{\mathcal{O}}\leq\Lambda^\IR_{\max}$. Keeping the captured fraction fixed provides a consistent criterion for comparing the effectiveness of low-mode averaging across different setups. Based on the analysis of the twisted-mass spectral density presented above, and on the arguments developed below, we argue that maintaining a fixed $\Lambda^\IR_{\mathcal{O}}$ requires the number of deflated eigenvectors to scale as
\begin{equation}\label{eq:IR_scaling}
N^{\IR}_{\rm ev} \;\propto\; \Lambda^\IR_{\mathcal{O}}\;\propto\;\mu\, V \, .
\end{equation}
Thus, for a fixed deflation threshold, the number of eigenvectors required grows linearly with both the quark mass and the physical volume.

We illustrate this explicitly for two cases: quark loops and two-point correlation functions of quark bilinears. For a quark loop, the IR contribution reads as
\begin{equation}
L^{\IR}(t,\vb*{q},\Gamma)=\Tr \left[\proj{q}{t}\Gamma\,S^{\IR}\right] = \sum_{j=1}^{N_{\EV}^\IR} \frac{1}{\lambda_j} E_{jj}(t, \vb*{q}, \Gamma)
\quad\text{with}\quad
E_{ij}(t, \vb*{q}, \Gamma) = v_i^\dagger\,\proj{q}{t}\,{\Gamma}\,{v_j^{\phantom1}}\,.
\qq{and} 
\proj{p}{\tau}(\vb*{x},t) = \delta_{t\tau}e^{i\vb*{x}\cdot\vb*{p}} 
\end{equation}
Empirically, we observe that the magnitude of $E_{jj}(t, \vb*{q}, \Gamma)$ is approximately independent of the eigenvalue index $j$, assuming a value $\bar{E}(\vb*{q}, \Gamma)$. This is consistent with the orthonormality of the eigenvectors, which implies that $\sum_t E_{ij}(t,\vb*{0},\mathds{1}) = \delta_{ij}$ for any $i,j$.
Based on this observation, we approximate
\begin{equation}
L^{\IR}(t,\vb*{q},\Gamma)\;\simeq\; \bar{E}(\vb*{q}, \Gamma) \sum_{j=1}^{N_{\EV}^\IR} \frac{1}{\Lambda_j+i\mu}\;=\;\frac{\bar{E}(\vb*{q}, \Gamma)}{\mu} \sum_{j=1}^{N_{\EV}^\IR} \frac{e^{-i \arccot(x_j)}}{\sqrt{x_j^2+1}} \qquad\text{with}\qquad x_j = \frac{\Lambda_j}{\mu}\,.
\end{equation}
Consequently, the magnitude of $L^{\IR}$ remains approximately invariant under changes in $\mu$, provided that the number of eigenvalues included in the sum, $N_{\EV}^\IR$, scales linearly with $\mu$. 

\begin{figure}
\centering
\includegraphics[width=0.98\linewidth]{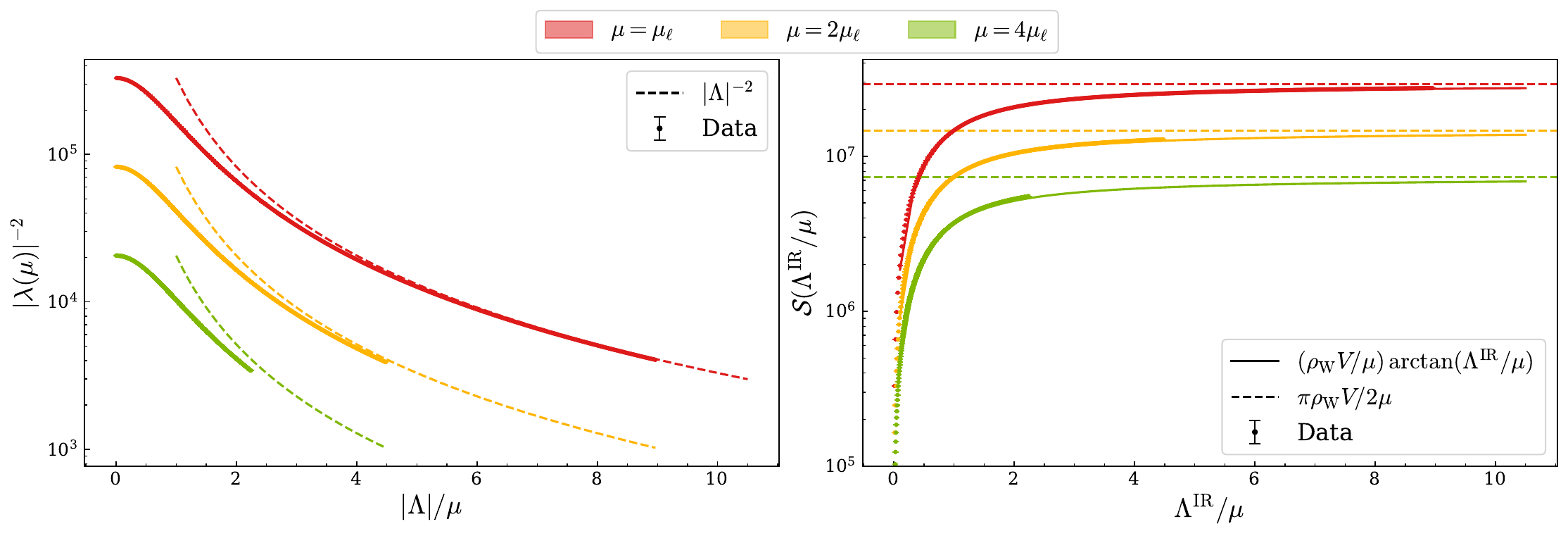}
\caption{\textit{Left:} Dependence on the quark mass $\mu$ of $|\lambda(\mu)|^{-2}$, which enters the sum over IR contributions to two-point quark-bilinear correlation functions in~\cref{eq:meson_corr_approx}. \textit{Right:} Sum of the contributions shown on the left over eigenvalues satisfying $\Lambda_i\leq\Lambda^{\IR}$, compared with the corresponding analytic prediction obtained by assuming a continuous spectrum as in~\cref{eq:sum_vs_int}. The data shown correspond to eigenvalues measured on the C80 ensemble, with uncertainties reflecting fluctuations across the ensemble.\label{fig:lambdamu_scal}}
\end{figure}

Conversely, for a two-point correlation function of quark bilinears, we have
\begin{equation}
    C^{\IR}_{\rm 2pt}(t,\vb*{p},\Gamma)\;=\; 
\frac{1}{T}\sum_{\tau=1}^T\Tr\left[
\proj{p}{\tau} \gfi\Gamma\, S^{\IR}\, \proj{p}{\tau+t} \Gamma\gfi\, (S^{\IR})^{\dagger} \right] \;=\; \frac{1}{T}\sum_{\tau=1}^T\sum_{i, j =1}^{N_{\EV}^\IR} \frac{1}{\lambda_i\,\lambda_j^*} E_{ij}(\tau+t, \vb*{p}, \Gamma\gamma_5) E_{ji}(\tau, \vb*{p}, \gamma_5\Gamma)\,,
\end{equation}
as introduced in more detail in~\cref{sec:mesons_LMA}. As demonstrated empirically in~\cref{sec:mesons_IR}, the convolution of eigenvector inner products, $E_{ij}E_{ji}$, depends primarily on the time separation $t$, the squared momentum $p^2$, the $\Gamma$-structure, and whether the eigenvalue indices either belong to the diagonal, $i=j$, or on their distance from the diagonal, $|i-j|/\sqrt{2}$, with only a weak dependence on the individual eigenvalue indices. Furthermore, the diagonal terms, depending on the type of correlation function, might be dominant, e.g., for the pseudoscalar-pseudoscalar one. Using these observations, we approximate the IR contribution assuming diagonal dominance with constant magnitude and write
\begin{equation}\label{eq:meson_corr_approx}
C^{\IR}_{\rm 2pt}(t,\vb*{p},\Gamma)\;\simeq\; \frac{\overline{\mathcal{K}}(t,\vb*{p},\Gamma)}{\mu^2}\sum_{j=1}^{N_{\EV}^\IR} \frac{1}{x_j^2+1} 
\quad\qq{with}\quad
\overline{\mathcal{K}}(t,\vb*{p},\Gamma) = \frac{1}{TN_{\EV}^\IR}\sum_{\tau=1}^T\sum_{j=1}^{N_{\EV}^\IR} E_{jj}(\tau{+}t, \vb*{p}, \Gamma\gamma_5)\, E_{jj}(\tau, \vb*{p}, \gamma_5\Gamma)\,,
\end{equation}
where $\overline{\mathcal{K}}$ denotes the average diagonal contribution. As in the case of quark loops, we find that $\mu\, C^{\IR}_{\rm 2pt}$ remains approximately invariant under changes in $\mu$, provided that the number of eigenvectors included in the IR contribution scales linearly with $\mu$.
Indeed, the sum can be approximated by an integral that admits a closed-form expression:
\begin{equation}\label{eq:sum_vs_int}
\mathcal{S}\left(\Lambda^\IR\right) \;\equiv\; \frac{1}{\mu^2} \sum_{j=1}^{N_{\EV}^\IR} \frac{1}{x_j^2+1} \;\simeq\;
\frac{\bar{\rho}_\W V}{\mu}\int_0^{\Lambda^{\IR}\!/\mu}\!\! \frac{1}{x^2+1}\, dx \;=\; \frac{\bar{\rho}_\W V}{\mu} \arctan\left(\frac{\Lambda^{\IR}}{\mu}\right)
\qquad\Longrightarrow\qquad
\mathcal{S}\left(\infty\right) 
\;=\; \frac{\pi\,\bar{\rho}_\W V}{2\mu}\,,
\end{equation}
where $\Lambda^{\IR}$ is the largest Wilson eigenvalue included in the sum, and the limit $\Lambda^{\IR}{\gg}\mu$ is taken on the right-hand side. In~\cref{fig:lambdamu_scal}, we show that the closed-form expression accurately describes the empirical sum over twisted-mass eigenvalues and that its approach to the asymptotic value is invariant under changes in $\mu$. Therefore, if we quantify the performance of LMA by the fraction of the full IR contribution captured by the truncated sum, namely
\begin{equation}
\frac{C^{\IR}_{\rm 2pt}}{C^{\IR,\infty}_{\rm 2pt}} = \frac{2}{\pi\,\mu\,\bar{\rho}_\W V}\sum_{j=1}^{N_{\EV}^\IR} \frac{1}{x_j^2+1}
\qquad\text{with}\qquad 
C^{\IR,\infty}_{\rm 2pt} = \frac{\overline{\mathcal{K}}(t,\vb*{p},\Gamma)}{\mu^2}\int_0^{\infty} \frac{\mu\, \bar{\rho}_\W V}{x^2+1}\, dx\,,
\end{equation}
this fraction remains approximately invariant under changes in $\mu$ only if the total number of eigenvalues, $N_{\EV}^{\IR}$, is scaled according to~\cref{eq:IR_scaling}, which compensates the factor $1/\mu$ in front of the sum. In~\cref{sec:mesons_IR}, we extend this derivation to general quark-bilinear two-point correlation functions, without assuming diagonal dominance.

\subsection{Discussion on Eq.~\texorpdfstring{\eqref{eq:IR_scaling}}{(22)}}
\label{sec:LMA_perf_remarks}
Although the arguments in the previous section have been derived for the twisted mass operator, where the mass shift explicitly accumulates eigenvalues, the conclusion is expected to be more general rather than specific to this regularisation. Indeed, since different lattice discretisations describe the same infrared physics, the density of modes relevant for low-energy observables should exhibit the same parametric dependence. Therefore, it is  natural to expect that the linear scaling with the multiplicatively renormalisable quark-mass parameter (here $\mu$) and the spacetime volume $V$ holds more generally across fermion formulations. For instance, in the Wilson case, the eigenvalues are not similarly clustered but still populate the region near the mass shift threshold~\cite{Luscher:2007se}. 
This leads to several important consequences as discussed below.

\begin{description}[font=\itshape]

\item[{LMA computational cost and memory usage grow as $\mu \,V^2/a^4$}] 
Given the scaling of $N^{\IR}_{\rm ev} \propto \mu V$ and considering that the cost of lattice operations scales at least with the number of lattice points $V/a^4$, the computational cost and memory footprint of exact LMA scale as $\mu\,V^2/a^4$. This $V^2$ scaling renders the method rapidly impractical on large volumes due to both larger resource requirements and poor parallel scaling of algorithms, in particular, multigrid solvers.
LMA is expected to further deteriorate as the quark mass increases, while remaining smaller than $\Lambda^{\IR}_{\max}$. In particular, if $N^{\IR}_{\rm ev}$ modes are required to achieve a given performance, for a given observable, at the light-quark mass $\mu_\ell$, maintaining the same efficiency would require roughly $\mu_s/\mu_\ell \approx 27$ times more modes at the strange mass. Given that already at the physical light-quark mass on a lattice with linear extent $L \simeq 5\,{\rm fm}$ several hundred modes are needed for a significant gain, extending exLMA to heavier quarks quickly becomes computationally prohibitive. 
These considerations imply that LMA is most effective at light-quark masses, where the number of required modes remains manageable, and the balance between cost and variance reduction is most favourable.

\item[{Contraction costs for all-to-all correlation functions scale with $(N^{\IR}_{\rm ev})^n$}] Contraction costs can also become rapidly prohibitive with the number of propagators, $n$, involved. While exact eigenvectors enable the construction of all-to-all correlation functions, the associated cost typically scales as $(N^{\IR}_{\rm ev})^n$, leading to a steep increase already for a moderate number $n$ of propagators---see e.g.~\cref{fig:runtime} for baryon contractions. As our results will  demonstrate, this scaling effectively renders all-to-all baryon correlators ($n=3$) already impractical for light-quark masses and moderate volumes, and, as volume or quark mass increases, fully exact all-to-all correlation functions become generically unfeasible, except for simpler observables, such as single-quark loops with $n=1$.
In practice, this limitation can be alleviated by replacing all-to-all contractions with stochastic estimators, since the signal typically saturates at moderate statistics. Other possible strategies not considered here include sparsening~\cite{Detmold:2019fbk, Li:2020hbj, Christian:2025kke}.

\item[{Gain of LMA versus stochastic approaches decreases with $1/\mu^2$}] Assuming that a stochastic propagator samples the Dirac spectrum approximately uniformly, then the fraction of infrared modes it captures increases with the quark mass, since the IR region relevant for long-distance physics grows proportionally to $\mu$ for a given spectrum. Stochastic estimators, therefore, become progressively more efficient at sampling the physically relevant modes as the mass increases. 
At the same time, the cost of LMA rises linearly with $\mu$, as the number of eigenmodes that must be treated explicitly increases accordingly. Thus, the two approaches scale in opposite manners, with LMA becoming more expensive, while stochastic estimators improve in quality. As a result, the relative advantage of LMA over stochastic methods is expected to rapidly decrease with increasing quark mass, with a scaling that can eventually be parametrised by $1/\mu^{2}$ to a first approximation. Consequently, for sufficiently heavy quarks, stochastic approaches become markedly more efficient than LMA.

\item[{LMA performance is limited by the heaviest quark mass in the correlation function}]
The previous remarks apply to correlators with a single quark mass. A more subtle situation arises for mixed correlators involving both light and heavy quarks, with masses $\mu_\ell$ and $\mu_h$, such as the kaon two-point function. Suppose one deflates $N^\IR_\ell$ modes that are determined to be sufficient for achieving a good performance at $\mu_\ell$, and does not scale this number for  the heavier propagator, i.e., instead of the larger value $N^\IR_h = (\mu_h/\mu_\ell)\, N^\IR_\ell$ one still uses $N^\IR_\ell$---this is particularly advantageous for the twisted mass discretisation, where the same eigenvectors can be used for all quark masses; see~\cref{eq:S_IR}. Then the corresponding propagators can be written as
\begin{equation}
S^{\LMA,\ell}_{\eta} = S^{\IR,\ell}_{\rm ex} + S^{\UV,\ell}_\eta \qq{and}
S^{\LMA,h}_{\eta} = S^{\IR,h}_{\rm ex/\eta} + S^{\UV,h}_\eta
= S^{\IR,h\subset \ell}_{\rm ex} + S^{\IR,h\not\subset \ell}_{\eta} + S^{\UV,h}_\eta\,,
\end{equation}
where the infrared and ultraviolet parts are separated and labelled according to whether they are treated exactly (ex) or stochastically ($\eta$). For the heavy propagator, the full infrared region of size $N^\IR_h$ is split into an exact part for the first $N^\IR_\ell$ modes, $S^{\IR,h\subset \ell}_{\rm ex}$, and a stochastic remainder, $S^{\IR,h\not\subset \ell}_{\eta}$.

Considering then the infrared contribution to a bilinear correlator, one obtains
\begin{equation}
C^{\IR,\ell h}_{\rm ex/\eta}
\equiv C(S^{\IR,\ell}_{\rm ex}, S^{\IR,h}_{\rm ex/\eta})
= C(S^{\IR,\ell}_{\rm ex}, S^{\IR,h\subset \ell}_{\rm ex})
+ C(S^{\IR,\ell}_{\rm ex},S^{\IR,h\not\subset \ell}_{\eta})
= C^{\IR,\ell(h\subset l)}_{\rm ex} + C^{\IR,\ell(h\not \subset l)}_{\eta}\,,
\end{equation}
where, in the first equality, the IR contraction is split into two contributions (one purely exact and one mixed); and then, in the second equality, the mixed exact–stochastic term is effectively rendered stochastic since we assume~\cref{eq:corr_defl} to hold. Consequently, only the light–light part of the correlator benefits from LMA, while the remaining contribution does not. Namely, using \cref{eq:corr_defl}, the full correlator can be written as
\begin{equation}
C^{\LMA,\ell h}_{\rm ex/\eta}
= C^{\LMA,\ell h}_{\eta} - C^{\IR,\ell(h\subset l)}_{\eta} + C^{\IR,\ell(h\subset l)}_{\rm ex}\,,
\end{equation}
making explicit that only the fully exact part is improved. This means that if all $N^\IR_h$ eigenvalues contribute approximately equally, the improved fraction of the correlator scales as $\mu_\ell/\mu_h$, since we expect $N^\IR_\ell\times N^\IR_h$ modes to contribute in $C^{\IR,\ell h}$, but only $N^\IR_\ell\times N^\IR_\ell$ have been deflated in $C^{\IR,\ell (h\subset l)}$. For example, for a kaon at physical quark masses, this yields a suppression of order $1/27$ of the noise reduction, significantly reducing the overall effects of LMA.

Once again, twisted-mass fermions allow us to make explicit the dependence on the number of eigenmodes. We consider a flavour-symmetrized quark-bilinear two-point function in which the light-quark mass is $\mu$ and the heavy-quark mass is $k\mu$. The infrared IR contribution can then be approximated as
\begin{equation}
    C^{\IR}_{\rm 2pt}(k; t,\vb*{p},\Gamma)\;\simeq\; \frac{\overline{\mathcal{K}}(t,\vb*{p},\Gamma)}{\mu^2}\sum_{j=1}^{N_{\EV}^\IR} \frac{x_j^2+k}{(x_j^2+1)(x_j^2+k^2)}
\end{equation}
Replacing the sum with a continuous integral, we obtain
\begin{equation}
\sum_{j=1}^{N_{\EV}^\IR} \frac{x_j^2+k}{(x_j^2+1)(x_j^2+k^2)} \;\simeq\; 
\int_0^{\Lambda^{\IR}}\!\!\!\frac{\bar{\rho}_{\W}V\,(x^2+k)}{\left( x^2+1\right)\left(x^2+k^2\right)}\, \dd{x} = \frac{\arctan(\Lambda^{\IR}) + \arctan(\Lambda^{\IR}\slash k)}{k+1} = \frac{\pi}{k+1} + \order{\frac{1}{\Lambda^{\IR}}}\,.
\end{equation}
We can now compare the case $k>1$ with the degenerate-mass case $k=1$, for which the two quark masses are equal. This yields
\begin{equation}
\frac{k+1}{2}\frac{C^{\IR}_{\rm 2pt}(k; t,\vb*{p},\Gamma)}{C^{\IR}_{\rm 2pt}(1; t,\vb*{p},\Gamma)} \;\simeq\; 
\frac{1}{2}\left(1+\frac{\arctan(\Lambda^{\IR}\slash k)}{\arctan(\Lambda^{\IR})}\right) = 1 - \frac{k-1}{\pi\Lambda^{\IR}} +\order{\frac{1}{(\Lambda^{\IR})^2}}\,.
\end{equation}
Thus, to maintain the same level of accuracy as the mass ratio $k$ increases, the number of IR eigenmodes must be increased accordingly. At leading order, this implies a scaling $N^{\IR}_{\EV}\propto (k-1)$ to matain performance constant.
\end{description}

The above considerations strongly indicate that, in modern lattice QCD applications, LMA is beneficial primarily for correlation functions composed exclusively of light-quark propagators, which thus define the focus of this work.

\subsection{On the Banks--Casher relation and the chiral condensate}
\label{sec:Banks}

\begin{figure}[tb]
\centering
\includegraphics[height=0.33\paperheight, valign=b]{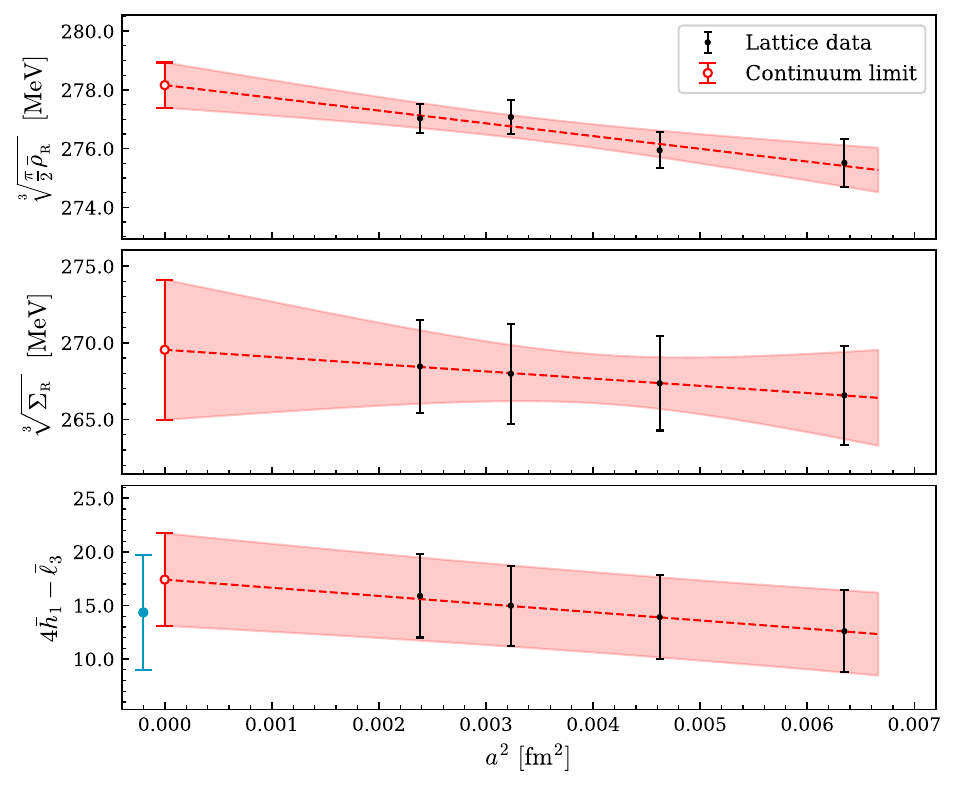}
\hspace{-10pt}
\includegraphics[height=0.33\paperheight, valign=b]{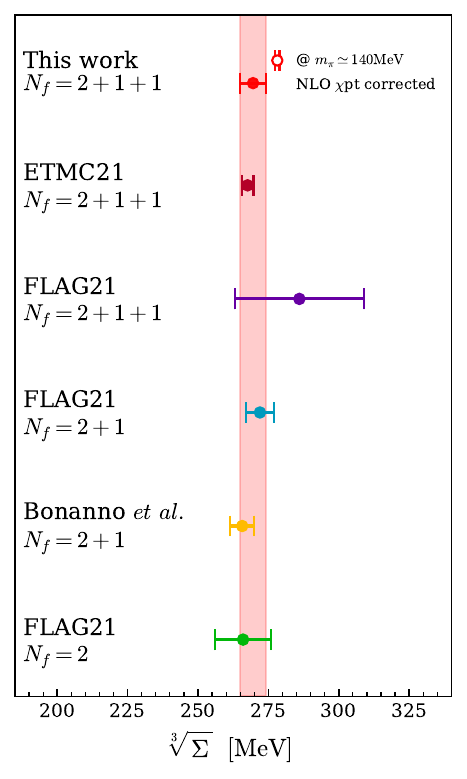}
\caption{\textit{Left panels:} Continuum limits of (\textit{top}) the renormalised density of low modes of the massless Wilson Dirac operator, $\bar{\rho}_\W$, on $140~\mev$ ensembles; (\textit{middle}) the chiral condensate $\Sigma$ determined from the density after correcting for the pion-mass dependence using $4\bar{h}_1-\bar{l}_3=14.3(5.4)$ with an arbitrary 30\% error (light blue dot in the bottom panel); and (\textit{bottom}) our determination of the low-energy constant combination $4\bar{h}_1-\bar{l}_3$, using as input the densities and the chiral condensate obtained in ETMC21~\cite{ExtendedTwistedMass:2021gbo}. \textit{Right panel:} Comparison of our determination of the chiral condensate with other recent results, ETMC21~\cite{ExtendedTwistedMass:2021gbo}, Bonanno \textit{et al.}~\cite{Bonanno:2023xkg}, and FLAG21 averages~\cite{FlavourLatticeAveragingGroupFLAG:2021npn, Cichy:2013gja, Alexandrou:2017bzk, Bazavov:2010yq, Borsanyi:2012zv, BMW:2013fzj, Boyle:2015exm, Cossu:2016eqs, ETM:2009ztk, Brandt:2013dua, Engel:2014eea}, renormalised in the $\overline{\mathrm{MS}}$ scheme at $2~\mathrm{GeV}$. The ETMC21 result~\cite{ExtendedTwistedMass:2021gbo} is extracted from the pion-mass dependence on the quark mass and is independent of the present determination, although obtained from an overlapping set of ensembles. 
}
\label{fig:condensate}
\end{figure}

As a by-product of the analysis presented in~\cref{sec:spectrum}, the extracted values of $\rho_{\R} \equiv Z_P \Bar{\rho}_{\W}$ enable a determination of the renormalised chiral condensate $\Sigma_{\R}$ via the Banks--Casher relation~\cite{Banks:1979yr}. The chiral condensate $\Sigma_{\R}$ is related to the disconnected contribution of the scalar insertion, which can be written as
\begin{equation}
\expval{\left(u\Bar{u} + d\Bar {d}\right)_{\R}} (\mu) \equiv
Z_P \expval{\left(u\Bar{u} + d\Bar {d}\right)_{\mathrm{sub}}} (\mu) =
-\frac{2 \mu}{V} \Tr[D^{-1}(+\mu)D^{-1}(-\mu)] = -\frac{2 \mu}{V} \sum_i \frac{1}{\Lambda^2_i+\mu^2}\,,
\end{equation}
where $\expval{\left(u\Bar{u} + d\Bar {d}\right)_{\mathrm{sub}}}$ denotes the properly subtracted (mixing with $\mathds{1}a^{-2}\mu$ and $\mathds{1}\mu^{3}$ removed) scalar light-quark density.

In the infinite-volume limit, the sum can be replaced by an integral. Assuming a constant spectral density up to a cutoff $\Lambda^{\IR}_{\max}$, one obtains
\begin{equation}    
\expval{\left(u\Bar{u} + d\Bar {d}\right)_{\R}} (\mu) \simeq 
-2\,Z_P\,\bar{\rho}_{\W}  \int_0^{\Lambda^{\IR}_{\max}}\!\!\! \frac{\mu}{\Lambda^2+\mu^2}\, \dd \Lambda =-\pi\,Z_P\,\bar{\rho}_{\W}+ \order{\frac{Z_P\,\mu}{\Lambda^{\IR}_{\max}}}\,,
\end{equation}
where we used
\begin{equation}
\int_0^{\Lambda^{\IR}_{\max}}\!\!\!\frac{\mu}{\Lambda^2+\mu^2}\, \dd\Lambda = \arctan{\frac{\Lambda^{\IR}_{\max}}{\mu}} = \frac{\pi}{2} + \order{\frac{\mu}{\Lambda^{\IR}_{\max}}}\,.
\end{equation}

The renormalised chiral condensate is then defined as
\begin{equation}
\Sigma_{\R} = 
-\frac{1}{2}\lim_{\mu\to 0}\expval{\left(\Bar{u}u + \Bar{d}d\right)_{\R}}(\mu) = \frac{\pi}{2}Z_P\,\bar{\rho}_{\W}^{\,0}\,,
\label{eq:BanksCasher}
\end{equation}
where $Z_P \bar{\rho}_{\W}^{\,0}$ denotes the spectral density in the chiral limit, i.e. for vanishing sea and valence quark masses. Owing to the relation between the spectral density of the twisted-mass operator and that of the massless Wilson operator, the extracted $\bar{\rho}_{\R}$ are already in the valence chiral limit, but correspond to a finite sea quark mass. 
Indeed, the Dirac operator spectrum is extracted from the simulation point, where $m_\pi\simeq140\,$MeV, while the lattice spacing, used as a conversion scale, is determined after the tuning of all bare parameters to match the FLAG/Edimburg definition of isoQCD~\cite{ExtendedTwistedMass:2026nbv}. In particular,  $a^{\smallcap{FLAG}}=(af_\pi)/f_{\pi}^{\smallcap{FLAG}}$.

Taking the continuum limit of $\bar{\rho}_{\R}$ and comparing with independent determinations of $\Sigma_{\R}$, we find values in the expected range; see~\cref{fig:condensate}, albeit in mild tension with our earlier result from ETMC21~\cite{ExtendedTwistedMass:2021gbo}. In that work, $\Sigma_{\R}$ was extracted using chiral perturbation theory relations for the pion mass across multiple ensembles, including heavier pion masses and a controlled extrapolation to the chiral limit. The observed discrepancy can therefore be attributed to residual sea-quark mass effects in the present determination, since other lattice errors appear to be well under control, including finite size corrections. Indeed, by comparing three significantly different volumes for the B ensemble yields consistent spectral densities, see~\cref{tab:spectrum}.

To quantify these residual sea-quark mass effects, one may employ the chiral expansion of the subtracted quark scalar operator vacuum expectation value~\cite{Gasser:1983yg},
\begin{equation}
\expval{\left(u\Bar{u} + d\Bar {d}\right)_{\R}}(\xi_\pi) = - 2\, \Sigma_{\R} \left(1+\xi_\pi (4\bar{h}_1-\bar{l}_3) +\mathcal{O}(\xi_\pi^2)\right)\qq{with}\xi_\pi=\frac{m^2_\pi}{16\pi^2f_\pi^2}\,,
\label{eq:scalar_density}
\end{equation}
where $\bar{l}_3=3.53(26)$~\cite{FlavourLatticeAveragingGroupFLAG:2021npn, ETM:2010cqp} is the next-to-leading-order low-energy constant (LEC) governing the quark-mass dependence of $m_\pi^2$, while $\bar{h}_1$ denotes a less constrained contact term entering scalar correlation functions. Owing to the relation between the scalar insertion and the spectral density via the trace of modes, this expression directly translates into
\begin{equation}
\bar{\rho}_{\R}(\xi_\pi) = Z_P \bar{\rho}_{\W}(\xi_\pi) = \frac{2\, \Sigma_{\R}}{\pi} \Big(1+\xi_\pi (4\bar{h}_1-\bar{l}_3) +\mathcal{O}(\xi_\pi^2)\Big)
\qquad\Longrightarrow\qquad 
4\bar{h}_1-\bar{l}_3 = \frac{1}{\xi_\pi}\left(\frac{\pi\,\bar{\rho}_{\R}}{2\,\Sigma_{\R}}-1\right)\,.
\end{equation}
This relation suggests two possible strategies. One may either adopt an external input for $\bar{h}_1$ to correct for pion-mass effects in the spectral density or, alternatively, determine $\bar{h}_1$ directly from our spectral-density data combined with an independent continuum limit determination of $\Sigma_{\R}$ from ETMC21~\cite{ExtendedTwistedMass:2021gbo} 

Regarding the first approach, an estimate for $H_1^r$ in three-flavour chiral perturbation theory at $\mu=0.77~\gev$ is available in Ref.~\cite{Amoros:1999dp}, determined in the large $N_c$ limit.  Converting this to the scale-independent $\bar{h}_1$~\cite{Gasser:1983yg}, and assigning a conservative uncertainty of $30\%$~\cite{BijensPrivate} to account for the theoretical uncertainty associated with the large-$N_c$ approximation, one obtains $\bar{h}_1 \simeq 4.5(1.3)$. The corresponding $4\bar{h}_1-\bar{l}_3$ value, represented as a blue dot in \cref{fig:condensate}, can then be used to correct the chiral condensate for chiral effects. The resulting continuum-limit result is
\begin{equation}
\sqrt[3]{\Sigma_{\R}}=269.5(0.7)_{\lambda}(0.6)_{a^2}(4.5)_{\chi pt} = 269.5(4.6)~\mev\qquad \text{in}~~N_f{=}2{+}1{+}1~\text{at 2 GeV scale}\,,
\label{eq:chiral_condensate}
\end{equation}
as shown in~\cref{fig:condensate}, with very good agreement with our previous ETMC21 determination. Here, $(\cdot)_{\lambda}$ denotes the uncertainty associated with the eigenvalue analysis, including statistical and systematic effects, as well as the contributions from the lattice spacing and the renormalisation constants, with these four sources contributing almost equally. The uncertainty associated with the continuum extrapolation, $(\cdot)_{a^2}$, is estimated as the increase in the error induced by the extrapolation itself, using data points for which the uncertainty on the chiral extrapolation is not included. The uncertainty associated with the chiral extrapolation, dominated by the LEC $\bar{h}_1$, is denoted by $(\cdot)_{\rm \chi pt}$ and is estimated as the remaining uncertainty that, when added in quadrature to the other uncertainties, gives the total uncertainty quoted on the right-hand side.

Following the second strategy instead, by comparing to ETMC21 to determine the chiral dependence, we find
\begin{equation}
4\bar{h}_1-\bar{l}_3 = 17.4(1.2)_{\lambda} (1.0)_{a^2} (4.4)_{\smallcap{\Sigma}} = 17.4(4.7) \qquad\Longrightarrow\qquad \bar{h}_1=5.2(1.2)\,,
\label{eq:hbar}
\end{equation}
where $(\cdot)_{\smallcap{\Sigma}}$ represents the uncertainty arising from the determination of $\Sigma_{\smallcap{R}}$ in ETMC21, while the splitting of uncertainties described above applies here as well.
\section{Results on Exact LMA for Quark Bilinear Correlation Functions}\label{sec:mesons}

In this section, we present results for two-point bilinear correlation functions computed on four physical-point ETMC ensembles, the parameters of which are listed in~\cref{tab:spectrum} and the statistics used are reported in~\cref{Tab:LMA_ensembles}. These correlators were originally generated to study the long-distance contribution to the muon anomalous magnetic moment~\cite{ETMC:2026mpp,ExtendedTwistedMass:2026nbv}, but were constructed for generic Dirac structures. This setup allows us to report more broadly on the achieved noise reduction across a range of physically relevant correlation functions.
\begin{table}[tbh!]
\centering
\begin{tabular}{lrcr}
\toprule
Ensemble & $N_U$ & $N^{\IR}_{\rm ev}$ & $N_\eta$  \\
\midrule
B64  & 1300 & 400 & 1024  \\
C80  & 590  & 530 & 1600  \\
D96  & 570  & 530 &  960  \\ 
E112 & 670  & 530 & 1344  \\ 
\bottomrule
\end{tabular}
\caption{Statistics for the two-point correlation functions using  the ETMC ensembles of~\cref{tab:spectrum},  where LMA noise-reduction techniques are  applied. The second column reports the number of configurations $N_U$, the third the number of low modes exactly deflated $N^{\IR}_{\rm ev}$ and used in the LMA setup to reconstruct the IR part, and the fourth the number of stochastic sources $N_\eta$ used to compute the full correlation function. The number of eigenvectors scales with the physical volume. The B64 ensemble has $L\simeq5.1$~fm, whereas C80, D96, and E112 have $L\simeq5.5$~fm. This accounts for the slightly larger number of modes for the latter three ensembles.}
\label{Tab:LMA_ensembles}%
\end{table}

\subsection{LMA of quark-bilinear correlation function}\label{sec:mesons_LMA}

We consider a generic meson two-point correlation function, computed using a backwards-running propagator via $\gfi$-hermiticity, $S=\gfi S^\dagger \gfi$\footnote{
In the case of TM fermions, one has $\gfi S^\dagger_{r}\gfi = S_{-r}$, where $r=\pm1$ is the Wilson parameter. In this section, we adopt the canonical quark basis, where the $f$-quark mass term is $\mu \Bar{f}f$ and the $f$-quark propagator reads $S_r = \exp{i\gfi\frac{\pi}{4}}\left[D_{\W}(r) + i\mu\gfi\right]^{-1}\! \exp{i\gfi\frac{\pi}{4}}$.
}, defined as
\begin{equation}\label{eq:corr_bil}
C(t_1,t_2,\vb*{p}_1,\vb*{p}_2,\Gamma_1,\Gamma_2) = 
\Tr\left[
\proj{p_2}{t_2} \gfi\Gamma_2\, S\, \proj{p_1}{t_1} \Gamma_1\gfi\, S^{\dagger}
\right]\qq{with}\,\proj{p}{\tau}(\vb*{x},t) = \delta_{t\tau}e^{i\vb*{x}\cdot\vb*{p}}, 
\end{equation}
where $x=(\vb*{x},t)$ denotes the spacetime coordinates. The operator $\proj{}{}\!\!$ acts on the lattice coordinates, projecting the propagator onto the timeslice $t_i$ and momentum $\vb*{p}_i$. While the two propagators $S$ may in general have different masses, here we restrict to light quarks, where the benefits from LMA are greatest as discussed in the previous section. For the twisted-mass operator, the propagators can be generated with either equal or opposite signs of the Wilson parameter $r=\pm1$ multiplying the light-quark twisted mass $\mu$, hence $\pm\mu$ in~\cref{eq:TMDirac}, corresponding either to twisted-mass (TM) or Osterwalder–Seiler (OS) regularisations. Both setups are tested, yielding comparable improvements. Therefore, results are presented for TM or OS regularisation, as specified below.

Regarding momentum boosts, we restrict to the zero-boost case, $\vb*{p}_1=\vb*{p}_2=\vb*{0}$, since non-zero momenta would require  dedicated inversions on momentum stochastic sources. Concerning the Dirac structures, several choices vanish at zero momentum, in particular those with unmatched spatial Dirac indices. Also, channels with pseudoscalar or scalar structures ($\Gamma_i=\gamma_5$ or ${\bf 1}$) rapidly saturate the stochastic noise, and the correlation functions are subject only to gauge noise. In these cases, no significant improvement from LMA is observed, as the correlator on a single configuration is already saturated with a small number of stochastic sources. In the following, we therefore restrict to non-vanishing vector, axial, and tensor structures, where a significant noise-to-signal ratio  is observed.

The correlation functions have been computed either stochastically or via all-to-all methods exploiting the exact low-mode eigenvectors,  as follows from~\cref{eq:corr_defl}. For the stochastic part, we employ momentum-projected stochastic sources with time and spin dilution,
\begin{equation}
    \eta_{\tau,\vb*{p},\mu} (x,\nu, a) = \delta_{t\tau}\delta_{\mu\nu}\,e^{\tfrac i2\vb*{x}\cdot\vb*{p}}\,\eta(\vb*{x},a) \qq{with}\frac{1}{N_\eta} \sum_{\eta} \eta^*(\vb*{x},a)\eta(\vb*{y},b) \approx \delta_{ab}\delta_{\vb*{x}\vb*{y}},
\end{equation}
where $\mu,\nu$ are spin indices, and $a,b$ colour indices. Subscripts label independent sources, each requiring a separate inversion. In practice, we invert over all four Dirac components $\mu$, while $\tau$ is varied together with a new stochastic sample $\eta$, and we take $\vb*{p}=\vb*{0}$. These sources allow us to approximate the term
\begin{equation}
    \proj{p_1}{t_1} \Gamma_1\gfi \approx \frac{1}{N_\eta} \sum_{\eta}\sum_{\mu\nu} \eta_{t_1,\vb*{p}_1,\mu}\; (\Gamma_1\gfi)_{\mu\nu}\;  \eta^\dagger_{t_1,\vb*{p}_1,\nu}
\end{equation}
in~\cref{eq:corr_bil} leading to
\begin{equation}\label{eq:stoch_corr_bil}
C_\eta(t_1,t_2,\vb*{p}_1,\vb*{p}_2,\Gamma_1,\Gamma_2) =  \sum_{\mu\nu}(\Gamma_1\gfi)_{\mu\nu}\,
\phi^\dagger_{\eta,t_1,\vb*{p}_1,\nu}\, \proj{p_2}{t_2} \gfi\Gamma_2\, \phi_{\eta,t_1,\vb*{p}_1,\mu}\qq{with}\phi_{\eta,t_1,\vb*{p}_1,\mu} = S \,\eta_{t_1,\vb*{p}_1,\mu}\,, 
\end{equation}
which corresponds to an inner product of the propagated stochastic sources $\phi_\eta$. 
The stochastic IR contribution is obtained analogously as
\begin{equation}
    C^\IR_\eta \qq{same as~\cref{eq:stoch_corr_bil} replacing $\phi_\eta$ with} \phi^\IR_{\eta,t_1,\vb*{p}_1,\mu} = S^\IR \,\eta_{t_1,\vb*{p}_1,\mu} =  \sum_{j=0}^{N^\IR_{\rm ev}} \frac{1}{\lambda_j}\,\gfi v_j v^\dagger_j\,\eta_{t_1,\vb*{p}_1,\mu}\,,
\end{equation}
where $\eta$ is exactly the same source, but propagated using only the IR part of the spectrum.

The remaining ingredient is the exact all-to-all IR contribution. Explicitly inserting the eigenmode decomposition of $S$ given in \cref{eq:S_IR}, in the correlation function yields
\begin{equation}
C^\IR(t_1,t_2,\vb*{p}_1,\vb*{p}_2,\Gamma_1,\Gamma_2) = 
\sum_{i,j}^{N^\IR_{\rm ev}}\frac{1}{\lambda_i\,\lambda_j^*}\Tr\left[
\proj{p_2}{t_2} \Gamma_2 \gfi\,v_iv_i^\dagger\, \proj{p_1}{t_1} \Gamma_1\gfi\, \,v_jv_j^\dagger
\right]\,. 
\end{equation}
From a computational perspective, it is more efficient to evaluate inner products of eigenvectors, obtaining
\begin{equation}\label{eq:corr_IR_bil}
C^{\IR}(t_1,t_2,\vb*{p}_1,\vb*{p}_2,\Gamma_1,\Gamma_2)
= \sum_{i, j =1}^{N_{\EV}^\IR} \frac{1}{\lambda_i\,\lambda_j^*} E_{ij}(t_1, \vb*{p}_1, \Gamma_1\gamma_5) E_{ji}(t_2, \vb*{p}_2, \gamma_5\Gamma_2)
\qq{with}
E_{ij}(t, \vb*{p}, \Gamma) = v_i^\dagger\,\proj{p}{t}\,{\Gamma}\,{v_j} 
\end{equation}
requiring only $N_{\EV}^\IR(N_{\EV}^\IR+1)/2$ independent contractions to be computed by exploiting
\begin{equation}
    E_{ji}(t, \vb*{p}, \Gamma) = E_{ij}^*(t, -\vb*{p}, \Gamma^\dagger)\,,
\end{equation}
and then assembling the all-to-all IR correlation functions during post-processing.

\subsection{The anatomy of the IR contribution}~\label{sec:mesons_IR}

\begin{figure}
\centering
\includegraphics[width=0.9\linewidth]{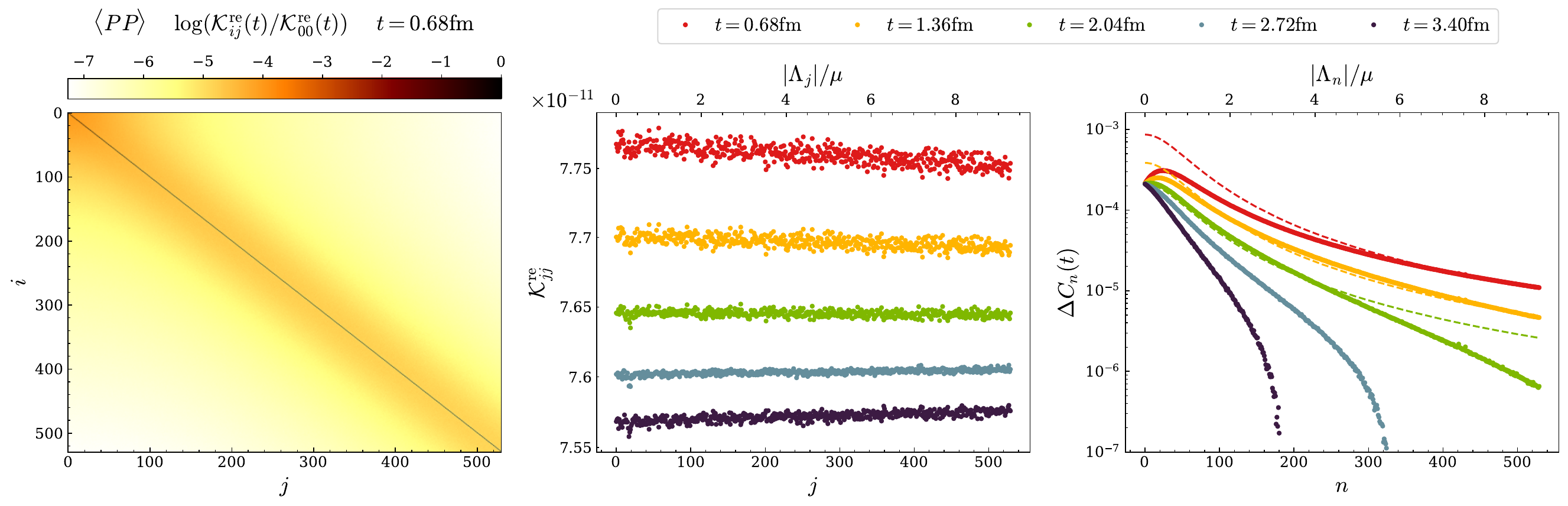}
\includegraphics[width=0.9\linewidth]{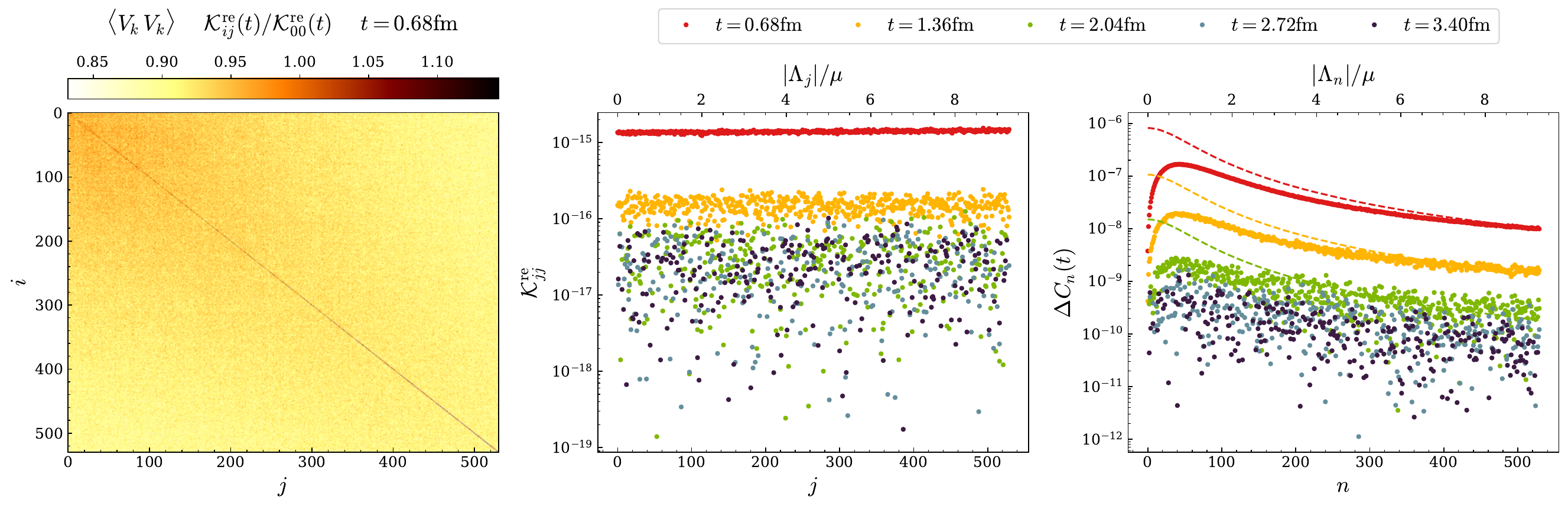}
\includegraphics[width=0.9\linewidth]{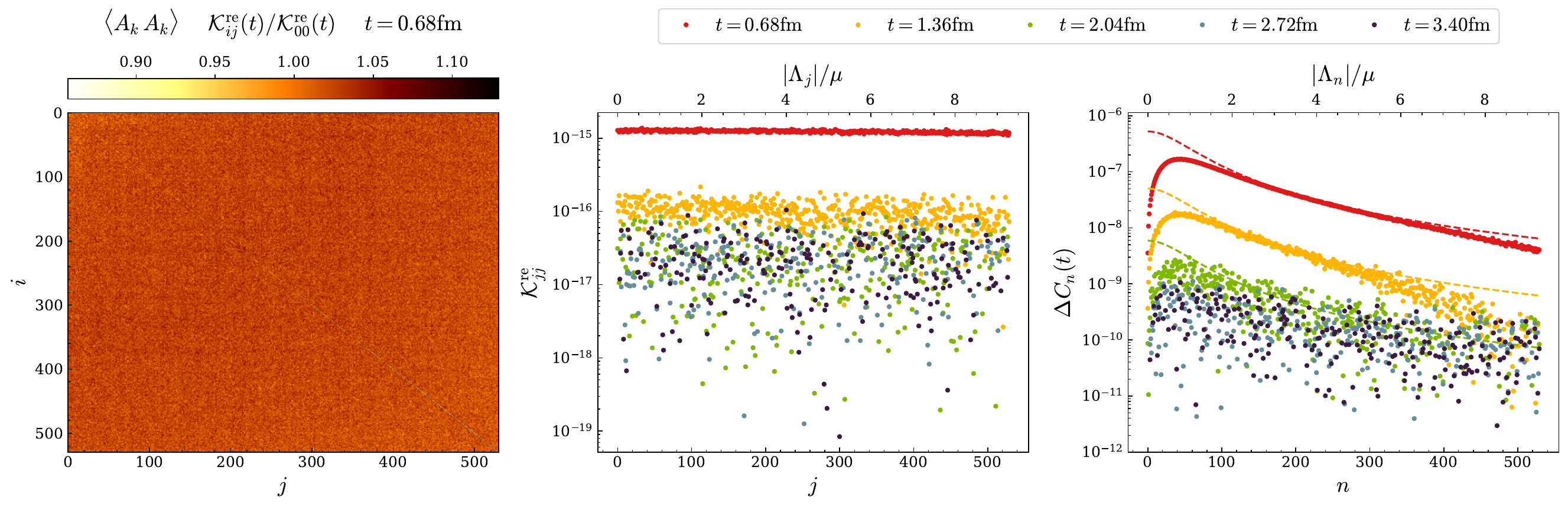}
\includegraphics[width=0.9\linewidth]{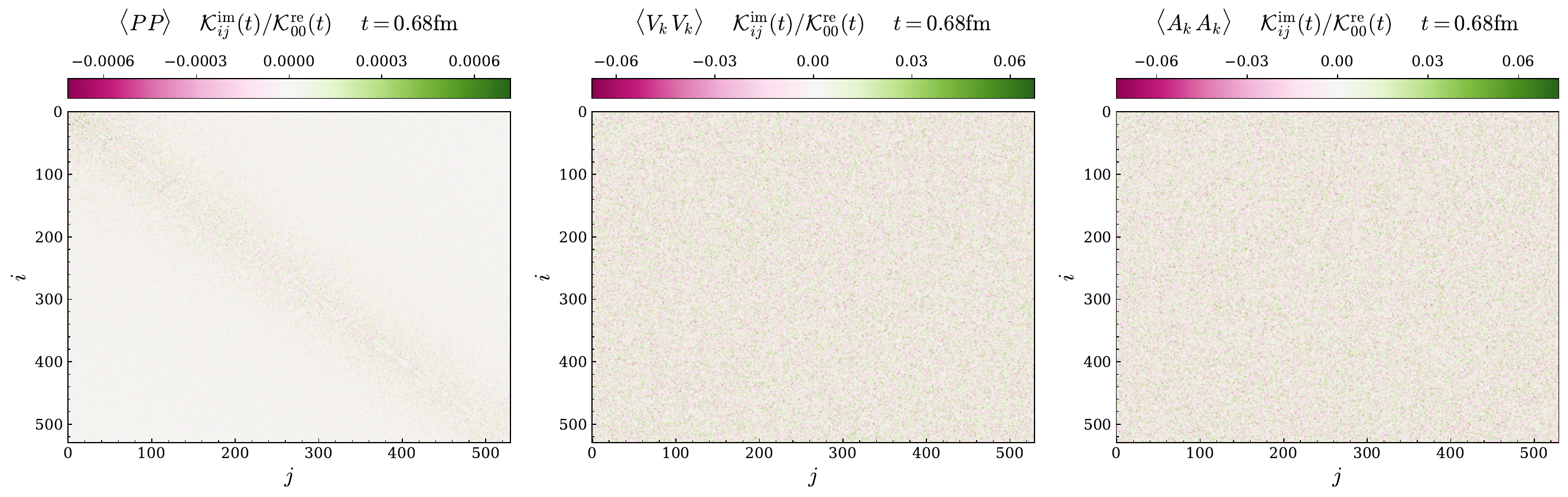}
\caption{\textit{Top three rows:} The pseudoscalar (first row), vector (second row), and axial (third row) correlation functions. From left to right, the \textit{leftmost column} shows the ensemble average of $\mathcal{K}^{\rm re}_{ij}/\mathcal{K}^{\rm re}_{00}$ at $t=0.68$,fm; the \textit{middle column} shows the diagonal elements $\mathcal{K}^{\rm re}_{jj}(t)$ at the times indicated in the legend; and the \textit{rightmost column} shows the incremental contributions $\Delta C^{\IR}_n(t)$, defined in~\cref{eq:meson_corr_IR_delta}, at the same times. In the latter panels, the dashed lines are $C/(\Lambda_n^2+\mu^2)$ with an arbitrary coefficient $C$ depicted to guide the eye. \textit{Bottom row:} From left to right, the three panels show the imaginary part, $\mathcal{K}^{\rm im}_{ij}/\mathcal{K}^{\rm re}_{00}$, at $t=0.68$,fm for the pseudoscalar, vector, and axial correlation functions, respectively.\label{fig:meson_IR_corr}}
\end{figure}

As anticipated in~\cref{sec:LMA_perf}, we now analyse the dependence of the contribution to the correlation function from each eigenvector pair $(i,j)$. We define
\begin{equation}
    \mathcal{K}_{ij}(t,\vb*{p},\Gamma) = \frac{1}{T}\sum_{\tau=1}^T E_{ij}(\tau{+}t, \vb*{p}, \Gamma\gamma_5)\, E_{ji}(\tau, \vb*{p}, \gamma_5\Gamma)\qquad\Longrightarrow\qquad C^{\IR}(t,\vb*{p},\Gamma) =  \sum_{i, j =1}^{N_{\EV}^\IR} \frac{1}{\lambda_i\,\lambda_j^*}\, \mathcal{K}_{ij}(t,\vb*{p},\Gamma)\,.
\end{equation}
In~\cref{fig:meson_IR_corr}, we show the ensemble averages of $\mathcal{K}_{ij}$ measured on the C80 ensemble for three types of correlation functions: pseudoscalar–pseudoscalar $\langle PP \rangle$, vector–vector $\langle V_kV_k \rangle$, and axial–axial $\langle A_kA_k \rangle$, with the latter two averaged over the spatial index $k$. We focus on the real part, $\mathcal{K}^{\rm re}$, since the imaginary part, $\mathcal{K}^{\rm im}$, shown in the bottom panels of~\cref{fig:meson_IR_corr}, fluctuates around zero and is at least two orders of magnitude smaller than $\mathcal{K}^{\rm re}_{00}$.

For the pseudoscalar channel, the diagonal contributions dominate, with the off-diagonal contributions already about four orders of magnitude smaller at $|i{-}j|=1$, consistent with the orthonormality of the eigenvectors. In contrast, the vector and axial channels exhibit a much weaker dependence on $(i,j)$, with contributions of comparable magnitude across eigenvector pairs. The vector channel shows a mild diagonal dominance, with off-diagonal contributions about 10–20\% smaller than the diagonal ones. In the axial channel, all contributions are comparable, exhibiting fluctuations of approximately $\pm10\%$ without a clear diagonal enhancement. This weak dependence on $(i,j)$ is further illustrated in the middle column of~\cref{fig:meson_IR_corr}, which shows the diagonal $\mathcal{K}^{\rm re}_{jj}$ for all three channels.

In the rightmost panels of~\cref{fig:meson_IR_corr}, we examine the resummation of the IR contribution as the number of included eigenvectors increases up to $N^{\IR}_{\EV}$. Each point represents the additional contribution arising from including one more eigenvector. Specifically, we define
\begin{equation}\label{eq:meson_corr_IR_delta}
    \Delta C^{\IR}_n(t,\vb*{p},\Gamma) = \sum_{i, j =1}^{n} \frac{1}{\lambda_i\,\lambda_j^*}\, \mathcal{K}_{ij}(t,\vb*{p},\Gamma) - \sum_{i, j =1}^{n-1} \frac{1}{\lambda_i\,\lambda_j^*}\, \mathcal{K}_{ij}(t,\vb*{p},\Gamma) \qquad\Longrightarrow\qquad C^{\IR}(t,\vb*{p},\Gamma) = \sum_{n=1}^{N^{\IR}_{\EV}}\Delta C^{\IR}_n(t,\vb*{p},\Gamma)\,,
\end{equation}
so that summing $\Delta C^{\IR}_n$ over $n$ up to $N^{\IR}_{\EV}$ recovers $C^{\IR}$. As discussed in~\cref{sec:LMA_perf}, the contribution is no longer independent of the eigenvalue range, as it is weighted by the eigenvalue-dependent coefficients. 
Expanding the sums in the definition of $\Delta C^{\IR}_n$, we find
\begin{align}
    \Delta C^{\IR}_n(t,\vb*{p},\Gamma) =\frac{1}{\mu^2} \frac{1}{x_n^2+1}\, \mathcal{F}_{n}(t,\vb*{p},\Gamma) ~~\qq{with}~~ \mathcal{F}_{n}(t,\!\vb*{p},\!\Gamma) &\equiv \sum_{j =1}^{n}(2-\delta_{nj})\,{\rm Re}\bigg(\frac{x_n+i}{x_j+i}\, \mathcal{K}_{nj}(t,\vb*{p},\Gamma)\bigg)\\
    &\simeq \sum_{j =1}^{n}(2-\delta_{nj})\, \frac{x_jx_n+1}{x_j^2+1} \,\mathcal{K}_{nj}^{\rm re}(t,\vb*{p},\Gamma)\,,\label{eq:F_n_approx}
\end{align}
where $x_j=\Lambda_j/\mu$ was already introduced in~\cref{sec:LMA_perf}. In the approximation of~\cref{eq:F_n_approx}, we assume that $\mathcal{K}$ is dominated by its real part. The rightmost panels of~\cref{fig:meson_IR_corr} also show that the suppression of $\Delta C^{\IR}_n$ with increasing $n$ is consistent with $C/(x^2_n+1)$ (dashed lines), indicating that $\mathcal{F}_{n}$ is approximately constant on a wide region of the considered spectrum. This leads us to the same conclusions as those drawn in~\cref{sec:LMA_perf_remarks}, where only diagonal contributions, i.e. $\mathcal{F}_n\approx\mathcal{K}_{nn}^{\rm re}$, were considered. Here this approximation has been lifted.
Consequently, maintaining a fixed LMA performance requires scaling the number of included eigenvectors, $N^{\IR}_{\EV}$, with the quark-mass parameter $\mu$, which acts as a uniform shift of the eigenvalue spectrum.

\subsection{Error reduction with the number of modes}

The most critical aspect in optimising LMA performance is determining how many modes to deflate. As discussed in~\cref{sec:spectrum}, for the twisted mass operator, one needs at least as many modes as those within the region $[\mu,\sqrt{2}\mu]$, since all eigenvalues in this interval are of comparable magnitude. For the B64 ensemble, which is  used for the testing in this section, this corresponds to roughly 45 modes at the simulated light-quark mass, as reported in~\cref{tab:spectrum}.

\begin{figure}[htb]
\centering
\includegraphics[width=1.0\textwidth]{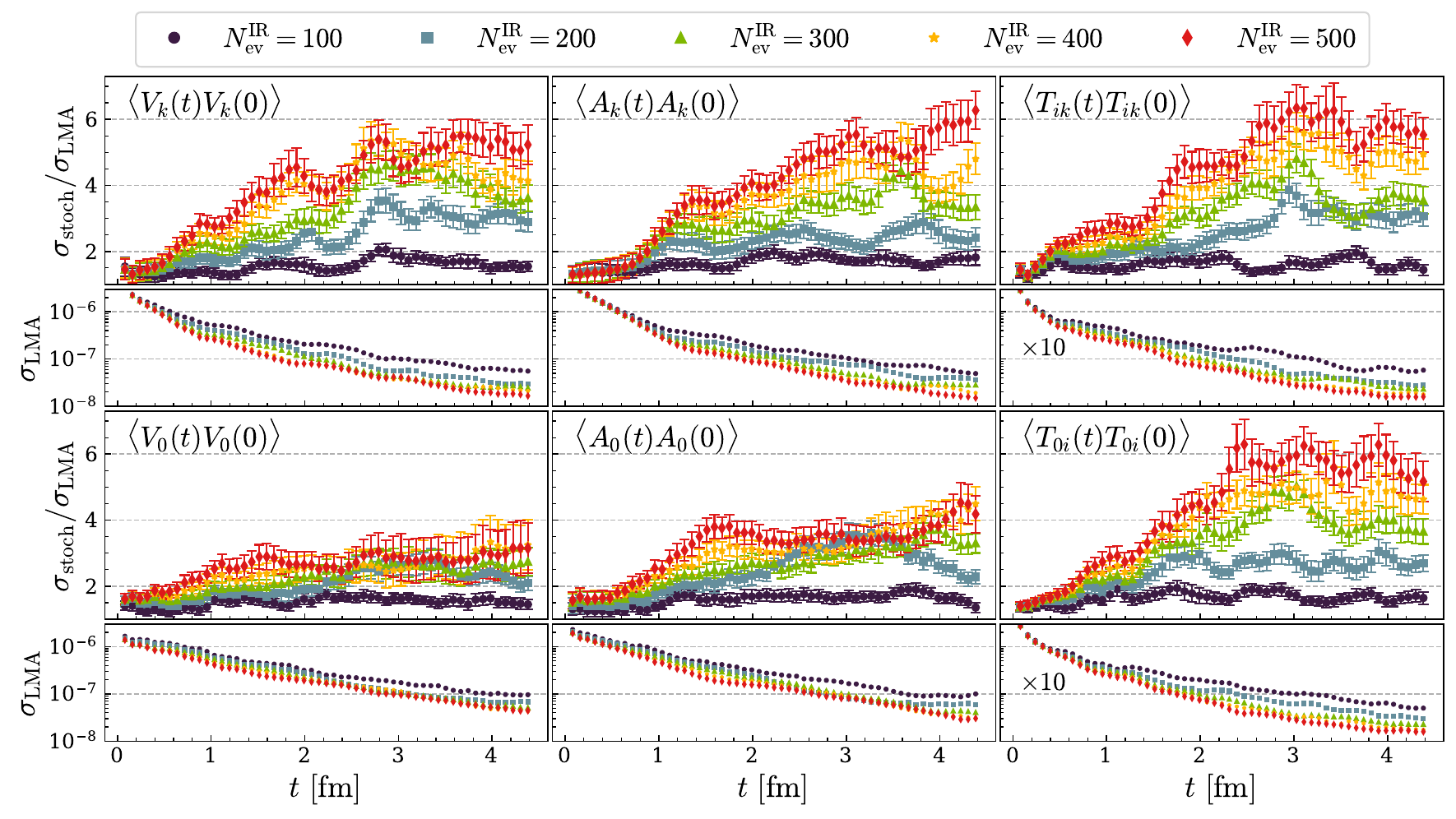}
\caption{Gain in the signal-to-noise ratio defined in \cref{Eq:gain}, computed for the deflated two-point light-quark flavour non-singlet correlators on a $64^3 \times 128$ lattice with spatial extent $L = 5.1~\mathrm{fm}$. Each panel corresponds to a different Dirac structure of the correlators. Different colours correspond to $N_{\mathrm{ev}^{\IR}} =100, 200, 300, 400, 500$ eigenvectors, using $N_U = 56$ gauge configurations and $N_\eta = 256$ stochastic sources. The bottom panels show the time dependence of the two-point function errors. The optimal choice is $N_{\mathrm{ev}}^{\IR}=400$, which yields a gain statistically compatible with that obtained using 500 eigenvectors while requiring fewer computational resources.}
\label{fig:gain_nvec}
\end{figure}

Results for different numbers of deflated modes are shown in~\cref{fig:gain_nvec} for a subset of about 50 configurations of the B64 ensemble, where the noise reduction is quantified as
\begin{align}
\frac{\sigma_{\mathrm{stoch}}(t)}{\sigma_{\mathrm{LMA}}(t)} \equiv \frac{{\rm Err}[C_\eta(t)]}{{\rm Err}[C^\LMA_\eta(t)]} \simeq \sqrt{\frac{{\rm Var}[C_\eta(t)]}{{\rm Var}[C^\LMA_\eta(t)]}}\, ,
\label{Eq:gain}
\end{align}
i.e., as the ratio between the statistical error of the fully stochastic correlator and that of the LMA-improved one, cf.~\cref{eq:corr_defl}. We observe that 100 eigenvectors, namely, more than twice the minimal estimate, are still insufficient to yield any noticeable improvement. A clear noise reduction appears only for significantly larger numbers of modes, predominantly at large Euclidean time separations. The error saturates if one uses around 400–500 eigenvectors for most correlation functions, as shown in~\cref{fig:gain_nvec}. The bottom panels show the time dependence of the two-point function errors. For Euclidean distances above 1~fm, we clearly observe a separation of the error magnitudes by increasing the number of deflated modes; however, going from $N_{\mathrm{ev}}^{\IR}=400$ to $N_{\mathrm{ev}}^{\IR}=500$ does not show a substantial improvement justifying the usage of larger computational resources. In particular, this is true for the vector–vector correlator that enters the study of the muon g-2. This motivates the choice of 400 modes used for the B64 ensemble.     
\begin{figure}[htb]
    \centering
    \includegraphics[width=0.9\textwidth]{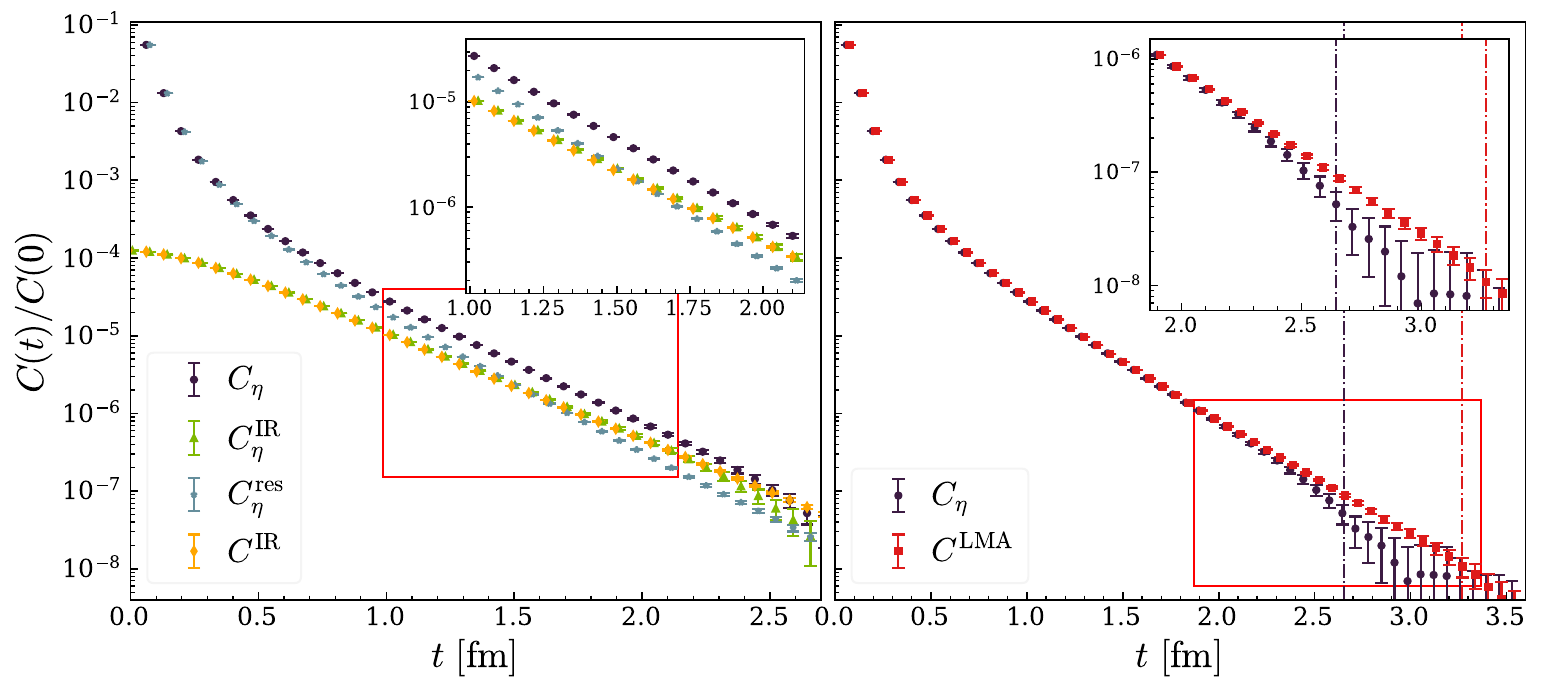}
    \caption{
    On the C80 ensemble, the comparison of the two-point vector-vector correlator in isoQCD. The left panel shows the $C_\eta$, $C_{\eta}^{\IR}$ and $C^{\IR}$ contributions to $C_{\eta}^{\LMA}$ of \cref{eq:corr_defl}, as well the residual correlator $C_{\eta}^{\mathrm{res}}$ of \cref{eq:corr_residual}. In the small box, the crossover region between $C_{\eta}^{\mathrm{res}}$ and $C^{\IR}$is highlighted. In the right panel, the direct comparison between $C_\eta$ and $C_{\eta}^{\LMA}$ is shown. The dashed vertical lines represent the time slice where $30\%$ relative error is reached. The small box highlights the region $t \ge 2~\fm$.
    }
    \label{fig:corr_gain_plot}
    \end{figure}
Scaling with the volume then leads to about 530 modes for the C80, D96, and E112 ensembles, which have a $\sim7\%$ larger linear extent, corresponding to a $\sim32\%$ increase in volume. 

\begin{figure}[htb]
\centering
\includegraphics[width=1.0\textwidth]{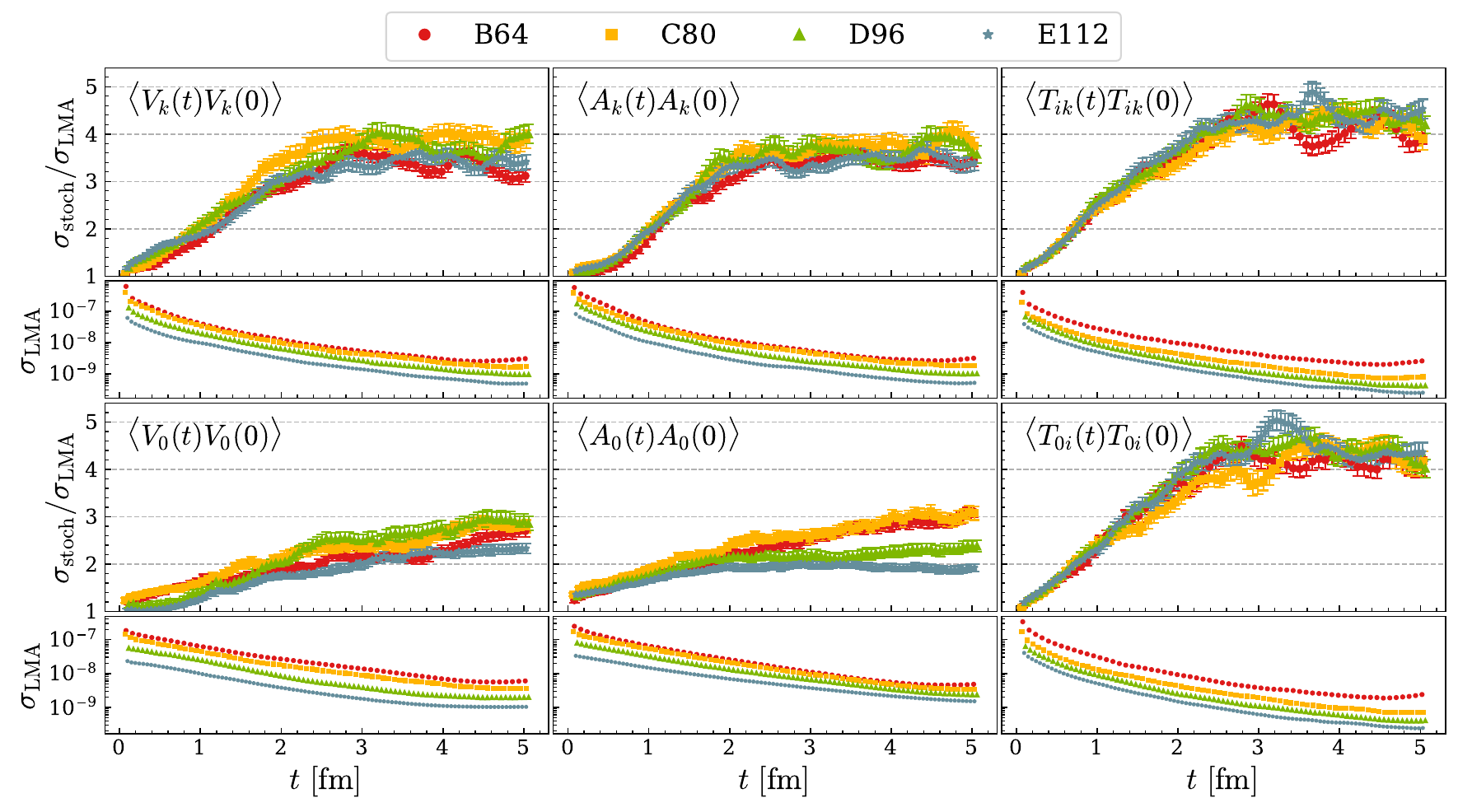}
\caption{The gain in the signal-to-noise ratio in light-quark flavour non-singlet two-point correlation functions, as defined in \cref{Eq:gain}, is shown for the B64, C80, D96 and E112 ensembles. The gain reached is above 3.5 for all the ensembles, except for the correlators involving the time components of the vector and axial-vector. The bottom panels show the  error time dependence of the improved two-point correlation functions.}
\label{fig:gain_ens}
\end{figure}

Results for all ensembles are depicted in~\cref{fig:gain_ens}. Several remarks are in order here:
\begin{description}
    \item[\textit{Dependence on Euclidean time and number of modes}] The observed behaviour as a function of the number of modes and Euclidean time can be understood by examining the different contributions entering $C^\LMA_\eta$ in~\cref{eq:corr_defl}. This is illustrated in the left panel of~\cref{fig:corr_gain_plot}, where, for the two-point vector-vector correlator, we show the fully stochastic correlator $C_\eta(t)$, its IR part $C^\IR_\eta(t)$, and their difference (the residual),
    \begin{equation}
        C^{\rm res}_\eta(t) = C_\eta(t)-C^\IR_\eta(t)\,,
        \label{eq:corr_residual}
    \end{equation}
    and the all-to-all IR contribution $C^\IR(t)$. At sufficiently large Euclidean time separations, a cross-over between the IR and residual contributions is observed, around 1.5\,fm for this case, indicating that beyond this point the full correlator becomes dominated by the IR part. It is also worth noting that for such time separations, the fully UV part of the correlator is expected to be highly suppressed, making the residual correlator effectively the mixed $\UV-\IR$ contribution to the correlator. As shown in the figure, both the signal and its error are subleading w.r.t. the IR ones. The location and even the existence of this cross-over depend on the number of deflated modes. With too few modes, the IR part lacks sufficient information to describe the ground state and never dominates. With enough modes, the asymptotic state is fully captured by the IR contribution, and as more modes are included, the cross-over shifts to shorter Euclidean times as excited states are progressively incorporated. Based on the results shown in~\cref{fig:gain_nvec} for the B64 ensemble, we conclude that around 400-500 modes are necessary to fully describe the ground state and reach maximal gain at large Euclidean time.
    \item[\textit{Asymptotic noise reduction at large Euclidean time}] The asymptotic gain at large Euclidean time is determined by how much more precisely the IR contribution is computed compared to its stochastic estimate. In the present setup, where all-to-all correlators are used, $C^\IR(t)$ is known exactly up to gauge noise. By contrast, the stochastic estimate $C^\IR_\eta(t)$ is dominated by the stochastic noise and its error depends on the number of noise sources. In the regime where errors of the stochastic estimate scale ideally, one expects
    \begin{equation}
        G^\IR(t) = \frac{{\rm Err}[C^\IR_\eta(t)]}{{\rm Err}[C^\IR(t)]} \simeq \frac{1}{\sqrt{N_\eta}}\sqrt{\frac{{\rm Var}[C^\IR_\eta(t)]}{{\rm Var}[C^\IR(t)]}}\,,
    \end{equation}
    so that the gain decreases as the number of stochastic sources is increased. Consequently, the number of stochastic sources should be chosen as small as possible while still ensuring that the residual contribution is under control, thereby maximising the gain in the IR-dominated region while controlling the stochastic noise in the UV region. This observation also explains the difference in the asymptotic gain between~\cref{fig:gain_nvec} and~\cref{fig:gain_ens}: in the former, a smaller number of stochastic sources was used, yielding an error reduction of about 4.5, whereas the final production, performed according to~\cref{Tab:LMA_ensembles}, results in an error reduction of about 3.5 for the vector-vector case.
    \item[\textit{Dependence on the Dirac structure}] We observe that all correlation functions affected by a significant noise-to-signal ratio exhibit very similar noise-reduction properties and, in particular, the vector--vector, axial--axial, and tensor--tensor channels. Exceptions are correlation functions built from scalar, pseudoscalar, and the temporal components of the vector and axial currents. As discussed above, correlation functions involving scalar or pseudoscalar operators saturate very rapidly with stochastic noise, eliminating any gain from the LMA approach.
    For $\langle V_0(t)V_0(0)\rangle$ and $\langle A_0(t)A_0(0)\rangle$, we do observe an improvement, although it is less pronounced than for the spatial components and does not saturate at large Euclidean time (see~\cref{fig:gain_ens}). This suggests that a larger number of modes would be required to fully capture the correlator.
    Finally, $\langle A_0(t)A_0(0)\rangle$ exhibits a distinctive behaviour, namely the improvement decreases significantly as the continuum limit is approached. Our interpretation is that the stochastic noise is decreasing towards the continuum limit, thereby diminishing the relative gain from LMA. This is consistent with the fact that the $\langle A_0(t)A_0(0)\rangle$ asymptotically has  the pseudoscalar pion as its ground state; as the continuum limit is approached, this correlator increasingly resembles a pseudoscalar correlation function, for which no signal-to-noise problem is present.    
\end{description}

\section{Results on Multigrid LMA for Quark Bilinear Correlation Functions}\label{sec:mesons_mg}

We now move to results produced using multigrid LMA (mgLMA), which belongs to the class of inexact LMA approaches~\cite{Luscher:2007se}. Here, the setup cost is reduced by avoiding the exact computation of eigenvectors, while exploiting local coherence to capture multiple low modes simultaneously. A successful inexact deflation strategy, therefore, offers clear advantages for LMA, as it targets both a reduction in setup cost and a decrease in the number of required vectors, with the additional prospect of improved scaling with the volume and the quark mass. In particular, algebraic multigrid (AMG) has emerged as one of the most effective realisations of inexact deflation and is nowadays widely used as a preconditioner in solvers for many fermion discretisations at the physical point~\cite{Brannick:2007ue,Babich:2010qb,Frommer:2013fsa,Alexandrou:2016izb,Brannick:2014vda,Brower:2020xmc,Brower:2018ymy}. Therefore, it is natural to expect that its success as a preconditioner can translate directly into an effective LMA strategy. However, as we will discuss in this section, our experience indicates that this connection is not straightforward   and requires further investigation.

\subsection{Multigrid operators and their properties}

In algebraic multigrid, a coarse operator $D_c$ is defined in terms of prolongation $P$ and restriction $R$ operators as
\begin{equation}
D_c \equiv R D P \,,
\end{equation}
where $D$ denotes the fine operator. In lattice QCD, $D_c$ is also referred to as the \textit{little Dirac operator}~\cite{Luscher:2007se}. While this construction generalises straightforwardly to multiple levels, here we restrict the discussion to the finest and coarsest levels only. Accordingly, $D_c$ denotes the operator on the coarsest grid, and $R$ and $P$ represent the composite restriction and prolongation operators mapping directly between the finest and coarsest grids.

To define multigrid-based projectors, one typically requires that $R$ and $P$ preserve the identity,
\begin{equation}
\mathbbm{1}_c \equiv R \mathbbm{1} P\,,
\end{equation}
with $\mathbbm{1}_c$ the identity on the coarse space. This leads to the following projectors acting on the fine level:
\begin{equation}
\mathbb{P} \equiv PR\,,\qquad\mathbb{P}_{D,r} \equiv PD_c^{-1}RD\,,\qquad\mathbb{P}_{D,l} \equiv DPD_c^{-1}R\,,\qq{and}\mathbb{P}_{D,lr} \equiv \sqrt{D}PD_c^{-1}R\sqrt{D}\,,
\end{equation}
The operator $\mathbb{P}$ simply projects the identity onto the coarse subspace, while the remaining operators incorporate the coarse inverse and define suitable projections for the quark propagator. For instance, using $\mathbb{P}_{D,r}$ one can decompose the propagator $S = D^{-1}$ as
\begin{equation}
    S = \mathbb{P}_{D,r} S + (\mathbbm{1}-\mathbb{P}_{D,r}) S \qquad\text{with}\qquad \mathbb{P}_{D,r} S = PD_c^{-1}RD D^{-1} = PD_c^{-1}R \equiv S^\MG\,.
\end{equation}
An equivalent representation follows from the other projectors, finding
\begin{equation}
S-S^\MG = (\mathbbm{1}-\mathbb{P}_{D,r}) S = S(\mathbbm{1}-\mathbb{P}_{D,l}) = \sqrt{S}(\mathbbm{1}-\mathbb{P}_{D,lr})\sqrt{S}\,.
\end{equation}
The connection with exact deflation becomes explicit if $R$ and $P$ are chosen as the left and right IR eigenvector matrices of $D$, respectively:
\begin{equation}
R=U^\IR\,,\qquad P=V^\IR\,,\qq{and} \Lambda^\IR \equiv U^\IR DV^\IR\qquad\Longrightarrow\qquad \mathbb{P} = V^\IR U^\IR\,,\qquad D_c = \Lambda^\IR\,,\qq{and} S^\MG=S^\IR\,.
\end{equation}
Thus, the multigrid propagator $S^\MG$ approaches the IR propagator $S^\IR$ to the extent that $P$ and $R$ accurately capture the low-mode subspace. When specialising this general AMG framework to lattice QCD, additional symmetry-preserving structures are typically imposed:
\begin{itemize}
    \item To preserve the sparsity of the Dirac operator, $P$ and $R$ act on space-time aggregates of the lattice, ensuring that $D_c$ retains a nearest-neighbor structure. This is also in line with the property of \textit{local coherence}~\cite{Luscher:2007se}, namely, eigenvectors restricted to aggregates still maintain significant overlap with the low-mode subspace.

    \item To preserve $\gamma_5$-hermiticity and related symmetries, $P$ and $R$ are constructed to act separately on left- and right-handed spin components, enforcing $\gamma_5$-compatibility:
    \begin{equation}
        \gamma_{5,c} = R \gamma_5 P\,,\qquad \gamma_{5,c} R = R \gamma_5\,,\qq{and}\gamma_5 P = P \gamma_{5,c}\,.
    \end{equation}
    This allows one to relate left and right eigenspaces and consistently choose
    \begin{equation}
        R = P^\dagger \qquad\Longrightarrow\qquad D_c = P^\dagger D P \,.
    \end{equation}
    It also follows that $\gamma_5$-hermiticity is preserved at the coarse level,
    \begin{equation}
        D_c^\dagger \quad=\quad P^\dagger D^\dagger P \quad=\quad P^\dagger \gamma_5 D \gamma_5 P \quad=\quad \gamma_{5,c} P^\dagger D P \gamma_{5,c} \quad=\quad \gamma_{5,c} D_c \gamma_{5,c}\,,
    \end{equation}
    allowing the definition of a Hermitian coarse operator
    \begin{equation}
        Q_c \equiv D_c \gamma_{5,c} = Q_c^\dagger \,.
    \end{equation}

    \item The $\gamma_5$-compatibility is particularly crucial for twisted-mass fermions~\cite{Alexandrou:2016izb}, as it ensures that the twisted-mass term retains its linear form at all multigrid levels:
    \begin{equation}
    D_c(\mu) \equiv P^\dagger (D_{\W} + i\mu \gamma_5) P = P^\dagger D_\W P + i\mu \gamma_{5,c} = D_{\W, c} + i\mu \gamma_{5,c}\,,
    \end{equation}
    where $D_{\W}$ denotes the massless Wilson operator. The coarse operator, therefore, inherits the same structure as the fine twisted-mass operator, i.e., the twisted mass $\mu$ provides a low-mode cutoff, while the eigenvectors remain those of the massless Wilson operator. In practice, this allows a prolongation operator $P$ constructed at a given $\mu$ to be reused for both signs and arbitrary values of $\pm \mu$, which is advantageous when employing multigrid as a preconditioner for twisted-mass simulations and observables~\cite{Alexandrou:2016izb,Alexandrou:2018wiv,Bacchio:2017pcp}.
\end{itemize}

\subsection{From multigrid preconditioning to multigrid LMA}

The properties discussed above closely mirror those of exact LMA, but within a more general multigrid framework. It is important to identify, however, where the construction of multigrid preconditioning deviates from the construction of a multigrid-based LMA approach. In particular:
\begin{description}
\item[\textit{Multigrid preconditioning is dynamic}] A central feature of multigrid solvers for lattice QCD is their adaptivity. In a typical multigrid correction, the inverse of $D_c$ is computed only to very low accuracy, and this approximate correction is then combined with a smoother. Additionally, the solver dynamically cycles among levels, often via K-cycles, using nested Krylov methods to determine whether to recurse further to coarser grids or return to finer levels. This adaptivity optimises the computational effort by adjusting the coarse-level work according to the current condition number and convergence state. As a result, the effective multigrid correction operator is not fixed, but evolves during the solve. Such strategies necessitate combining the preconditioner with a \textit{flexible} Krylov solver.

\item[\textit{LMA requires static operators}] In contrast, LMA demands a strictly static construction where all sources of adaptivity must be removed so that the multigrid-based IR propagator defines a fixed operator.
In particular, if the inverse of $D_c$ is considered, this should be computed to high precision, rather than approximately with an adaptive approach.
This is especially challenging for twisted-mass fermions, where the coarse operator is typically highly ill-conditioned; even achieving a relatively loose tolerance (e.g.\ a $10^{-1}$ relative residual) can already require  hundreds of iterations~\cite{Alexandrou:2016izb}.
A practical resolution is to perform exact deflation on the coarsest grid, which is the strategy adopted here. We define
\begin{equation}\label{eq:S_MG_IR}
    S^{\MG,\IR} = P S^\IR_c P^\dagger \qquad\text{with}\qquad S^\IR_c =  \sum_{j=0}^{N^\MG_{\rm ev}} \frac{1}{\lambda_{c,j}}\,\gamma_{5,c} v_{c,j} v^\dagger_{c,j} \qq{and} Q_c v_{c,j} = \lambda_{c,j}v_{c,j}\,
\end{equation}
so that the exact LMA construction is carried out on the coarsest grid. This significantly reduces the cost, both because of the smaller system size and because fewer eigenvectors are required, with the prolongation operator effectively lifting the coarse IR subspace to the fine level. 

\item[\textit{Different hyper-parameters and tuning}] As a consequence, the set of hyperparameters and their tuning strategy differs substantially from standard multigrid preconditioning. Since LMA excludes any dynamic components, only parameters entering the definition of the coarse operator remain relevant, i.e., the aggregation size between levels, the number of null vectors, and the number of IR modes treated exactly on the coarse grid, $N^\MG_{\rm ev}$. As we will see, in practice, the optimal values of these parameters also differ significantly from the corresponding ones that are optimal in multigrid preconditioning.
Additionally, the realisation of optimal multigrid LMA will differ significantly with the kind of fermion discretisation, in the same way multigrid preconditioning does for Wilson~\cite{Brannick:2007ue,Babich:2010qb,Frommer:2013fsa}, twisted-mass~\cite{Alexandrou:2016izb}, overlap~\cite{Brannick:2014vda}, domain-walls~\cite{Brower:2020xmc}, and staggered~\cite{Brower:2018ymy} fermions. Thus, each of these requires dedicated studies.

\end{description}

\subsection{Multigrid LMA of quark-bilinear correlation functions}

\begin{figure}[tbh]
    \centering
    \includegraphics[width=0.49\linewidth]{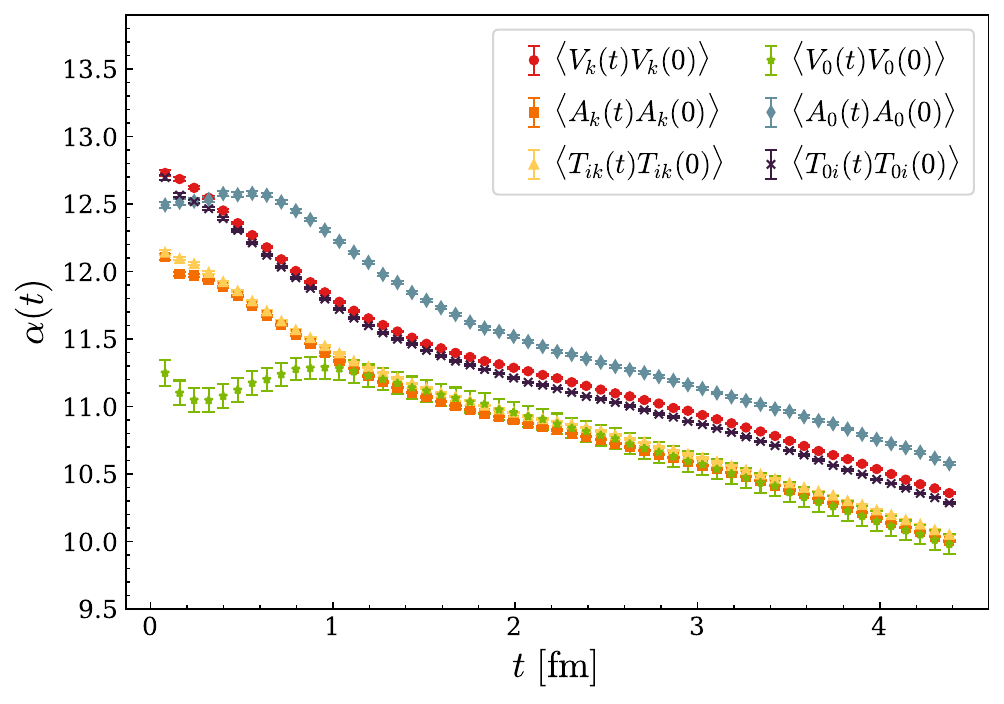}
    \includegraphics[width=0.49\linewidth]{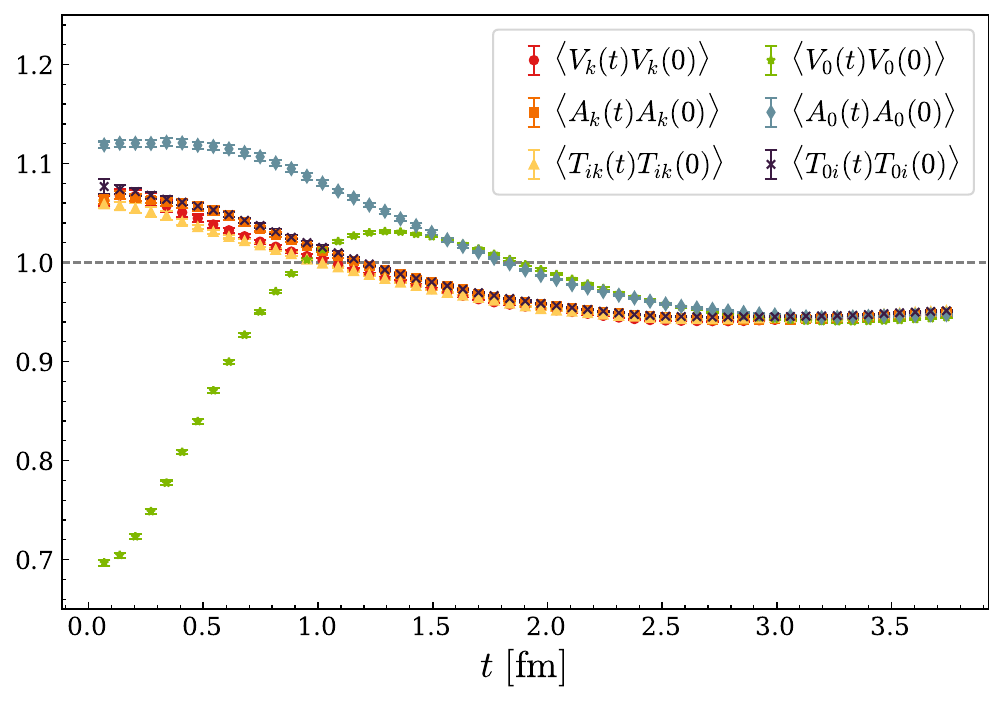}
    \caption{\textit{Left}: Values of control-variate parameter $\alpha(t)$ used to rescale the $C^\MG(t)$ correlation function to improve its overlap with the standard correlation function $C(t)$ up to an overall observable and time-dependent normalisation. Its value is determined according to \cref{eq:corr_defl_MG} and is provided for each quark bilinear current shown in~\cref{fig:mgGain}. \textit{Right}: The control-variate parameter $\alpha(t)$ computed on the C80 ensemble in the exact LMA setup where \cref{eq:corr_defl} holds.} 
    \label{fig:mgAlpha}
\end{figure}

As given in~\cref{eq:S_MG_IR}, our strategy combines multigrid operators with exact deflation to construct the propagator $S^{\MG,\IR}$. A first consequence of this choice is that the identity in~\cref{eq:corr_defl} no longer holds. This is due to the fact that $S^{\MG,\IR}$ involves a nested projection, whereas the identity would apply to either the full $S^{\MG}$ or the exact $S^{\IR}$.
Nevertheless, \cref{eq:corr_defl} can still be used as a guiding principle to define an improved estimator by embedding it in a control variates framework:
\begin{equation}
C^{\LMA,\MG}_\eta (t) \equiv C_\eta(t) + \alpha(t)\Big( C^\MG_{\smallcap{impr.}}(t) - C^\MG_\eta(t)\Big)\,,\qquad
\alpha(t) = \frac{{\rm Cov}[C_\eta(t), C^\MG_\eta(t)]}{{\rm Var}[C^\MG_\eta(t)]}\,.
\label{eq:corr_defl_MG}
\end{equation}
Here $\alpha(t)$ is chosen to minimise the variance of $C^{\LMA,\MG}_\eta(t)$ at each time slice. In this formulation, $C^\MG_{\smallcap{impr.}}(t)$ acts as a low-noise control observable, while $C^\MG_\eta(t)$ provides a stochastic estimator with the same expectation value. Their difference therefore has vanishing mean but non-trivial correlation with $C_\eta(t)$, enabling a variance reduction without introducing bias: when $C^\MG_\eta(t)$ is strongly correlated with $C_\eta(t)$, a substantial variance reduction can be achieved up to the limit of perfect correlation, where the stochastic estimators cancel. In a setup where~\cref{eq:corr_defl} holds, one expects $\alpha(t)=1$. However, as shown in the right panel of \cref{fig:mgAlpha}, this is not exactly true, and we find that the optimised  $\alpha(t)$ deviates from unity by less than $10\%$ in the large time limit. This would lead to a further reduction in the statistical error, which is, though, only about $2\%$ and thus overall negligible. 

In the case of mgLMA, since the stochastic source is coarsened before applying the IR component of the multigrid propagator, the coefficient is expected to reflect the reduced dimensionality.
The resulting values of $\alpha(t)$ for the results presented below are shown in the left panel of~\cref{fig:mgAlpha}. Their relatively large magnitude is consistent with this expectation, while exhibiting a mild, non-trivial time dependence, as well as a weak dependence on the observable considered. Overall, the behaviour is smooth and quite reasonable.

As in exact LMA, the two additional correlation functions in~\cref{eq:corr_defl_MG} are built from the deflated propagator. The correlator $C^\MG_\eta(t)$ has the same contraction structure and stochastic sources $\eta$ as $C_\eta(t)$, but with the propagator replaced by $S^{\MG,\IR}_\eta$, while $C^\MG_{\smallcap{impr.}}(t)$ is evaluated at higher statistics. Several strategies are possible for the latter. Ideally, one would compute an all-to-all correlator, but already for bilinear correlators, the all-to-all construction becomes demanding:
\begin{equation}\label{eq:corr_MG_all}
C^\MG_{\smallcap{all\text{-}to\text{-}all}} (t_1,t_2,\vb*{p}_1,\vb*{p}_2,\Gamma_1,\Gamma_2) = 
\Tr\left[
\proj{p_2}{t_2} \Gamma_2\, P\, S^\IR_c P^\dagger\, \proj{p_1}{t_1} \Gamma_1 \,P\, S^\IR_c P^\dagger
\right]\,, 
\end{equation}
as it requires constructing the coarse operator $P^\dagger \proj{p}{t} \Gamma P$ for each choice of time, momentum, and Dirac structure.

A possible improvement is to design the prolongation operator $P$ to be compatible with quark bilinear structures, which is particularly natural at zero momentum. If $P$ does not aggregate in the time direction and is fully compatible with spin structures (i.e.\ $\Gamma$-compatible), one obtains
\begin{equation}
    P^\dagger\, \proj{0}{t}\,{\Gamma}\, P = \projc{0}{t}\,{\Gamma_c}\,.
\end{equation}
In this case, one can follow the same strategy as in~\cref{eq:corr_IR_bil} and write
\begin{equation}\label{eq:corr_MG_all_2}
C^\MG_{\smallcap{all\text{-}to\text{-}all}}(t_1,t_2,\Gamma_1,\Gamma_2)
= \sum_{j, k =1}^{N_{{\rm ev},c}^\IR} \frac{1}{\lambda_{c,j}\,\lambda_{c,k}^*} E^\MG_{jk}(t_1, \Gamma_1\gamma_5)\, E^\MG_{kj}(t_2, \gamma_5\Gamma_2)
\qq{with}
E^\MG_{jk}(t, \Gamma) = v_{c,j}^\dagger\,\projc{0}{t}\,{\Gamma_c}\,{v_{c,k}} \,.
\end{equation}
While promising and computationally efficient for bilinear observables, this approach requires additional implementation effort and sacrifices generality, and is therefore not pursued here. Instead, we approximate $C^\MG_{\smallcap{impr.}}(t)$ through high-statistics sampling,
\begin{equation}
    C^\MG_{\smallcap{impr.}}(t) = \frac{1}{N^\MG_{\eta}} \sum_\eta^{N^\MG_{\eta}} C^\MG_\eta(t),
\end{equation}
leveraging the efficient application of the coarse-level correction via QUDA together with optimised accumulated contractions.

\subsection{Using QUDA for multigrid LMA}

Let us now emphasise that the efficient testing and implementation of this multigrid LMA strategy is greatly facilitated by the use of the QUDA library~\cite{Clark:2009wm,Babich:2011np,Clark:2016rdz}. QUDA provides key capabilities, including efficient setup and application of multigrid operators, as well as deflation of low modes on the coarse grid. In practice, the multigrid propagator is obtained by invoking the multigrid preconditioner with a specific choice of parameters: no smoother iterations are applied, and a single V-cycle is performed, corresponding to a direct traversal to the coarsest grid and back without intermediate iterations. On the coarsest level, a deflated solver is employed with zero iterations, such that only the infrared (IR) component of the operator is effectively applied. This yields an extremely efficient application of the preconditioner, requiring only a fraction of a second per propagator. For the setup phase, instead, standard strategies are used.
An additional feature of QUDA exploited here is the use of the even–odd preconditioned operator, rather than the standard Dirac operator, at all levels of the multigrid hierarchy.

\subsection{Results on the B96 ensemble}

The motivation for employing multigrid LMA on large-volume ensembles stems from the unfavourable $V^2$ scaling of exact LMA. To illustrate this, we consider representative numbers from this work.
For the B64 ensemble, production runs typically use 4 or 8 GPU nodes, depending on workload and memory constraints. With exact LMA, however, storing the full set of $400$ eigenvectors already requires running on 16 nodes. While this corresponds to a factor of 2–4 increase in node count, the overall scaling remains close to ideal at this volume, since the evaluation of deflated correlation functions is still dominated by full-size operator applications using the strategy discussed in~\cref{sec:LMA}. As a result, the overhead associated with the larger node count remains manageable for B64.

The situation changes dramatically for larger volumes. The B96 ensemble, with approximately five times the volume, is normally executed on $32$ GPU nodes. Scaling exact deflation to this case would require roughly five times more eigenvectors to maintain comparable performance, i.e.\ about $2000$ modes. Combined with the increased lattice size, this leads to a prohibitive memory footprint, namely, storing these eigenvectors would require on the order of $950$ GPUs with 64~GB memory, corresponding to at least $256$ nodes in a realistic setup. This represents an order-of-magnitude increase over the baseline node count and would severely impact both resource usage and parallel efficiency, effectively negating the computational advantage of exact LMA.

In contrast, multigrid LMA achieves a satisfactory noise reduction with significantly reduced resources. In this work, we obtain a moderate improvement for the B96 ensemble using only 54 nodes. This requires a careful retuning of the multigrid setup. In particular, standard multigrid parameters are insufficient. As in exact LMA, a large number of modes is needed to accurately capture the low-energy subspace. In the multigrid context, this is controlled by the aggregation strategy, which determines the size of the coarse grid and, thus, the dimensionality of the coarse operator relative to the fine Dirac operator. 

The parameters used are summarised in~\cref{tab:mgLMA}. Conventional multigrid preconditioning typically employs a three-level hierarchy with aggregation factors of $4^4$ between the fine and first coarse level and $2^4$ between subsequent levels (up to minor variations, as in the present case, where the lattice size $96$ favours aggregation factors of $3$). For LMA, however, we find that a much finer aggregation is required. In particular, the overall aggregation between the fine and coarsest level is reduced by a factor of $20$, leading to a correspondingly larger coarse operator. Despite this, the coarse operator remains smaller than the fine operator by a factor of $48$ once spin aggregation and the use of $32$ null vectors are taken into account. On this enlarged coarse operator, we compute $6000$ low modes to construct the multigrid-based IR subspace. 

\begin{table}[tbh]
    \centering
    \begin{tabular}{lcccccc}
    \toprule
         Type & Cycle & Smoother & L1 aggreg. & L2 aggreg. & $N^\MG_{\mathrm{col}}$ & $N^\MG_{\mathrm{ev}}$\\
    \midrule
        mgLMA & V-cycle & No & $4{\times}4{\times}2{\times}1$ & $2{\times}2{\times}2{\times}1$ & 32 & 6000\\
        Precond.~~ & K-cycle & Yes & $3{\times}3{\times}3{\times}6$ & $2{\times}2{\times}2{\times}4$ & 24 & 4000\\
    \bottomrule
    \end{tabular}\hspace{1cm}
    \begin{tabular}{lccc}
    \toprule
        Ensemble & $N_U$ & $N^\MG_\eta$ & $N_\eta$\\
    \midrule
        B96 & 490 & 11200 & 768\\
    \bottomrule
    \end{tabular}
    \caption{\textit{Left table:} comparison of the relevant AMG parameters after tuning for its application in LMA and as a preconditioner. Columns two through seven indicate ii) the multigrid cycle used, iii) whether a smoother is employed, iv) the aggregation size between the fine and first coarse level, v) the aggregation size between the first and second coarse levels, vi) the number of null vectors used ($N^\MG_{\mathrm{col}}$) on all levels, and vii) the number of deflated modes on the coarsest level. \textit{Right table:} statistics used for the results shown on the B96 ensemble in terms of the number of gauge configurations $N_U$, the number of stochastic sources for the improved estimator of the multigrid correlation function $N^\MG_\eta$, and the number of stochastic sources for the full correlator $N_\eta$.}
    \label{tab:mgLMA}
\end{table}

From our tests, the most critical parameter in this setup is the aggregate size, which determines the coarse lattice resolution. If the aggregates are too large, no significant noise reduction is observed, which is analogous to retaining too few low modes in standard LMA. Conversely, overly small aggregates lead to costs comparable to exact LMA, eliminating any practical advantage. A careful tuning is therefore required. By contrast, we observe only a mild dependence on the number of low modes. The choice of 6000 modes should be regarded as conservative, corresponding to the maximum number that can be accommodated without degrading performance, i.e. within the available device memory.

It is also worth noticing that the resulting multigrid setup differs substantially from the one used for preconditioning and is also significantly less efficient if used for that purpose. Therefore, we again split the production of correlation functions into two independent runs: one performed using the standard node count and the multigrid setup optimised for preconditioning, used to produce the full correlation functions $C_\eta$; and another using the modified setup employed for the multigrid-based correlation functions $C^\MG_{\smallcap{impr.}}$ and $C^\MG_{\eta}$.
\begin{figure}[tbh]
    \centering
    \includegraphics[width=1.0\linewidth]{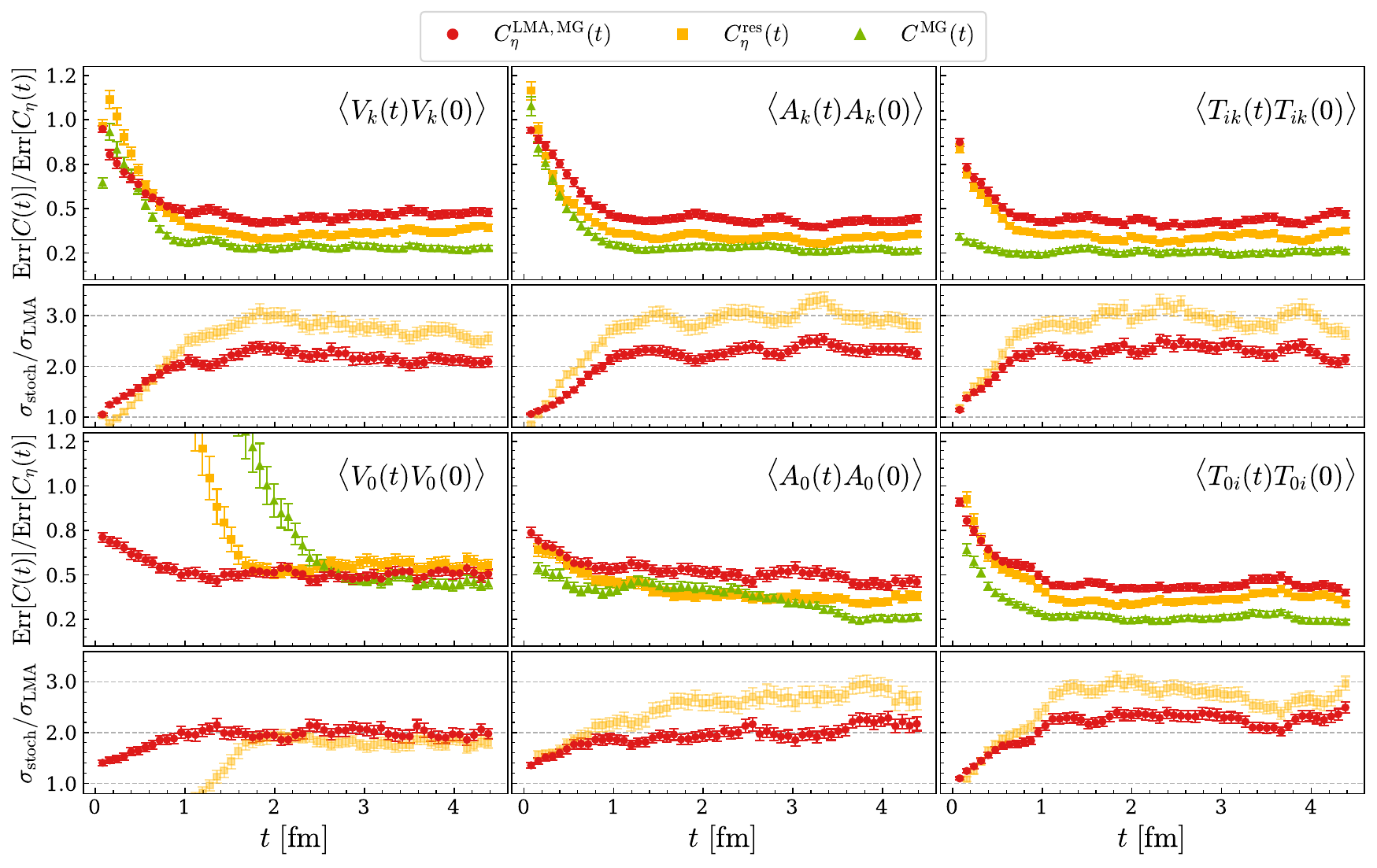}
    \caption{Error reduction (top panels) and gain (bottom panels) in the statistical uncertainty of the correlation functions relative to the standard stochastic estimate $C_\eta$. The red curves correspond to the full mgLMA-improved correlator, while the orange and green curves show the contributions of its two components; see~\cref{eq:corr_defl_MG_2}.}
    \label{fig:mgGain}
\end{figure}

Results for this setup are shown in~\cref{fig:mgGain}, where we focus on the statistical error of the various correlators entering $C^{\LMA,\MG}_\eta (t)$, defined in~\cref{eq:corr_defl_MG}. The remaining two contributions are
\begin{equation}\label{eq:corr_defl_MG_2}
   C^{\LMA,\MG}_\eta (t)\equiv C^{{\rm res},\MG}_\eta(t)+C^{\MG}(t) \qq{with} C^{{\rm res},\MG}_\eta(t) \equiv C_\eta(t) - \alpha(t)\, C^{\MG,\IR}_\eta(t)\qq{and}    C^{\MG}(t) \equiv  \alpha(t)\, C^{\MG,\IR}_{\smallcap{impr.}}(t)\,,
\end{equation}
which sum to $C^{\LMA,\MG}_\eta (t)$. Assuming these two contributions are uncorrelated, i.e. in the regime where stochastic noise dominates, the quadratic sum of their individual errors should reproduce the statistical error of $C^{\LMA,\MG}_\eta (t)$. This assumption holds well in all cases except at short Euclidean times, where stochastic noise saturates, and for $\langle V_0(t)V_0(0)\rangle$, which constitutes a special case due to the conserved current.

Guided by this, the number of sources (11200 vs 768) for the improved MG estimator is chosen such that its error remains smaller than that of the corresponding residual contribution. In this way, the residual term dominates the total error while maintaining a balanced computational cost for the MG component. In~\cref{fig:mgGain}, we also display the noise reduction for both the total and residual components, where the former reflects the achieved gain (about 2 in all cases), while the latter indicates the potential asymptotic improvement (about 3 in most cases) as the MG estimator error is further reduced. Within the range of stochastic sources considered here, no saturation of the MG estimator noise is observed, leaving room for further error reduction, for instance, by increasing the number of sources or by employing all-to-all estimators as in~\cref{eq:corr_MG_all}.
Regarding the latter, as can be seen in \cref{tab:mgLMA}, the current choice of aggregation does not coarsen in the time direction and would thus enable an efficient evaluation of the all-to-all contribution in~\cref{eq:corr_MG_all_2}, provided that a fully $\gamma$-compatible (i.e., spin-diluted) multigrid setup is employed, which is not currently the case. Our findings strongly support pursuing this approach for quark bilinear correlation functions, as it would allow a substantial improvement in noise reduction (from a factor of about 2 to about 3, comparing the red and orange points in~\cref{fig:mgGain}) at significantly lower cost to compute the improved estimator, since the high-statistics multigrid correlation function would not need to be computed.

\subsection{Comparison with the analysis of Ref.~\texorpdfstring{\cite{Gruber:2024cos}}{\texttt{Gruber:2024cos}}}

A mgLMA approach was also introduced in Ref.~\cite{Gruber:2024cos}. While the two approaches share the same basic idea of exploiting the local coherence of the low-mode subspace to construct a hierarchy of reduced-dimensional operators, there are several important differences. First, our implementation is designed and optimized for ensembles at the physical pion mass and for the large physical volumes relevant for current precision calculations, whereas Ref.~\cite{Gruber:2024cos} demonstrated the method on $N_f=2$ non-perturbatively $O(a)$-improved Wilson fermions at $m_\pi\simeq270$~MeV and volumes up to $L\simeq4$~fm. Second, we formulate the variance reduction in terms of a control variate based on the multigrid approximation to the low-mode contribution, as in~\cref{eq:corr_defl_MG}. Besides sharing all the advantages of \cref{eq:corr_defl} discussed in \cref{sec:LMA_remarks}, this avoids the explicit evaluation of all individual terms generated by inserting the multigrid decomposition into the quark-line contraction. In particular, for a contraction involving $n_p$ propagators and a hierarchy with $N_\ell$ levels, an explicit decomposition can generate a number of contributions that grows as $N_\ell^{n_p}$. The control-variate formulation instead requires the evaluation of the full correlator together with the corresponding multigrid estimator, namely two terms only, substantially simplifying the implementation.

A further difference concerns the use of exact deflation. In Ref.~\cite{Gruber:2024cos}, exact low modes of the fine-grid Dirac operator are used to construct the block-localised basis at the first level of the hierarchy, with the subsequent coarse spaces obtained recursively from this basis. In our approach, the multigrid prolongator is constructed using the AMG procedure, and exact low-mode averaging is subsequently performed on the coarsest operator. The resulting coarse-grid low-mode contribution is therefore used as an approximation to the infrared part of the fine-grid propagator rather than as an explicit decomposition of the latter into all multigrid levels. This distinction is particularly useful in the present application, where the number of fine-grid low modes required for conventional LMA becomes prohibitively large.
\def\etap{{x_{\!s}}}

\section{Results on Exact LMA for Nucleon two- and three-point functions}
\label{sec:baryon_results}

Returning to the exact LMA, we present and discuss results for nucleon two- and three-point correlation functions. After a brief introduction of the correlators, we examine the exact all-to-all contribution, which proves prohibitively expensive due to the large number of eigenvectors required, and thus focus on high-statistics estimates. We then report results for both two- and three-point functions, analysing the gain from LMA and its dependence on the choice of current as well as on the temporal separation.

\subsection{Nucleon two- and three-point functions with LMA}

The nucleon interpolating field commonly used is
\begin{equation}
\chi_N^{\gamma}(x)=\epsilon^{abc} \left[u(x)_a^T \mathcal{C} \gamma_5 d(x)_b\right] u(x)_c^{\gamma}\,,
\end{equation}
where $u$ and $d$ are the up- and down-quark interpolating fields, $\mathcal{C}$ is the charge conjugation matrix and $\epsilon$ is the totally antisymmetric Levi-Civita tensor. After performing the Wick contractions, the nucleon two-point function with generic spatial momenta $\vb*{p}$, reads as
\begin{align}
C_N(t, \vb*{p})\equiv &
\Tr{\mathcal{P}_{\gamma\gamma'}\expval{\chi_N^{\gamma}(t,\vb*{p})\,\bar\chi_N^{\gamma'}(0,\vb*{p})}} \nonumber\\
& = \sum_{\vb*{x}, \vb*{y}} e^{i\left(\vb*{x}-\vb*{y}\right)\cdot\vb*{p}} \, 
\mathcal{P}_{\gamma\gamma'}\, \mathcal{W}_{\alpha\beta}^{abc}\, \mathcal{W}_{\alpha'\beta'}^{a'b'c'}\,
S_d(x,y)_{bb'}^{\beta\beta'} \left[S_u(x,y)_{aa'}^{\alpha\alpha'}S_u(x,y)_{cc'}^{\gamma\gamma'}-S_u(x,y)_{ac'}^{\alpha\gamma'}S_u(x,y)_{ca'}^{\gamma\alpha'}\right]
\,,
\label{eq:interpolator}
\end{align}
where the unpolarized positive-parity projector $\mathcal{P}$ and the tensor $\mathcal{W}$ are defined by 
\begin{equation}\label{eq:nucl_W_P}
\mathcal{W}_{\alpha\beta}^{abc} \equiv \epsilon^{abc}\,\cgfi_{\alpha\beta} \qq{and} \mathcal{P} = \frac{\mathds{1} + \gamma_0}{2}\,.
\end{equation}
Inserting the LMA propagator of \cref{eq:prop_defl} yields a purely stochastic UV contribution, an exact IR contribution, and six additional mixed IR--UV terms. As shown in \cref{app:proof}, when point sources are used, the purely stochastic and mixed contributions combine to reproduce the standard correlator with the stochastic IR correlator subtracted. Consequently, \cref{eq:corr_defl} holds, and only the IR correlator needs to be computed in addition to the standard one.

\begin{figure}[tbh]
    \centering
    \includegraphics[width=0.65\linewidth]{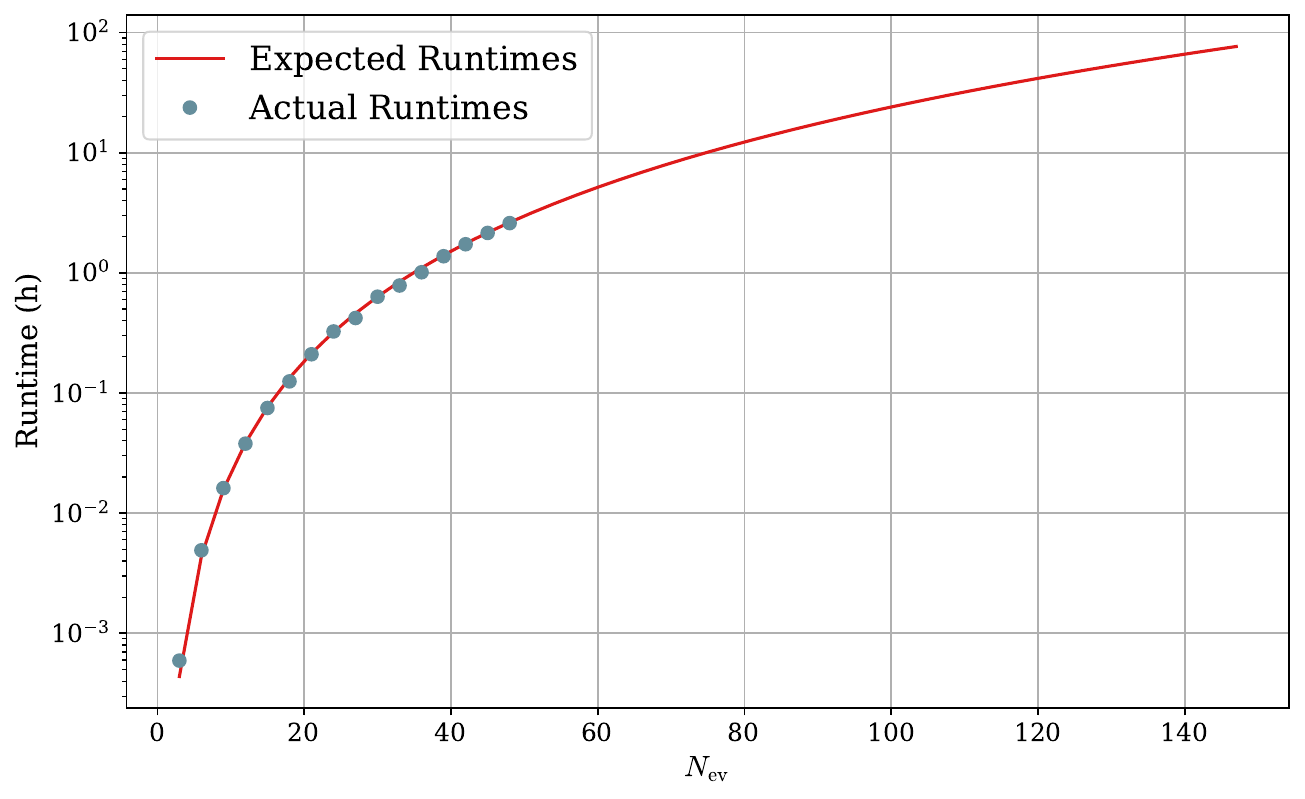}
    \caption{The measured (grey-blue dots) and predicted according to \cref{eq:cIR_bar_scal} (red solid line) runtime for the exact IR correlators of \cref{eq:correee} as a function of the number of eigenvectors $N_{\EV}^{\IR}$ on the B48 ensemble ($V/a^4 = 48^3\times96$) on a single node.}
    \label{fig:runtime}
\end{figure}

The evaluation of the exact all-to-all IR correlator turns to be computationally demanding. Here we outline the most efficient strategy we have identified. We begin by defining the Fourier-transformed contraction of three eigenvectors over colour and two spin indices,
\begin{align}
E_{ijk}^\gamma(t,\vb*{p})=
\sum_{\vb*{x}} e^{i\vb*{x}\cdot\vb*{p}}\,
\mathcal{W}_{\alpha\beta}^{abc}\,
v_i(\vb*{x},t)_a^{\alpha}
v_j(\vb*{x},t)_b^{\beta}
v_k(\vb*{x},t)_c^{\gamma}\,,
\label{eq:Eijk}
\end{align}
and its conjugate
\begin{align}
E_{ijk}^\gamma(t,\vb*{p})^{\star}=
\sum_{\vb*{x}} e^{-i\vb*{x}\cdot\vb*{p}}\,
\mathcal{W}_{\alpha\beta}^{abc}\,
v_i^{\dagger}(\vb*{x},t)_a^{\alpha}
v_j^{\dagger}(\vb*{x},t)_b^{\beta}
v_k^{\dagger}(\vb*{x},t)_c^{\gamma}\,,
\end{align}
with repeated indices implicitly summed. The exact IR correlator can then be written as
\begin{align}
C_{N}^{\IR}(t,\vb*{p})= & \;(\gamma_5 \mathcal{P})_{\gamma\gamma'}\!\!\sum_{i,j,k=1}^{N_{\EV}^\IR} \frac{1}{\lambda_i\lambda_j^*\lambda_k}  
 \left[
E_{ijk}^\gamma(t_0+t,\vb*{p})^{\star} - 
E_{kji}^\gamma(t_0+t,\vb*{p})^{\star}
\right]E_{ijk}^{\gamma'}(t_0,\vb*{p})
\,.
\label{eq:corrEE}
\end{align}
We first note that the quantity $E_{ijk}^\gamma$ is symmetric under the exchange of $i\leftrightarrow j$, having
\begin{equation}
    E_{ijk}^\gamma(t, \vb*{p}) = E_{jik}^\gamma(t, \vb*{p}) \qquad\Longrightarrow\qquad N^{E}_{\smallcap{dof}} = \frac{\left(N_{\EV}^\IR\right)^2 (N_{\EV}^\IR + 1)}{2}\,,
\end{equation}
and, thus, with $N^{E}_{\smallcap{dof}}$ independent entries. 
This number can be further reduced by noticing that in $C^\IR_N$ we need an antisymmetric contribution under the exchange of $i\leftrightarrow k$, namely
\stepcounter{footnote}
\footnotetext[\value{footnote}]{We thank the anonymous referee for suggesting the relations on the right-hand-side of~\cref{eq:cIR_bar_sym} leading to an improved scaling in~\cref{eq:cIR_bar_scal}.}
\begin{equation}\label{eq:cIR_bar_sym}
\Tilde{E}_{ijk}^\gamma(t, \vb*{p}) \equiv E_{ijk}^{\gamma}(t,\vb*{p}) - E_{kji}^{\gamma} (t,\vb*{p})=-\Tilde{E}_{kji}^\gamma(t, \vb*{p})
\qquad \text{respecting\footnotemark[\value{footnote}]} \qquad
\Tilde{E}_{ijk} + \Tilde{E}_{jki} + \Tilde{E}_{kij} = 0\,.
\end{equation}
By using these two relations, the number of independent entries reduces to 
\begin{equation}\label{eq:cIR_bar_scal}
N^{\tilde{E}}_{\smallcap{dof}} = \frac{N_{\EV}^\IR \left(\left(N_{\EV}^\IR\right)^2 - 1\right)}{3}\,.
\end{equation}
Indeed,
\begin{align}
    C_N^\IR(t,\vb*{p})
    =&\;(\gamma_5 \mathcal{P})_{\gamma\gamma'}\!\!
      \sum_{i,j=1}^{N_{\EV}^\IR} \sum_{k=1}^i 
      \frac{1}{\lambda_i\lambda_j^\star\lambda_k}\Big(
       \big[E_{ijk}^\gamma(t_0+t,\vb*{p})^\star - E_{kji}^\gamma(t_0+t,\vb*{p})^\star\big]E_{ijk}^{\gamma'}(t_0,\vb*{p}) \\
    &\qquad\qquad\qquad\qquad\qquad\qquad + 
       \big[E_{kji}^\gamma(t_0+t,\vb*{p})^\star - E_{ijk}^\gamma(t_0+t,\vb*{p})^\star\big]E_{kji}^{\gamma'}(t_0,\vb*{p})
       \Big) \nonumber \\
    = & \;(\gamma_5 \mathcal{P})_{\gamma\gamma'}\!\!
      \sum_{i,j=1}^{N_{\EV}^\IR} \sum_{k=1}^i
      \frac{1}{\lambda_i\lambda_j^\star\lambda_k}\,
      \tilde{E}_{ijk}^\gamma(t_0+t,\vb*{p})^\star\,\tilde{E}_{ijk}^{\gamma'}(t_0,\vb*{p})\label{eq:correee} \,.
\end{align}
This offers, to our knowledge, the most computationally efficient approach for all-to-all baryon correlation functions, requiring $N^{\tilde{E}}_{\smallcap{dof}}$ contractions, but nevertheless scaling with $\propto\left(N_{\EV}^\IR\right)^3$.
This cubic scaling of the contractions renders the computational cost prohibitive for large volumes, such as the one considered in this work, where at least $N_{\EV}^\IR \approx 500$ are required. Indeed, as shown in \cref{fig:runtime}, the runtime to compute the exact IR correlators already approaches one day of wall time for $N_{\EV}^{\IR} \approx 90$. Thus, all-to-all correlation functions are realistically unfeasible via this approach.

A simple way to address this issue is to still estimate the IR contribution stochastically, but with higher precision. To this end, we adopt the same strategy used for the standard correlator, evaluating both the full correlator and the IR part from point sources, while increasing the number of sources $N^\IR_{\etap}$ used for the latter. This leads to the definition
\begin{equation}
    C_{\smallcap{impr.}}^{\IR} = \frac{1}{N^\IR_{\etap}} \sum_\etap^{N^\IR_{\etap}} C^\IR_\etap(t)\,.
\end{equation}
The resulting LMA improved correlation function is then
\begin{align}
C_{\etap}^{\LMA}(t) & \equiv
C_{\etap}^{\mathrm{res}}(t) + C_{\smallcap{impr.}}^{\IR}(t)
\qq{with} C_{\etap}^{\mathrm{res}}(t) \equiv C_{\etap}(t) - C_{\etap}^{\IR}(t)\,.
\label{eq:fullCorr}
\end{align}

We restrict ourselves to point sources, as these are the preferred choice for the computation of three-point correlation functions that we also analyse in this work. In particular, they allow one to freely inject a momentum boost at the source, and, when combined with sequential inversions through a fixed sink, enable efficient calculations of form factors and related observables. We therefore define the nucleon three-point function with momentum transfer $\vb*{q}$ as 
\begin{align}
C_N^{3pt}(t_{\rm ins}, t_s; \Gamma, \mathcal{P}^{\Gamma}, \vb*{q}) & \equiv 
\Tr{
\mathcal{P}^{\Gamma}_{\gamma\gamma'} \expval{\chi_N^{\gamma}\left(t_s, \vb*{p}+\vb*{q}\right)\,\mathcal{J}_{\Gamma}\left(t_{\rm ins}, \vb*{q}\right)\,\Bar{\chi}_N^{\gamma'}\left(0, \vb*{p}\right)}},
\end{align}
where the current $\mathcal{J}_{\Gamma}$ is defined by
\begin{equation}
\mathcal{J}_{\Gamma}(z) = \Bar{u}(z)\Gamma u(z) \pm \Bar{d}(z)\Gamma d(z) \qq{with} \Gamma = \mathds{1}, \gfi\gmu, \sigma_{\mu\nu}
\end{equation}
corresponding  to scalar, axial, and tensor operator insertions, respectively, and $\mathcal{P}^{\Gamma}$ depends on the Dirac structure. We refer to the $u+d$ combination as isoscalar and to the $u-d$ combination as isovector. After renormalisation and in the large-time limit, the ratio of three-point to two-point functions yields the nucleon charges $g_{\Gamma}$
\begin{equation}
    \frac{C_N^{3pt}(t_{\rm ins}, t_s; \Gamma)}{C_N(t_s)} \xrightarrow[a\ll t_{\rm ins} \ll t_s]{} g_{\Gamma}\,,
\end{equation}
at zero momentum transfer $\vb*{q}=\vb*{0}$, and more generally, at finite $\vb*{q}$, to form factors.

\subsection{Error reduction in two- and three-point functions}

To investigate the performance of LMA for baryon correlators, we  generate results using the B48 and B64 ensembles (see \cref{tab:spectrum} for the simulation parameters). The maximal statistics collected are reported in \cref{tab:LMA_baryon}. Since these are test runs rather than production runs, the number of configurations is limited. However, we use production-level statistics for both the eigenvectors and the number of sources.
\begin{table}[tbh]
\centering
\begin{tabular}{ccccc}
\toprule
     & $N_U$ & $N_{\mathrm{ev}}^{\IR}$ & $N_{\etap}$ & $N_{\etap}^{\IR}$ \\
\midrule
B48     & 20 & 1500 & 100 & 2000 \\
B64     & 50 & 1000 & 100 & 2000 \\
\bottomrule
\end{tabular}
\caption{Maximal statistics collected for the baryonic two- and three-point correlation functions using the ETMC ensembles of~\cref{tab:spectrum}. The second column reports the number of configurations $N_U$, the third the maximum number of low modes exactly deflated $N_{\mathrm{ev}}^{\IR}$ and used in \cref{eq:fullCorr}, the fourth and fifth the maximum number of point sources $N_{\etap}$ and $N_{\etap}^{\IR}$ used to compute residual and IR improved contribution, respectively.}
\label{tab:LMA_baryon}
\end{table}

\begin{figure}[tbh]
\centering
\includegraphics[width=0.9\linewidth]{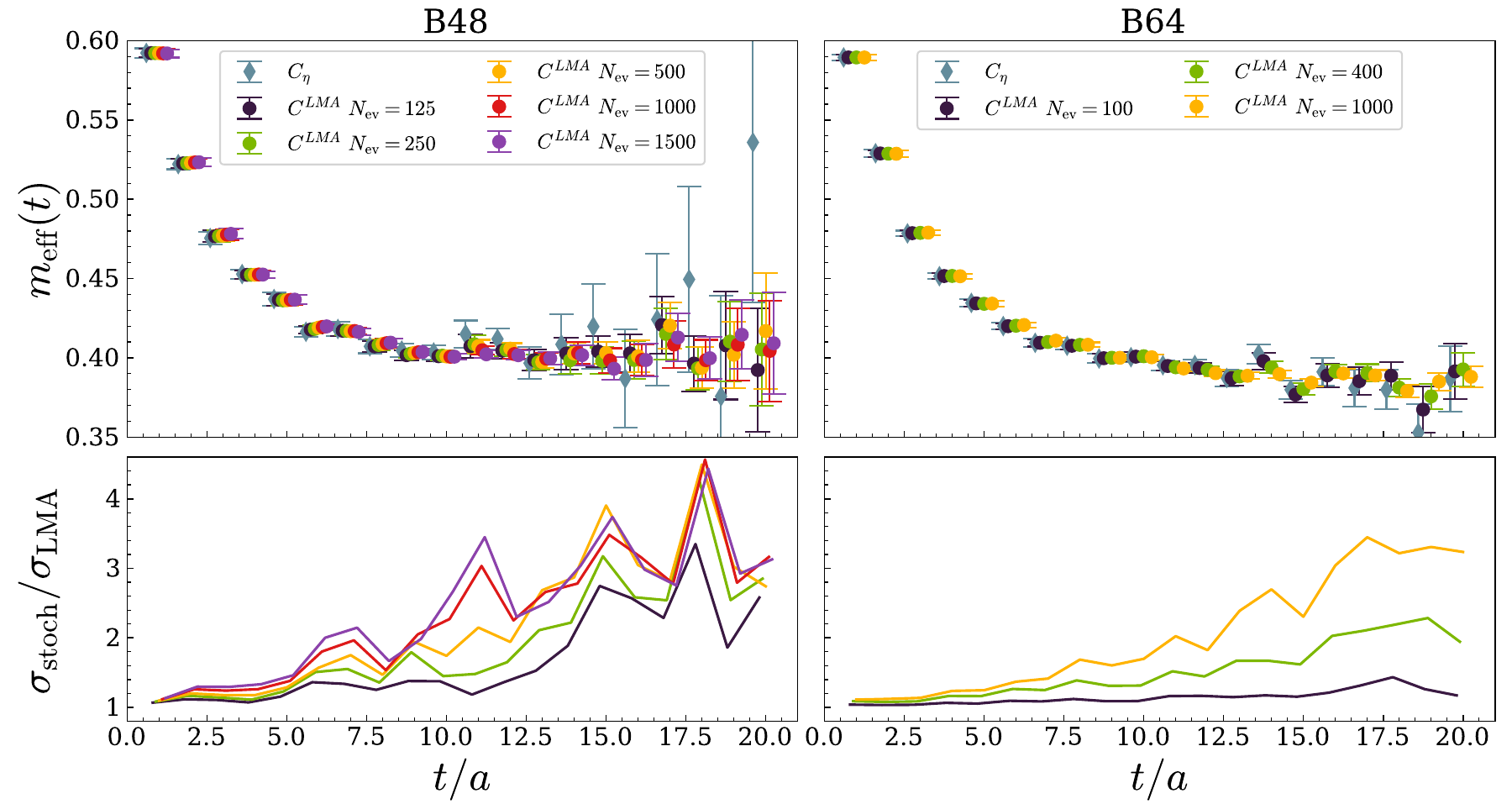}
\caption{\textit{Top}: Comparison of the nucleon effective mass extracted from standard correlators ($C_\etap$) and LMA-improved ones ($C^{\LMA}_{\etap}$). Each colour corresponds to a different number ($N_{\mathrm{ev}}^{\IR}$) of eigenvectors used for the infrared part. 
\textit{Bottom}: Error gain (see \cref{Eq:gain}) of the effective mass at each time slice for the B48 (left) and B64 (right) ensembles.}
\label{fig:effmass}
\end{figure}

In \cref{fig:effmass}, we present the LMA-improved nucleon effective mass for the B48 (left panel) and B64 (right panel) ensembles at maximal statistics, varying the number of eigenvectors. For the B48 ensemble, the gain saturates at approximately $N_{\EV}^{\IR} \approx 500$, which, when scaled with the physical volume, would correspond to about 1600 eigenvectors for B64. However, due to computational limitations, we restrict ourselves to a maximum of 1000 eigenvectors for the B64 ensemble, which nonetheless still yields a significant improvement. Importantly, we highlight that the saturation occurs at roughly four times more eigenvectors than in the meson case, indicating that a substantially larger number of modes is required to efficiently resolve nucleon observables compared to the corresponding mesons.

\begin{figure}[tbh]
    \centering
    \includegraphics[width=0.75\linewidth]{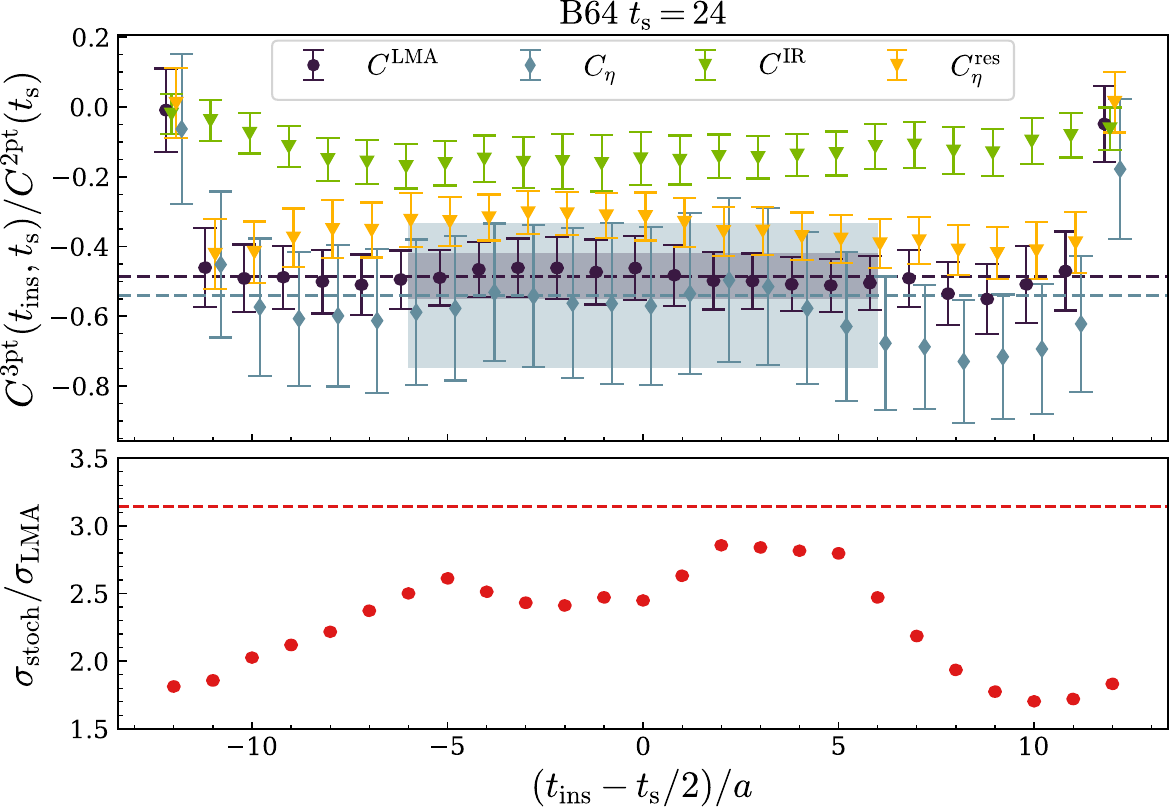}
    \caption{\textit{Top}: Comparison between the standard (grey-blue points) and LMA-improved (dark-violet points) ratios of three- to two-point functions (see \cref{eq:fullCorr}) for the nucleon isoscalar axial current at $t_s=24$. The improved infrared (light-green points) and residual (yellow points) contributions are also shown. The horizontal dashed line and bands represent the plateau fit results for the two datasets. 
    \textit{Bottom}: Error gain at each time slice (red points), with the dashed horizontal line indicating the gain of the fitted result.}
    \label{fig:3pt}
\end{figure}

Based on this observation for the B64 ensemble, we study LMA-improved three-point functions using $N_{\EV}^{\IR} = 1000$. We consider source–sink time separations $t_s/a{=}$16 and 24, corresponding to $t_s{=}$1.3\,fm and 1.9\,fm, respectively. While essentially no improvement is observed at the smallest separation, a clear reduction in statistical noise emerges at the largest one.
In \cref{fig:3pt}, we compare the standard and LMA-improved ratios of three- to two-point functions at $t_s/a{=}$24 that lead to the isoscalar axial charge. In the improved case, we observe a reduction of the statistical uncertainty by approximately a factor of 2.5 in the plateau region, with an even larger improvement in the corresponding plateau fit result.
The infrared and residual contributions, shown separately in~\cref{fig:3pt}, exhibit comparable statistical uncertainties. This indicates that the chosen number of eigenmodes achieves an approximately optimal balance between the infrared and residual sectors. Further reduction of only one component would therefore lead to diminishing returns, as the total uncertainty would become dominated by the other. 

\subsection{Computational gain of LMA in nucleon three-point functions}

We now quantify the efficiency of the LMA procedure for baryons by taking computational costs into account. To this end, we define the cost ratio
\begin{equation}
R_\mathrm{\mathcal{C}} 
\equiv 
\dfrac{\mathcal{C}^{\smallcap{std}}}{\mathcal{C}^{\LMA}} = \cfrac{N_{\etap}\mathcal{C}_{\etap}}{N_{\etap}\mathcal{C}_{\etap}+N_{\etap}^{\IR}\mathcal{C}^{\IR}_{\etap}}=
\left[1 + \frac{\mathcal{C}^{\IR}_{\etap}}{\mathcal{C}_{\etap}}\cdot \frac{N_{\etap}^{\IR}}{N_{\etap}}\right]^{-1}\qq{with}N_{\etap}^{\IR}>N_{\etap}\qq{and}\mathcal{C}^{\IR}_{\etap}<\mathcal{C}_{\etap}\,,
\label{eq:cost}
\end{equation}
where $\mathcal{C}_{\etap}$ and $\mathcal{C}^{\IR}_{\etap}$ denote the per-source costs of computing the standard correlator and its infrared component, respectively.
In this expression, we neglect the cost associated with the computation of eigenvectors, assuming that it is amortised over their reuse in multiple observables. Consequently, it does not enter the cost comparison for a single measurement. The ratio, therefore, depends only on the relative cost of constructing the two correlator contributions. While the contraction costs are identical, the propagator part is significantly cheaper in the IR case, since it is directly computed from the known eigenvectors.
In our numerical estimates, based on our costs, we take $\mathcal{C}^{\IR}_{\etap}/\mathcal{C}_{\etap}=0.2$, i.e. the standard correlator is assumed to be five times more expensive than the IR component per source.

Next, we define the error gain as
\begin{equation}
G_\mathcal{E}(t)\equiv \frac{{\rm Err}[C^{\smallcap{std}}(t)]}{{\rm Err}[C^\LMA(t)]} \simeq \sqrt{\cfrac{
\dfrac{{\rm Var}[C_\etap(t)]}{N_{\etap}}
}{\dfrac{{\rm Var}[C_\etap^{\smallcap{res}}(t)]}{N_{\etap}}+
\dfrac{{\rm Var}[C_\etap^\IR(t)]}{N_{\etap}^{\IR}}
    }
    }\,,
\end{equation}
assuming that the stochastic noise, i.e., the variance between independent point sources, dominates over gauge noise.
Furthermore, assuming that the infrared and residual contributions are weakly correlated,
\begin{equation}
    {\rm Var}[C_\etap(t)] = {\rm Var}[C_\etap^\IR(t)] + {\rm Var}[C_\etap^{\smallcap{res}}(t)]\,,\quad\qq{we obtain} G^2_\mathcal{E}(t)\simeq\left[1 - \frac{{\rm Var}[C_\etap^\IR(t)]}{{\rm Var}[C_\etap(t)]}\cdot \frac{N_{\etap}^{\IR}-N_{\etap}}{N_{\etap}^{\IR}}\right]^{-1}\,.
\end{equation}
The error gain is therefore controlled by the ratio  ${\rm Var}[C_\etap^\IR(t)]/{\rm Var}[C_\etap(t)]$, and is consequently time-dependent, observable-dependent, and sensitive to the number of eigenvectors included in the infrared sector. By analysing $G^2_\mathcal{E}(t)$ in our data as a function of $N_{\etap}^{\IR}$ fixing $N_{\etap}$ to the largest available statistics, we extract this ratio for several nucleon charges at $t_s/a=16$ and~24. The results are reported in~\cref{tab:var_charges}, showing that the error gain almost saturates at 100\% for certain charges at $t_s/a=24$, while it is much reduced for other charges or at shorter distances.

In the regime where stochastic noise dominates over gauge noise, $G^2_\mathcal{E}(t)$ represents directly the increase in statistics required to achieve the same precision using the standard approach. This allows us to quantify the overall computational gain as
\begin{equation}
G_{\mathcal{C}}\left(N_{\etap}, N_{\etap}^{\IR}\right) = G_\mathcal{E}^2\left(N_{\etap}, N_{\etap}^{\IR}\right) \cdot R_\mathcal{C}\left(N_{\etap}, N_{\etap}^{\IR}\right)\,.
\label{eq:costgain}
\end{equation}

\begin{table}[tbh]
\centering
\setlength{\tabcolsep}{0.8em}
\renewcommand{\arraystretch}{1.8}
\aboverulesep=0ex % Solution part 1 of 3
\belowrulesep=0ex % Solution part 1 of 3
\begin{tabular}{l|cccccc}
\toprule \rule{0pt}{1.1EM}
  $\frac{{\rm Var}[C_\etap^\IR(t_s)]}{{\rm Var}[C_\etap(t_s)]}$   & $g_A^{u-d}$ & $g_A^{u+d}$ & $g_S^{u-d}$ & $g_S^{u+d}$ & $g_T^{u-d}$ & $g_T^{u+d}$ \\[2pt]
\midrule
   $t_s/a=16$  & 50.7(1.1)\% & 42.3(1.2)\% & 78.5(1.7)\% & 82.7(1.1)\% & 32.6(1.2)\% & 23.4(1.1)\% \\
   $t_s/a=24$  & 87.6(1.0)\% & 99.2(0.7)\% & 99.0(0.9)\% & 99.8(0.7)\% & 55.5(2.1)\% & 71.5(2.0)\% \\
\bottomrule
\end{tabular}
\caption{Percentage contribution of the IR correlation variance to the total correlation variance for the various three-point functions entering the charge determinations reported in the different columns, evaluated at $t_s/a=16$ and 24 with $t_{\rm ins}=t_s/2$.}
\label{tab:var_charges}
\end{table}
\begin{figure}
\centering
\includegraphics[width=0.85\linewidth]{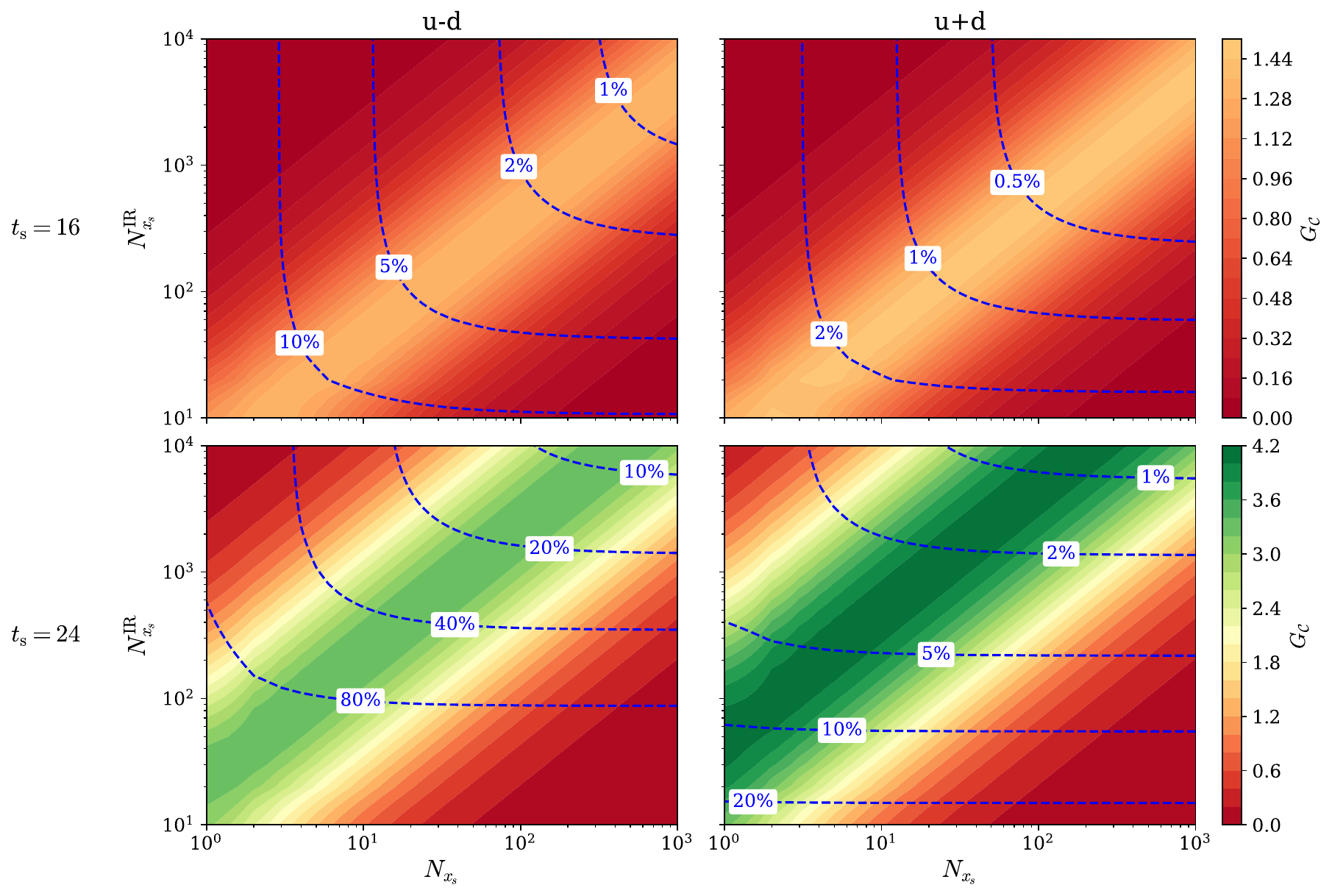}
\caption{The cost-gain $G_\mathcal{C}$ of \cref{eq:costgain} for the isovector (left panels) and isoscalar (right panels) combination of $g_S$. In the top panels, the source-sink time separation is fixed at $t_s/a=16$, while in the bottom panels, it is $t_s/a=24$. The blue dashed lines represent the $N_{\etap}$ and $N_{\etap}^{\IR}$ dependence of the expected relative error when computing correlators on roughly a thousand gauge configurations.}
\label{fig:gS_costgain}
\end{figure}

\begin{figure}
\includegraphics[width=0.8\linewidth]{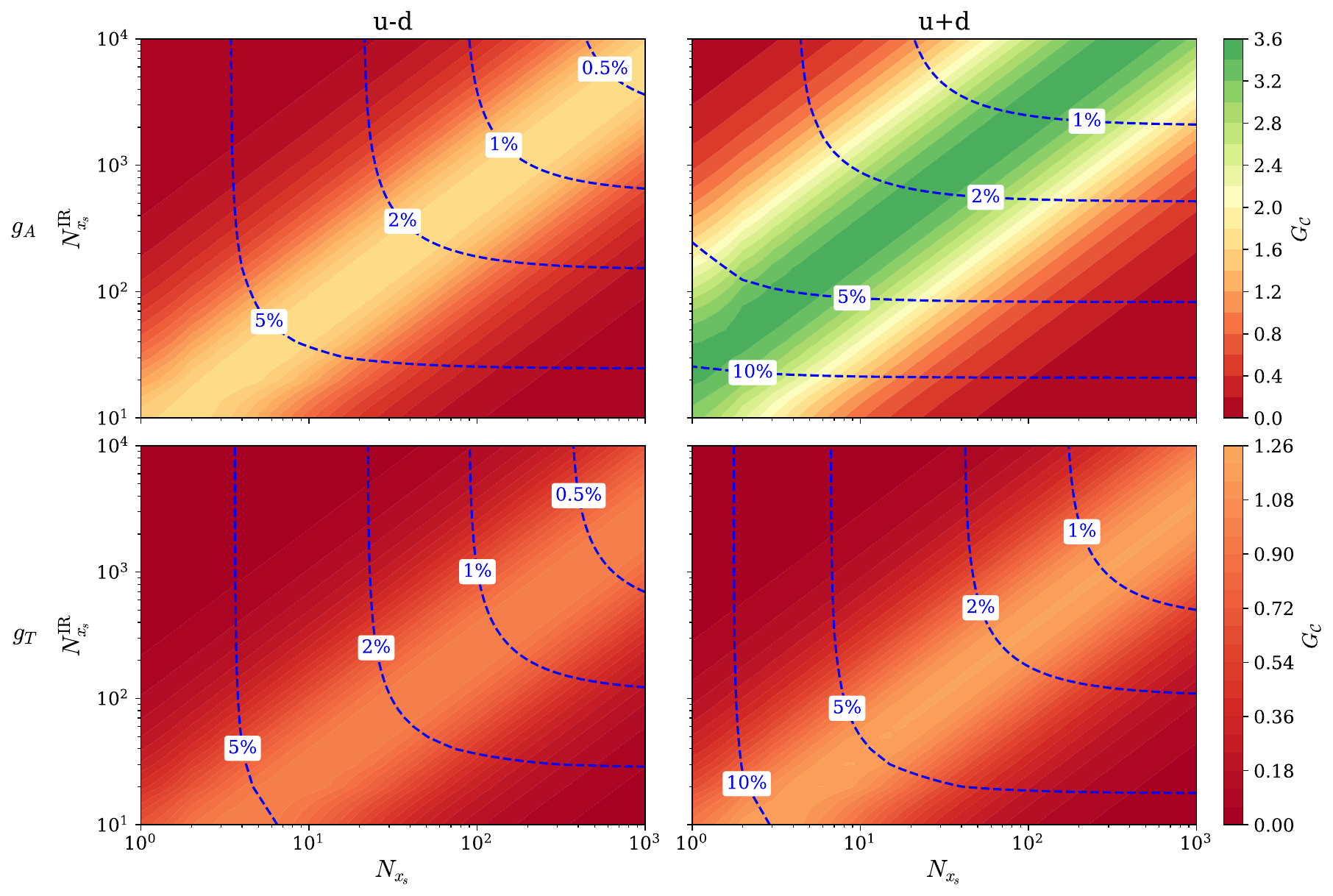}
\caption{The cost-gain $G_\mathcal{C}$ of \cref{eq:costgain} for the isovector (left panels) and isoscalar (right panels) combination of $g_A$ (top panels) and $g_T$ (bottom panels) at source-sink time separation $t_s/a=24$. The blue dashed lines indicate relative error thresholds as in \cref{fig:gS_costgain}.}
\label{fig:costgain24}
\end{figure}

We analyse this gain for the three-point correlation functions of axial, scalar, and tensor operator insertions, considering both isovector and isoscalar contributions. In \cref{fig:gS_costgain}, we present the cost-gain defined in \cref{eq:costgain} for the scalar operator at source-sink separations $t_s = 16$ and $t_s = 24$. The blue dashed lines represent the $N_{\etap}$ and $N_{\etap}^{\IR}$ dependence of the expected relative error when computing correlators on roughly a thousand gauge configurations. As expected, we find that the cost-gain increases with increasing $t_s$, while, conversely, the statistical precision at fixed $N_{\etap}$ and $N_{\etap}^{\IR}$ deteriorates.
The corresponding results for the axial and tensor operators are shown in \cref{fig:costgain24} only for $t_s/a=24$. In these cases, the cost-gain is reduced compared to the scalar insertion, indicating that the efficiency gain is observable-dependent. In all cases, the isoscalar combination  exhibits the most pronounced improvement with respect to  the isovector one.

\section{Conclusions}\label{sec:conclusion}

In this work, we have demonstrated that low-mode averaging (LMA) on physical-point ensembles yields a consistent noise reduction by a factor of three to five relative to standard stochastic estimators at large source–sink time separations for observables dominated by stochastic noise. This improvement translates into an order-of-magnitude reduction in the required statistics and, when setup and solver costs are taken into account, to a substantial overall decrease in computational expense.

At the same time, the magnitude of the gain and the optimal implementation of LMA are strongly observable dependent. In addition, they  depend on the lattice physical  volume and, potentially, on the quark mass, although the latter was not explored here, as all results are obtained at the physical point. Consequently, systematic tuning is required for each application. In particular, we find that baryonic correlators require significantly more low modes than mesonic ones. Furthermore, different operators  benefit to varying degrees, and the choice between multigrid-based and exact LMA depends sensitively on the cost–benefit balance for a given setup, especially after accounting for  volume scaling and memory constraints.

Beyond these numerical results, which are necessarily tied to the specific setup of our calculations and therefore partly qualitative, we emphasise three general conclusions of this work:
\begin{itemize}
    \item As presented in \cref{sec:LMA}, and extended to the multigrid formulation in \cref{sec:mesons_mg}, we motivate the application of LMA based on the generic estimator
    \begin{equation}\label{eq:corr_defl_var}
    C^\LMA_\eta (t) \equiv C_\eta(t) + \alpha(t)\Big( C^\IR_{\smallcap{impr.}}(t) - C^\IR_\eta(t)\Big)\qq{with} \alpha(t) = \frac{{\rm Cov}[C_\eta(t), C^\IR_\eta(t)]}{{\rm Var}[C^\IR_\eta(t)]}\,.
    \end{equation}
    Here, $C_\eta$ denotes the standard stochastic correlator, $C^\IR_\eta$ the corresponding correlator with propagators restricted to the IR sector, and $C^\IR_{\smallcap{impr.}}$ an improved estimator of the same IR contribution, ideally computed all-to-all or, more generally, at higher statistics. The control-variate coefficient $\alpha(t)$ becomes essential whenever \cref{eq:corr_defl} is not satisfied, and in particular in the multigrid LMA setup, where evaluating correlators on a coarser grid modifies their normalisation and must be accounted for. Alternatively, one may set $\alpha(t)=1$. This formulation offers several computational advantages, which are discussed in \cref{sec:LMA_remarks}.
    
    \item As presented in \cref{sec:scaling},  the spectrum of the massless Wilson operator is analysed, demonstrating a very smooth and stable spectrum for the considered physical quark mass ensembles. From chiral perturbation theory relations, the chiral condensate is extracted in the chiral limit for $N_f{=}2{+}1{+}1$ at a renormalisation scale of 2 GeV, with the result  given in \cref{eq:chiral_condensate}. From its pion-mass dependence, we also determine the low-energy constant $\bar{h}_1$ in two-flavour chiral perturbation theory for the first time from lattice QCD, with the corresponding value reported in \cref{eq:hbar}.

    \item As also discussed in \cref{sec:scaling}, the twisted-mass fermion discretisation exhibits particularly useful properties for the application of LMA. In this formulation, the spectrum depends trivially on the quark mass, since the operator at any mass value shares exactly the same eigenvectors. This allows us to predict the mass dependence of the spectrum straightforwardly and, crucially, to reuse the same eigenvectors for propagators at arbitrary quark masses. As a result, the cost of computing the IR component of correlation functions can be significantly reduced: once the necessary eigenvector inner products have been computed, they can be recombined with different eigenvalues without additional eigenvector calculations. A disadvantage, however, is that computing eigenvectors numerically at heavy quark mass requires essentially the same effort as at light quark mass, since in both cases one is effectively computing the eigenvectors of the same massless Wilson operator. This issue has also been identified previously in the context of multigrid solvers, where twisted-mass fermions were found to be unique compared to Wilson fermions because of the severe ill-conditioning of the coarse operator. Twisted-mass fermions therefore provide both an advantageous setup for LMA and a distinctive framework to which it is difficult to extend conclusions about the applicability of LMA made for other discretisations.
\end{itemize}

\section*{Acknowledgments}

We thank the QUDA developers, in particular Kate Clark and Evan Weinberg, for useful discussions on the multigrid preconditioning software.
We thank all members of the ETM Collaboration for the most enjoyable collaboration. 
We thank in particular Giuseppe Gagliardi, Marco Garofalo, and Bartosz Kostrzewa for the exchange of ideas, technical info and feedback that were valuable in developing the LMA methodology within the framework of the ETMC $g{-}2$ project. 
We also thank Johan Bijnens for providing references and comments on $\bar{h}_1$.

C.~A and  A.~E, and S.~B. acknowledge, respectively, support from the projects EXCELLENCE/0524/0459 (IMAGE-N) and EXCELLENCE/0524/0017 (MuonHVP), co-financed by the European Regional Development Fund and the Republic of Cyprus through the Research and Innovation Foundation within the framework of the Cohesion Policy Programme “THALIA 2021-2027”.
R.~F. and F.~M. are supported by the Italian Ministry of University and Research (MUR) under the grant PNRR-M4C2-I1.1-PRIN 2022-PE2 Non-perturbative aspects of fundamental interactions, in the Standard Model and beyond F53D23001480006 funded by E.U.- NextGenerationEU. 
C.S. is supported under the AQTIVATE EJD from the European Union’s research and innovation programme under the Marie Skłodowska-Curie Doctoral Networks action and Grant Agreement No 101072344.
We gratefully acknowledge CINECA and the EuroHPC Joint Undertaking for granting access to the Leonardo Supercomputer. Computing time on Leonardo Booster was allocated through the Extreme Scale Access Call (grant EHPC-EXT-2024E01-027), and additional GPU resources were provided under the INFN-LQCD123 initiative.
We acknowledge the Swiss National Supercomputing Centre (CSCS) access to Alps through the Chronos programme under project ID CH15. 
The authors gratefully acknowledge the Gauss Centre for Supercomputing e.V. (\href{https://www.gauss-centre.eu/}{www.gauss-centre.eu}) for funding this project by providing computing time on the GCS  Supercomputer JUWELS \cite{JUWELS} at J\"ulich Supercomputing Centre (JSC).

\appendix
\section{Demonstration of validity of~\texorpdfstring{\cref{eq:corr_defl}}{Eq.~(6)} for the correlation functions considered in this work}
\label{app:proof}

\subsection{Quark loops}

Consider a generic quark loop of the form
\begin{equation}
    L(t,\vec{p},\Gamma) = \Tr \left[\proj{q}{t}\Gamma\,S\right]
    \label{eq:loop_def}\,,
\end{equation}
where $\proj{q}{t}$ projects onto a momentum boost $\vb*{q}$ and time-slice $t$ and $\Gamma$ is an arbitrary Dirac matrix. In this case, the validity of \cref{eq:corr_defl} is immediate, but already illustrates the mechanism at work. Indeed, since for quark loops the observable depends linearly on the propagator, employing $S^\LMA_\eta$ from  \cref{eq:impr_prop} into \cref{eq:loop_def} and exploiting the linearity of the trace immediately yields
\begin{equation}
    L_{\eta}^{\LMA} = L_\eta - L_{\eta}^{\IR} + L^{\IR}\,,
\end{equation}
which is \cref{eq:corr_defl} applied to \cref{eq:loop_def}.
Here, each loop on the right-hand side is computed as the trace of the corresponding propagator $S_\eta$, $S_\eta^{\IR}$, or $S^{\IR}$.

\subsection{Meson two-point correlation functions with stochastic sources}

Consider a generic meson two-point correlation function, computed using a backward-going propagator obtained via $\gfi$-hermiticity, $S=\gfi S^\dagger \gfi$\footnote{
In the case of TM fermions, one needs to recall that $\gfi S^\dagger_{r}\gfi = S_{-r}$, where $r=\pm1$ is the Wilson parameter.
}, defined as
\begin{equation}
M(t_1,t_2,\vb*{p_1},\vb*{p_2},\Gamma_1,\Gamma_2) = 
\Tr\Big[\;
\proj{p_2}{t_2} \gfi\Gamma_2\, \;S\; \proj{p_1}{t_1} \Gamma_1\gfi\, \;S^{\dagger}\;
\Big] \,.
\label{eq:meson_corr}
\end{equation}
A standard approach to estimating this correlator is to introduce a set of stochastic sources $\eta$, as defined in Eqs.~(\ref{eq:stoch_sources_loc}--\ref{eq:stoch_sources}), to approximate the identity. The resulting estimator is
\begin{equation}
    M_\eta \equiv\;
\Tr\Big[\;
\proj{p_2}{t_2} \gfi\,\Gamma_2 \;S\,H_{\eta}\; \proj{p_1}{t_1} \Gamma_1\,\gfi \;S^{\dagger}\; \Big]\qquad\Longrightarrow\qquad \lim_{N_\eta\to\infty}\frac{1}{N_\eta} \sum_\eta M_\eta = M\,.
\end{equation}
In a commonly used low-mode averaging (LMA) implementation, one instead exploits the exact knowledge of $S^{\IR}$ to construct the estimator
\begin{equation}
    M_{\eta}^{\LMA} \equiv M^{\UV}_\eta + M^\smallcap{\rm mix1}_\eta + M^\smallcap{\rm mix2}_\eta + M^{\IR}\,,\qquad\Longrightarrow\qquad \lim_{N_\eta\to\infty}\frac{1}{N_\eta} \sum_\eta M_\eta^{\LMA} = M\,,
\end{equation}
where the four contributions are defined as
\begin{equation}
\renewcommand{\arraystretch}{2.5}
\setlength{\arraycolsep}{2pt}
\begin{array}{l >{\displaystyle}l >{\displaystyle}l >{\displaystyle}c >{\displaystyle}l >{\displaystyle}c >{\displaystyle}l}
&M^{\UV}_\eta & \equiv\; 
\Tr\Big[\;
\proj{p_2}{t_2} \gfi\,\Gamma_2 &S^{\UV}\,H_{\eta}\proj{p_1}{t_1}& \Gamma_1\,\gfi &\big(S^{\UV}\big)^{\dagger}& \Big] \\
&M^\smallcap{\rm mix1}_\eta & \equiv\; 
\Tr\Big[\;
\proj{p_2}{t_2} \gfi\,\Gamma_2 &S^{\UV}\,H_{\eta}\proj{p_1}{t_1}&  \Gamma_1\,\gfi &\big(S^{\IR}\big)^{\dagger}& \Big] \\
&M^\smallcap{\rm mix2}_\eta & \equiv\; 
\Tr\Big[\;
\proj{p_2}{t_2} \gfi\,\Gamma_2 &S^{\IR}& \Gamma_1\,\gfi &\big(S^{\UV}\,\proj{-p_1}{t_1}H_{\eta}\big)^{\dagger}& \Big] \\
&M^{\IR} & \equiv\; 
\Tr\Big[\;
\proj{p_2}{t_2} \gfi\,\Gamma_2 &S^{\IR}\proj{p_1}{t_1}&  \Gamma_1\,\gfi &\big(S^{\IR}\big)^{\dagger}
&\Big]\,.
\end{array}
\end{equation}
Each contribution containing $S^{\UV}$ includes a stochastic estimator $H_\eta$ for the ultraviolet component, whereas the purely infrared contribution is computed exactly using an all-to-all representation of $S^{\IR}$.

As stated in~\cref{eq:corr_defl}, this LMA estimator can equivalently be written as
\begin{equation}
\boxed{M_{\eta}^{\LMA} = M_\eta - M_{\eta}^{\IR} + M^{\IR}}\qquad\text{with}\qquad M^{\IR}_\eta \equiv\; 
\Tr\Big[\;
\proj{p_2}{t_2} \gfi\,\Gamma_2 \;S^{\IR}\,H_{\eta}\; \proj{p_1}{t_1} \Gamma_1\,\gfi \;\big(S^{\IR}\big)^{\dagger}\; \Big]\,,
\label{eq:meson_state}
\end{equation}
where this identity holds for each individual stochastic source, rather than only in the limit $N_\eta\to\infty$, provided that $H_\eta$ satisfies
\begin{equation}\label{eq:stoch_conditions}
 H_\eta^\dagger = H_\eta\qquad\text{and}\qquad [H_\eta\proj{p_1}{t_1},\Gamma_1\gfi] = 0\,.
\end{equation}
These conditions are satisfied by individual spin-diluted $\mathbb{Z}_n$-noise sources, as well as by point sources.

\paragraph{Proof}

To establish \cref{eq:meson_state}, we show that
\begin{equation}
    M_\eta - M_{\eta}^{\IR} = M^{\UV}_\eta + M^\smallcap{\rm mix1}_\eta + M^\smallcap{\rm mix2}_\eta \qquad\Longrightarrow\qquad M_\eta = M^{\UV}_\eta + M^\smallcap{\rm mix1}_\eta + M^\smallcap{\rm mix2}_\eta + M_{\eta}^{\IR}
\end{equation}
Expanding $M_\eta$ using $S=S^\IR+S^\UV$ yields the desired decomposition, except that the term corresponding to $M^\smallcap{\rm mix2}_\eta$ initially appears with $H_\eta$ in a different position. Explicitly,
\begin{equation}
    M_\eta - M^{\UV}_\eta - M^\smallcap{\rm mix1}_\eta - M_{\eta}^{\IR} \;=\; \Tr\Big[\;
\proj{p_2}{t_2} \gfi\,\Gamma_2\; S^{\IR}\,H_{\eta}\; \proj{p_1}{t_1} \Gamma_1\,\gfi\; \big(S^{\UV}\big)^{\dagger}\;\Big]\,,
\end{equation}
However, using the conditions in~\cref{eq:stoch_conditions}, we can commute $H_\eta\proj{p_1}{t_1}$ through $\Gamma_1\gfi$ and use $(H_\eta\proj{p_1}{t_1})^\dagger=\proj{-p_1}{t_1}H_\eta$ to recover precisely $M^\smallcap{\rm mix2}_\eta$. This completes the proof.

\subsection{Baryon two-point correlation functions with point sources}

The final example we consider is a baryon two-point correlation function, for which we show that \cref{eq:corr_defl} holds when point sources are employed.

For simplicity, we consider a baryon correlation function of the form
\begin{align}
B(t_x,t_y,\vb*{p},\mathcal{P},\mathcal{W}) = \sum_{\vb*{x}, \vb*{y}} e^{i\left(\vb*{x}-\vb*{y}\right)\cdot\vb*{p}} \, 
\mathcal{P}_{\gamma\gamma'}\, \mathcal{W}_{\alpha\beta}^{abc}\, \mathcal{W}_{\alpha'\beta'}^{a'b'c'}\,
S(x,y)_{aa'}^{\alpha\alpha'}S(x,y)_{bb'}^{\beta\beta'}S(x,y)_{cc'}^{\gamma\gamma'}
\,,
\label{eq:baryon_corr}
\end{align}
where $\mathcal{P}$ denotes a projector and $\mathcal{W}$ is antisymmetric in both spin and colour indices. For a positive-parity spin-$1/2$ baryon, these tensors take the form given in \cref{eq:nucl_W_P}.

We emphasise that \cref{eq:baryon_corr} is not completely generic. In particular, after performing the Wick contractions, additional terms involving different index orderings and quark-flavour structures may arise, as in the nucleon interpolator of \cref{eq:interpolator}. Nevertheless, the expression above is sufficiently representative for discussing the validity of \cref{eq:corr_defl}. We also note that at the moment, we have not yet employed stochastic sources or point sources.

Let us now discuss the effect of employing $S^\LMA_\etap=S^\UV_\etap+S^\IR$ in \cref{eq:baryon_corr}, where the $\UV$ part is computed using a point-source with coordinates $\etap$. This will then produce eight contracted terms built from the following sets of quark propagators,
\begin{equation}
    C_{1}(S^\IR, S^\IR, S^\IR)\,,\qquad C_{2,3,4}(S^\IR, S^\IR, S^\UV_\etap)\,,\qquad C_{5,6,7}(S^\IR, S^\UV_\etap, S^\UV_\etap)\,,\qq{or} C_{8}(S^\UV_\etap, S^\UV_\etap, S^\UV_\etap)\,.
\end{equation}
The fully point-source correlation function, i.e. where $S_\etap=S^\UV_\etap+S^\IR_\etap$ is used, has exactly the same structure, whereas $S^\IR$ is replaced by $S^\IR_\etap$. Thus, to demonstrate that \cref{eq:corr_defl} holds for baryon correlation functions, namely
\begin{equation}
    B^\LMA_\etap = B_\etap -B^\IR_\etap+B^\IR\qq{with} B^\IR_\etap=C_1(S^\IR_\etap,S^\IR_\etap,S^\IR_\etap)\qq{and}
    B^\IR=C_1(S^\IR,S^\IR,S^\IR)\,,
\end{equation}
we need to proceed similarly to the meson case, showing that the mixed IR--UV terms are the same in both cases, namely 
\begin{equation}
    C_{i}(S^\IR_\etap, S^\IR_\etap, S^\UV_\etap)=C_{i}(S^\IR, S^\IR, S^\UV_\etap)\qq{and}C_{i}(S^\IR_\etap, S^\UV_\etap, S^\UV_\etap)=C_{i}(S^\IR, S^\UV_\etap, S^\UV_\etap)\,.
\end{equation}
This is straightforward to show because the presence of at least one point source makes the sum over source coordinates collapse to a single point in any of such contractions.
Let us show it explicitly for one case, considering e.g.
\begin{equation}
C(S^\IR, S^\IR, S^\UV_\etap) 
=  \sum_{\vb*{x}, \vb*{y}} e^{i\left(\vb*{x}-\vb*{y}\right)\cdot\vb*{p}} \,
\mathcal{P}_{\gamma\gamma'}\, \mathcal{W}_{\alpha\beta}^{abc}\, \mathcal{W}_{\alpha'\beta'}^{a'b'c'}\,
S^\IR(x,y)_{aa'}^{\alpha\alpha'}
S^\IR(x,y)_{bb'}^{\beta\beta'}
S^\UV_\etap(x,y)_{cc'}^{\gamma\gamma'}\,.
\end{equation}
Since $S^\UV_\etap(x,y)$ is non-zero only at $y=x_s$, then the sum over $\vb*{y}$ collapses to the coordinate $x_s$ for all propagators, yielding to the equality above.

\bibliography{LMA_refs.bib,otherref.bib}

@article{Walker-Loud:2019cif,
    author = "Walker-Loud, Andr{\'e} and others",
    title = "{Lattice QCD Determination of $g_A$}",
    eprint = "1912.08321",
    archivePrefix = "arXiv",
    primaryClass = "hep-lat",
    reportNumber = "RIKEN-iTHEMS-Report-19, LLNL-PROC-800060",
    doi = "10.22323/1.317.0020",
    journal = "PoS",
    volume = "CD2018",
    pages = "020",
    year = "2020"
}

@article{Bali:2023sdi,
    author = {Bali, Gunnar S. and Collins, Sara and Heybrock, Simon and L{\"o}ffler, Marius and R{\"o}dl, Rudolf and S{\"o}ldner, Wolfgang and Weish{\"a}upl, Simon},
    collaboration = "RQCD",
    title = "{Octet baryon isovector charges from Nf=2+1 lattice QCD}",
    eprint = "2305.04717",
    archivePrefix = "arXiv",
    primaryClass = "hep-lat",
    doi = "10.1103/PhysRevD.108.034512",
    journal = "Phys. Rev. D",
    volume = "108",
    number = "3",
    pages = "034512",
    year = "2023"
}

@article{Djukanovic:2024krw,
    author = "Djukanovic, Dalibor and von Hippel, Georg and Meyer, Harvey B. and Ottnad, Konstantin and Wittig, Hartmut",
    title = "{Improved analysis of isovector nucleon matrix elements with Nf=2+1 flavors of O(a) improved Wilson fermions}",
    eprint = "2402.03024",
    archivePrefix = "arXiv",
    primaryClass = "hep-lat",
    reportNumber = "MITP-24-014",
    doi = "10.1103/PhysRevD.109.074507",
    journal = "Phys. Rev. D",
    volume = "109",
    number = "7",
    pages = "074507",
    year = "2024"
}

@article{Alexandrou:2024ozj,
    author = "Alexandrou, C. and Bacchio, S. and Finkenrath, J. and Iona, C. and Koutsou, G. and Li, Y. and Spanoudes, G.",
    title = "{Nucleon charges and {\ensuremath{\sigma}}-terms in lattice QCD}",
    eprint = "2412.01535",
    archivePrefix = "arXiv",
    primaryClass = "hep-lat",
    doi = "10.1103/PhysRevD.111.054505",
    journal = "Phys. Rev. D",
    volume = "111",
    number = "5",
    pages = "054505",
    year = "2025"
}

@article{FlavourLatticeAveragingGroupFLAG:2024oxs,
    author = "Aoki, Y. and others",
    collaboration = "Flavour Lattice Averaging Group (FLAG)",
    title = "{FLAG review 2024}",
    eprint = "2411.04268",
    archivePrefix = "arXiv",
    primaryClass = "hep-lat",
    reportNumber = "CERN-TH-2024-192, FERMILAB-PUB-24-0785-T",
    doi = "10.1103/nfzp-p5dn",
    journal = "Phys. Rev. D",
    volume = "113",
    number = "1",
    pages = "014508",
    year = "2026"
}

@article{Jang:2023zts,
    author = "Jang, Yong-Chull and Gupta, Rajan and Bhattacharya, Tanmoy and Yoon, Boram and Lin, Huey-Wen",
    collaboration = "Precision Neutron Decay Matrix Elements (PNDME)",
    title = "{Nucleon isovector axial form factors}",
    eprint = "2305.11330",
    archivePrefix = "arXiv",
    primaryClass = "hep-lat",
    reportNumber = "Los Alamos LA-UR-23-25225",
    doi = "10.1103/PhysRevD.109.014503",
    journal = "Phys. Rev. D",
    volume = "109",
    number = "1",
    pages = "014503",
    year = "2024"
}

@article{Djukanovic:2022wru,
    author = "Djukanovic, Dalibor and von Hippel, Georg and Koponen, Jonna and Meyer, Harvey B. and Ottnad, Konstantin and Schulz, Tobias and Wittig, Hartmut",
    title = "{Isovector axial form factor of the nucleon from lattice QCD}",
    eprint = "2207.03440",
    archivePrefix = "arXiv",
    primaryClass = "hep-lat",
    reportNumber = "MITP-22-053",
    doi = "10.1103/PhysRevD.106.074503",
    journal = "Phys. Rev. D",
    volume = "106",
    number = "7",
    pages = "074503",
    year = "2022"
}

@article{Alexandrou:2023qbg,
    author = "Alexandrou, Constantia and Bacchio, Simone and Constantinou, Martha and Finkenrath, Jacob and Frezzotti, Roberto and Kostrzewa, Bartosz and Koutsou, Giannis and Spanoudes, Gregoris and Urbach, Carsten",
    collaboration = "Extended Twisted Mass",
    title = "{Nucleon axial and pseudoscalar form factors using twisted-mass fermion ensembles at the physical point}",
    eprint = "2309.05774",
    archivePrefix = "arXiv",
    primaryClass = "hep-lat",
    doi = "10.1103/PhysRevD.109.034503",
    journal = "Phys. Rev. D",
    volume = "109",
    number = "3",
    pages = "034503",
    year = "2024"
}

@article{Djukanovic:2023beb,
    author = "Djukanovic, Dalibor and von Hippel, Georg and Meyer, Harvey B. and Ottnad, Konstantin and Salg, Miguel and Wittig, Hartmut",
    title = "{Electromagnetic form factors of the nucleon from Nf=2+1 lattice QCD}",
    eprint = "2309.06590",
    archivePrefix = "arXiv",
    primaryClass = "hep-lat",
    reportNumber = "MITP-23-044",
    doi = "10.1103/PhysRevD.109.094510",
    journal = "Phys. Rev. D",
    volume = "109",
    number = "9",
    pages = "094510",
    year = "2024"
}

@article{Tsuji:2023llh,
    author = "Tsuji, Ryutaro and Aoki, Yasumichi and Ishikawa, Ken-Ichi and Kuramashi, Yoshinobu and Sasaki, Shoichi and Sato, Kohei and Shintani, Eigo and Watanabe, Hiromasa and Yamazaki, Takeshi",
    collaboration = "PACS",
    title = "{Nucleon form factors in Nf=2+1 lattice QCD at the physical point: Finite lattice spacing effect on the root-mean-square radii}",
    eprint = "2311.10345",
    archivePrefix = "arXiv",
    primaryClass = "hep-lat",
    reportNumber = "UTHEP-783, UTCCS-P-149, HUPD-2307, YITP-23-143",
    doi = "10.1103/PhysRevD.109.094505",
    journal = "Phys. Rev. D",
    volume = "109",
    number = "9",
    pages = "094505",
    year = "2024"
}

@article{Alexandrou:2025vto,
    author = "Alexandrou, Constantia and Bacchio, Simone and Koutsou, Giannis and Prasad, Bhavna and Spanoudes, Gregoris",
    title = "{Proton and neutron electromagnetic form factors in the continuum limit using lattice QCD ensembles with physical pion masses}",
    eprint = "2507.20910",
    archivePrefix = "arXiv",
    primaryClass = "hep-lat",
    doi = "10.1103/tt39-n1df",
    journal = "Phys. Rev. D",
    volume = "113",
    number = "11",
    pages = "114524",
    year = "2026"
}

@article{ExtendedTwistedMassCollaborationETMC:2022sta,
    author = "Alexandrou, Constantia and others",
    collaboration = "Extended Twisted Mass Collaboration (ETMC)",
    title = "{Probing the Energy-Smeared R Ratio Using Lattice QCD}",
    eprint = "2212.08467",
    archivePrefix = "arXiv",
    primaryClass = "hep-lat",
    doi = "10.1103/PhysRevLett.130.241901",
    journal = "Phys. Rev. Lett.",
    volume = "130",
    number = "24",
    pages = "241901",
    year = "2023"
}

@article{Evangelista:2023fmt,
    author = "Evangelista, Antonio and Frezzotti, Roberto and Tantalo, Nazario and Gagliardi, Giuseppe and Sanfilippo, Francesco and Simula, Silvano and Lubicz, Vittorio",
    collaboration = "Extended Twisted Mass",
    title = "{Inclusive hadronic decay rate of the {\ensuremath{\tau}} lepton from lattice QCD}",
    eprint = "2308.03125",
    archivePrefix = "arXiv",
    primaryClass = "hep-lat",
    doi = "10.1103/PhysRevD.108.074513",
    journal = "Phys. Rev. D",
    volume = "108",
    number = "7",
    pages = "074513",
    year = "2023"
}

@article{ExtendedTwistedMass:2025tpc,
    author = "Margari, Francesca and others",
    collaboration = "Extended Twisted Mass",
    title = "{Smeared $R$-ratio in isospin symmetric QCD with Low Mode Averaging}",
    eprint = "2502.03187",
    archivePrefix = "arXiv",
    primaryClass = "hep-lat",
    doi = "10.22323/1.466.0446",
    journal = "PoS",
    volume = "LATTICE2024",
    pages = "446",
    year = "2025"
}

@article{ExtendedTwistedMass:2024myu,
    author = "Alexandrou, Constantia and others",
    collaboration = "Extended Twisted Mass",
    title = "{Inclusive Hadronic Decay Rate of the {\ensuremath{\tau}} Lepton from Lattice QCD: The u{\textasciimacron}s Flavor Channel and the Cabibbo Angle}",
    eprint = "2403.05404",
    archivePrefix = "arXiv",
    primaryClass = "hep-lat",
    doi = "10.1103/PhysRevLett.132.261901",
    journal = "Phys. Rev. Lett.",
    volume = "132",
    number = "26",
    pages = "261901",
    year = "2024"
}

@article{DeSantis:2025qbb,
    author = "De Santis, Alessandro and others",
    title = "{Inclusive semileptonic decays of the Ds meson: A first-principles lattice QCD calculation}",
    eprint = "2504.06063",
    archivePrefix = "arXiv",
    primaryClass = "hep-lat",
    reportNumber = "HIP-2025-9/TH",
    doi = "10.1103/3cxg-k322",
    journal = "Phys. Rev. D",
    volume = "112",
    number = "5",
    pages = "054503",
    year = "2025"
}

@article{Borsanyi:2020mff,
    author = "Borsanyi, Sz. and others",
    title = "{Leading hadronic contribution to the muon magnetic moment from lattice QCD}",
    eprint = "2002.12347",
    archivePrefix = "arXiv",
    primaryClass = "hep-lat",
    doi = "10.1038/s41586-021-03418-1",
    journal = "Nature",
    volume = "593",
    number = "7857",
    pages = "51--55",
    year = "2021"
}

@article{Djukanovic:2024cmq,
    author = "Djukanovic, Dalibor and von Hippel, Georg and Kuberski, Simon and Meyer, Harvey B. and Miller, Nolan and Ottnad, Konstantin and Parrino, Julian and Risch, Andreas and Wittig, Hartmut",
    title = "{The hadronic vacuum polarization contribution to the muon g {\ensuremath{-}} 2 at long distances}",
    eprint = "2411.07969",
    archivePrefix = "arXiv",
    primaryClass = "hep-lat",
    reportNumber = "CERN-TH-2024-196, MITP-24-080",
    doi = "10.1007/JHEP04(2025)098",
    journal = "JHEP",
    volume = "04",
    pages = "098",
    year = "2025"
}

@article{Boccaletti:2024guq,
    author = "Boccaletti, A. and others",
    title = "{Hybrid calculation of hadronic vacuum polarization in muon g {\ensuremath{-}} 2 to 0.48{\%}}",
    eprint = "2407.10913",
    archivePrefix = "arXiv",
    primaryClass = "hep-lat",
    doi = "10.1038/s41586-026-10449-z",
    journal = "Nature",
    volume = "653",
    number = "8114",
    pages = "373--377",
    year = "2026"
}

@article{FermilabLatticeHPQCD:2024ppc,
    author = "Bazavov, Alexei and others",
    collaboration = "Fermilab Lattice, HPQCD,, MILC",
    title = "{Hadronic Vacuum Polarization for the Muon g-2 from Lattice QCD: Long-Distance and Full Light-Quark Connected Contribution}",
    eprint = "2412.18491",
    archivePrefix = "arXiv",
    primaryClass = "hep-lat",
    reportNumber = "FERMILAB-PUB-24-0957-T, FERMILAB-PUB-24-0957-T",
    doi = "10.1103/d583-yhfs",
    journal = "Phys. Rev. Lett.",
    volume = "135",
    number = "1",
    pages = "011901",
    year = "2025"
}

@article{Moningi:2025dlf,
    author = "Moningi, Vaishakhi and Aubin, Christopher and Blum, Thomas and Golterman, Maarten and Jin, Luchang and Peris, Santiago",
    title = "{Progress on computing the hadronic vacuum polarization contribution to the muon anomalous magnetic moment with staggered fermions}",
    eprint = "2602.24200",
    archivePrefix = "arXiv",
    primaryClass = "hep-lat",
    doi = "10.22323/1.466.0247",
    journal = "PoS",
    volume = "LATTICE2024",
    pages = "247",
    year = "2025"
}

@article{ETMC:2026mpp,
    author = "Evangelista, Antonio and Bacchio, Simone and De Santis, Alessandro and Frezzotti, Roberto and Gagliardi, Giuseppe and Garofalo, Marco and Maio, Lorenzo and Margari, Francesca and Pittler, Ferenc and Romiti, Simone",
    collaboration = "Extended Twisted Mass (ETMC)",
    title = "{An update on the HVP contribution to g{\ensuremath{\mu}}{\ensuremath{-}}2g{\ensuremath{\mu}}{\ensuremath{-}}2 in isoQCD from ETMC}",
    eprint = "2603.03120",
    archivePrefix = "arXiv",
    primaryClass = "hep-lat",
    doi = "10.22323/1.518.0196",
    journal = "PoS",
    volume = "LATTICE2025",
    pages = "196",
    year = "2026"
}

@article{ExtendedTwistedMass:2026nbv,
    author = "Alexandrou, C. and others",
    collaboration = "Extended Twisted Mass",
    title = "{The hadronic-vacuum-polarization contribution to the muon anomalous magnetic moment in isospin-symmetric lattice QCD with twisted-mass fermions}",
    eprint = "2610.22224",
    archivePrefix = "arXiv",
    primaryClass = "hep-lat",
    month = "10",
    year = "2026"
}

@article{Neff:2001zr,
    author = "Neff, H. and Eicker, N. and Lippert, T. and Negele, John W. and Schilling, K.",
    title = "{On the low fermionic eigenmode dominance in QCD on the lattice}",
    eprint = "hep-lat/0106016",
    archivePrefix = "arXiv",
    doi = "10.1103/PhysRevD.64.114509",
    journal = "Phys. Rev. D",
    volume = "64",
    pages = "114509",
    year = "2001"
}

@article{Giusti:2004yp,
    author = "Giusti, Leonardo and Hernandez, P. and Laine, M. and Weisz, P. and Wittig, H.",
    title = "{Low-energy couplings of QCD from current correlators near the chiral limit}",
    eprint = "hep-lat/0402002",
    archivePrefix = "arXiv",
    reportNumber = "BI-TP-2004-03, CPT-2003-P-4622, DESY-04-009, FTUV-04-0203, IFIC-04-04, MPP-2004-6",
    doi = "10.1088/1126-6708/2004/04/013",
    journal = "JHEP",
    volume = "04",
    pages = "013",
    year = "2004"
}

@article{DeGrand:2004qw,
    author = "DeGrand, Thomas A. and Schaefer, Stefan",
    title = "{Improving meson two point functions in lattice QCD}",
    eprint = "hep-lat/0401011",
    archivePrefix = "arXiv",
    reportNumber = "COLO-HEP-497",
    doi = "10.1016/j.cpc.2004.02.006",
    journal = "Comput. Phys. Commun.",
    volume = "159",
    pages = "185--191",
    year = "2004"
}

@article{DeGrand:2005vb,
    author = "DeGrand, Thomas A. and Schaefer, Stefan",
    title = "{Chiral properties of two-flavor QCD in small volume and at large lattice spacing}",
    eprint = "hep-lat/0506021",
    archivePrefix = "arXiv",
    doi = "10.1103/PhysRevD.72.054503",
    journal = "Phys. Rev. D",
    volume = "72",
    pages = "054503",
    year = "2005"
}

@article{Giusti:2005sx,
    author = "Giusti, Leonardo and Necco, Silvia",
    editor = "Michael, Christopher",
    title = "{Low-mode averaging for baryon correlation functions}",
    eprint = "hep-lat/0510011",
    archivePrefix = "arXiv",
    doi = "10.22323/1.020.0132",
    journal = "PoS",
    volume = "LAT2005",
    pages = "132",
    year = "2006"
}

@article{Bali:2009hu,
    author = "Bali, Gunnar S. and Collins, Sara and Schafer, Andreas",
    title = "{Effective noise reduction techniques for disconnected loops in Lattice QCD}",
    eprint = "0910.3970",
    archivePrefix = "arXiv",
    primaryClass = "hep-lat",
    doi = "10.1016/j.cpc.2010.05.008",
    journal = "Comput. Phys. Commun.",
    volume = "181",
    pages = "1570--1583",
    year = "2010"
}

@article{Blum:2012uh,
    author = "Blum, Thomas and Izubuchi, Taku and Shintani, Eigo",
    title = "{New class of variance-reduction techniques using lattice symmetries}",
    eprint = "1208.4349",
    archivePrefix = "arXiv",
    primaryClass = "hep-lat",
    reportNumber = "RBRC-967",
    doi = "10.1103/PhysRevD.88.094503",
    journal = "Phys. Rev. D",
    volume = "88",
    number = "9",
    pages = "094503",
    year = "2013"
}

@article{Shintani:2014vja,
    author = "Shintani, Eigo and Arthur, Rudy and Blum, Thomas and Izubuchi, Taku and Jung, Chulwoo and Lehner, Christoph",
    title = "{Covariant approximation averaging}",
    eprint = "1402.0244",
    archivePrefix = "arXiv",
    primaryClass = "hep-lat",
    doi = "10.1103/PhysRevD.91.114511",
    journal = "Phys. Rev. D",
    volume = "91",
    number = "11",
    pages = "114511",
    year = "2015"
}

@article{Luscher:2007se,
    author = "Luscher, Martin",
    title = "{Local coherence and deflation of the low quark modes in lattice QCD}",
    eprint = "0706.2298",
    archivePrefix = "arXiv",
    primaryClass = "hep-lat",
    reportNumber = "CERN-PH-TH-2007-096",
    doi = "10.1088/1126-6708/2007/07/081",
    journal = "JHEP",
    volume = "07",
    pages = "081",
    year = "2007"
}

@article{xQCD:2010pnl,
    author = "Li, A. and others",
    collaboration = "xQCD",
    title = "{Overlap Valence on 2+1 Flavor Domain Wall Fermion Configurations with Deflation and Low-mode Substitution}",
    eprint = "1005.5424",
    archivePrefix = "arXiv",
    primaryClass = "hep-lat",
    reportNumber = "UK-10-01",
    doi = "10.1103/PhysRevD.82.114501",
    journal = "Phys. Rev. D",
    volume = "82",
    pages = "114501",
    year = "2010"
}

@article{Luscher:2007es,
    author = "Luscher, Martin",
    title = "{Deflation acceleration of lattice QCD simulations}",
    eprint = "0710.5417",
    archivePrefix = "arXiv",
    primaryClass = "hep-lat",
    reportNumber = "CERN-PH-TH-2007-200",
    doi = "10.1088/1126-6708/2007/12/011",
    journal = "JHEP",
    volume = "12",
    pages = "011",
    year = "2007"
}

@article{Brannick:2007ue,
    author = "Brannick, J. and Brower, R. C. and Clark, M. A. and Osborn, J. C. and Rebbi, C.",
    title = "{Adaptive Multigrid Algorithm for Lattice QCD}",
    eprint = "0707.4018",
    archivePrefix = "arXiv",
    primaryClass = "hep-lat",
    reportNumber = "BUHEP-07-05",
    doi = "10.1103/PhysRevLett.100.041601",
    journal = "Phys. Rev. Lett.",
    volume = "100",
    pages = "041601",
    year = "2008"
}

@article{Babich:2010qb,
    author = "Babich, R. and Brannick, J. and Brower, R. C. and Clark, M. A. and Manteuffel, T. A. and McCormick, S. F. and Osborn, J. C. and Rebbi, C.",
    title = "{Adaptive multigrid algorithm for the lattice Wilson-Dirac operator}",
    eprint = "1005.3043",
    archivePrefix = "arXiv",
    primaryClass = "hep-lat",
    doi = "10.1103/PhysRevLett.105.201602",
    journal = "Phys. Rev. Lett.",
    volume = "105",
    pages = "201602",
    year = "2010"
}

@article{Frommer:2013fsa,
    author = {Frommer, Andreas and Kahl, Karsten and Krieg, Stefan and Leder, Bj{\"o}rn and Rottmann, Matthias},
    title = "{Adaptive Aggregation-Based Domain Decomposition Multigrid for the Lattice Wilson--Dirac Operator}",
    eprint = "1303.1377",
    archivePrefix = "arXiv",
    primaryClass = "hep-lat",
    doi = "10.1137/130919507",
    journal = "SIAM J. Sci. Comput.",
    volume = "36",
    number = "4",
    pages = "A1581--A1608",
    year = "2014"
}

@article{Frommer:2013kla,
    author = "Frommer, A. and Kahl, K. and Krieg, S. and Leder, B. and Rottmann, M.",
    title = "{An adaptive aggregation based domain decomposition multilevel method for the lattice wilson dirac operator: multilevel results}",
    eprint = "1307.6101",
    archivePrefix = "arXiv",
    primaryClass = "hep-lat",
    month = "7",
    year = "2013"
}

@article{Alexandrou:2016izb,
    author = "Alexandrou, Constantia and Bacchio, Simone and Finkenrath, Jacob and Frommer, Andreas and Kahl, Karsten and Rottmann, Matthias",
    title = "{Adaptive Aggregation-based Domain Decomposition Multigrid for Twisted Mass Fermions}",
    eprint = "1610.02370",
    archivePrefix = "arXiv",
    primaryClass = "hep-lat",
    doi = "10.1103/PhysRevD.94.114509",
    journal = "Phys. Rev. D",
    volume = "94",
    number = "11",
    pages = "114509",
    year = "2016"
}

@article{Alexandrou:2018wiv,
    author = "Alexandrou, Constantia and Bacchio, Simone and Finkenrath, Jacob",
    title = "{Multigrid approach in shifted linear systems for the non-degenerated twisted mass operator}",
    eprint = "1805.09584",
    archivePrefix = "arXiv",
    primaryClass = "hep-lat",
    doi = "10.1016/j.cpc.2018.10.013",
    journal = "Comput. Phys. Commun.",
    volume = "236",
    pages = "51--64",
    year = "2019"
}

@article{Brannick:2014vda,
    author = {Brannick, James and Frommer, Andreas and Kahl, Karsten and Leder, Bj{\"o}rn and Rottmann, Matthias and Strebel, Artur},
    title = "{Multigrid Preconditioning for the Overlap Operator in Lattice QCD}",
    eprint = "1410.7170",
    archivePrefix = "arXiv",
    primaryClass = "hep-lat",
    doi = "10.1007/s00211-015-0725-6",
    journal = "Numer. Math.",
    volume = "132",
    number = "3",
    pages = "463--490",
    year = "2016"
}

@article{Brower:2020xmc,
    author = "Brower, Richard C. and Clark, M. A. and Howarth, Dean and Weinberg, Evan S.",
    title = "{Multigrid for chiral lattice fermions: Domain wall}",
    eprint = "2004.07732",
    archivePrefix = "arXiv",
    primaryClass = "hep-lat",
    doi = "10.1103/PhysRevD.102.094517",
    journal = "Phys. Rev. D",
    volume = "102",
    number = "9",
    pages = "094517",
    year = "2020"
}

@article{Brower:2018ymy,
    author = "Brower, Richard C. and Clark, M. A. and Strelchenko, Alexei and Weinberg, Evan",
    title = "{Multigrid algorithm for staggered lattice fermions}",
    eprint = "1801.07823",
    archivePrefix = "arXiv",
    primaryClass = "hep-lat",
    reportNumber = "FERMILAB-PUB-18-073-CD",
    doi = "10.1103/PhysRevD.97.114513",
    journal = "Phys. Rev. D",
    volume = "97",
    number = "11",
    pages = "114513",
    year = "2018"
}

@article{Bacchio:2017pcp,
    author = "Bacchio, Simone and Alexandrou, Constantia and Finkerath, Jacob",
    editor = "Della Morte, M. and Fritzsch, P. and G{\'a}miz S{\'a}nchez, E. and Pena Ruano, C.",
    title = "{Multigrid accelerated simulations for Twisted Mass fermions}",
    eprint = "1710.06198",
    archivePrefix = "arXiv",
    primaryClass = "hep-lat",
    doi = "10.1051/epjconf/201817502002",
    journal = "EPJ Web Conf.",
    volume = "175",
    pages = "02002",
    year = "2018"
}

@article{Finkenrath:2017cau,
    author = "Finkenrath, Jacob and Alexandrou, Constantia and Bacchio, Simone and Charalambous, Panagiotis and Dimopoulos, Petros and Frezzotti, Roberto and Jansen, Karl and Kostrzewa, Bartosz and Rossi, Giancarlo and Urbach, Carsten",
    editor = "Della Morte, M. and Fritzsch, P. and G{\'a}miz S{\'a}nchez, E. and Pena Ruano, C.",
    title = "{Simulation of an ensemble of $N_f=2+1+1$ twisted mass clover-improved fermions at physical quark masses}",
    eprint = "1712.09579",
    archivePrefix = "arXiv",
    primaryClass = "hep-lat",
    reportNumber = "DESY-17-165",
    doi = "10.1051/epjconf/201817502003",
    journal = "EPJ Web Conf.",
    volume = "175",
    pages = "02003",
    year = "2018"
}

@article{Alexandrou:2018egz,
    author = "Alexandrou, Constantia and others",
    title = "{Simulating twisted mass fermions at physical light, strange and charm quark masses}",
    eprint = "1807.00495",
    archivePrefix = "arXiv",
    primaryClass = "hep-lat",
    doi = "10.1103/PhysRevD.98.054518",
    journal = "Phys. Rev. D",
    volume = "98",
    number = "5",
    pages = "054518",
    year = "2018"
}

@article{Frommer:2021tqd,
    author = "Frommer, Andreas and Khalil, Mostafa Nasr and Ramirez-Hidalgo, Gustavo",
    title = "{A Multilevel Approach to Variance Reduction in the Stochastic Estimation of the Trace of a Matrix}",
    eprint = "2108.11281",
    archivePrefix = "arXiv",
    primaryClass = "math.NA",
    doi = "10.1137/21M1441894",
    journal = "SIAM J. Sci. Comput.",
    volume = "44",
    number = "4",
    pages = "A2536--A2556",
    year = "2022"
}

@article{Gruber:2024cos,
    author = "Gruber, Roman and Harris, Tim and Krstic Marinkovic, Marina",
    title = "{Multigrid low-mode averaging}",
    eprint = "2412.06347",
    archivePrefix = "arXiv",
    primaryClass = "hep-lat",
    doi = "10.1103/PhysRevD.111.074508",
    journal = "Phys. Rev. D",
    volume = "111",
    number = "7",
    pages = "074508",
    year = "2025"
}

@article{Frommer:2025wew,
    author = "Frommer, Andreas and Jimenez-Merchan, Jose and Knechtli, Francesco and Korzec, Tomasz and Ramirez-Hidalgo, Gustavo",
    title = "{Using orthogonal projectors in multigrid multilevel Monte Carlo for trace estimation in lattice QCD}",
    eprint = "2509.11424",
    archivePrefix = "arXiv",
    primaryClass = "hep-lat",
    month = "9",
    year = "2025"
}

@article{Banks:1979yr,
    author = "Banks, Tom and Casher, A.",
    title = "{Chiral Symmetry Breaking in Confining Theories}",
    reportNumber = "TAUP-785-79-REV, TAUP-785-79",
    doi = "10.1016/0550-3213(80)90255-2",
    journal = "Nucl. Phys. B",
    volume = "169",
    pages = "103--125",
    year = "1980"
}

@article{Leutwyler:1992yt,
    author = "Leutwyler, H. and Smilga, Andrei V.",
    title = "{Spectrum of Dirac operator and role of winding number in QCD}",
    reportNumber = "BUTP-92-10",
    doi = "10.1103/PhysRevD.46.5607",
    journal = "Phys. Rev. D",
    volume = "46",
    pages = "5607--5632",
    year = "1992"
}

@article{Giusti:2008vb,
    author = "Giusti, Leonardo and Luscher, Martin",
    title = "{Chiral symmetry breaking and the Banks-Casher relation in lattice QCD with Wilson quarks}",
    eprint = "0812.3638",
    archivePrefix = "arXiv",
    primaryClass = "hep-lat",
    reportNumber = "CERN-PH-TH-2008-239",
    doi = "10.1088/1126-6708/2009/03/013",
    journal = "JHEP",
    volume = "03",
    pages = "013",
    year = "2009"
}

@article{Frezzotti:2000nk,
    author = "Frezzotti, Roberto and Grassi, Pietro Antonio and Sint, Stefan and Weisz, Peter",
    collaboration = "Alpha",
    title = "{Lattice QCD with a chirally twisted mass term}",
    eprint = "hep-lat/0101001",
    archivePrefix = "arXiv",
    reportNumber = "CERN-TH-2000-384, MPI-PHT-2000-51, BICOCCA-FT-0027, NYU-TH-00-09-13",
    doi = "10.1088/1126-6708/2001/08/058",
    journal = "JHEP",
    volume = "08",
    pages = "058",
    year = "2001"
}

@article{Engel:2014eea,
    author = "Engel, Georg P. and Giusti, Leonardo and Lottini, Stefano and Sommer, Rainer",
    title = "{Spectral density of the Dirac operator in two-flavor QCD}",
    eprint = "1411.6386",
    archivePrefix = "arXiv",
    primaryClass = "hep-lat",
    doi = "10.1103/PhysRevD.91.054505",
    journal = "Phys. Rev. D",
    volume = "91",
    number = "5",
    pages = "054505",
    year = "2015"
}

@article{Detmold:2019fbk,
    author = "Detmold, W. and Murphy, D. J. and Pochinsky, A. V. and Savage, M. J. and Shanahan, P. E. and Wagman, M. L.",
    title = "{Sparsening algorithm for multihadron lattice QCD correlation functions}",
    eprint = "1908.07050",
    archivePrefix = "arXiv",
    primaryClass = "hep-lat",
    reportNumber = "MIT-CTP/5127, INT-PUB-19-026",
    doi = "10.1103/PhysRevD.104.034502",
    journal = "Phys. Rev. D",
    volume = "104",
    number = "3",
    pages = "034502",
    year = "2021"
}

@article{Li:2020hbj,
    author = "Li, Yuan and Xia, Shi-Cheng and Feng, Xu and Jin, Lu-Chang and Liu, Chuan",
    title = "{Field sparsening for the construction of the correlation functions in lattice QCD}",
    eprint = "2009.01029",
    archivePrefix = "arXiv",
    primaryClass = "hep-lat",
    doi = "10.1103/PhysRevD.103.014514",
    journal = "Phys. Rev. D",
    volume = "103",
    number = "1",
    pages = "014514",
    year = "2021"
}

@article{Christian:2025kke,
    author = "Christian, Sam and Detmold, William",
    title = "{Aspects of propagator sparsening in lattice QCD}",
    eprint = "2501.05404",
    archivePrefix = "arXiv",
    primaryClass = "hep-lat",
    reportNumber = "MIT-CTP/5826",
    doi = "10.1103/xfmp-c2d4",
    journal = "Phys. Rev. D",
    volume = "111",
    number = "11",
    pages = "114514",
    year = "2025"
}

@article{ExtendedTwistedMass:2021gbo,
    author = "Alexandrou, C. and others",
    collaboration = "Extended Twisted Mass",
    title = "{Quark masses using twisted-mass fermion gauge ensembles}",
    eprint = "2104.13408",
    archivePrefix = "arXiv",
    primaryClass = "hep-lat",
    doi = "10.1103/PhysRevD.104.074515",
    journal = "Phys. Rev. D",
    volume = "104",
    number = "7",
    pages = "074515",
    year = "2021"
}

@article{Bonanno:2023xkg,
    author = "Bonanno, Claudio and D'Angelo, Francesco and D'Elia, Massimo",
    title = "{The chiral condensate of N$_{f}$ = 2 + 1 QCD from the spectrum of the staggered Dirac operator}",
    eprint = "2308.01303",
    archivePrefix = "arXiv",
    primaryClass = "hep-lat",
    doi = "10.1007/JHEP11(2023)013",
    journal = "JHEP",
    volume = "11",
    pages = "013",
    year = "2023"
}

@article{FlavourLatticeAveragingGroupFLAG:2021npn,
    author = "Aoki, Y. and others",
    collaboration = "Flavour Lattice Averaging Group (FLAG)",
    title = "{FLAG Review 2021}",
    eprint = "2111.09849",
    archivePrefix = "arXiv",
    primaryClass = "hep-lat",
    reportNumber = "CERN-TH-2021-191, JLAB-THY-21-3528, FERMILAB-PUB-21-620-SCD-T",
    doi = "10.1140/epjc/s10052-022-10536-1",
    journal = "Eur. Phys. J. C",
    volume = "82",
    number = "10",
    pages = "869",
    year = "2022"
}

@article{Cichy:2013gja,
    author = "Cichy, Krzysztof and Garcia-Ramos, Elena and Jansen, Karl",
    title = "{Chiral condensate from the twisted mass Dirac operator spectrum}",
    eprint = "1303.1954",
    archivePrefix = "arXiv",
    primaryClass = "hep-lat",
    reportNumber = "DESY-13-038, HU-EP-13-08, SFB-CPP-13-17",
    doi = "10.1007/JHEP10(2013)175",
    journal = "JHEP",
    volume = "10",
    pages = "175",
    year = "2013"
}

@article{Alexandrou:2017bzk,
    author = "Alexandrou, Constantia and Athenodorou, Andreas and Cichy, Krzysztof and Constantinou, Martha and Horkel, Derek P. and Jansen, Karl and Koutsou, Giannis and Larkin, Conor",
    title = "{Topological susceptibility from twisted mass fermions using spectral projectors and the gradient flow}",
    eprint = "1709.06596",
    archivePrefix = "arXiv",
    primaryClass = "hep-lat",
    reportNumber = "DESY-17-139",
    doi = "10.1103/PhysRevD.97.074503",
    journal = "Phys. Rev. D",
    volume = "97",
    number = "7",
    pages = "074503",
    year = "2018"
}

@article{Bazavov:2010yq,
    author = "Bazavov, A. and others",
    editor = "Rossi, Giancarlo",
    title = "{Staggered chiral perturbation theory in the two-flavor case and SU(2) analysis of the MILC data}",
    eprint = "1011.1792",
    archivePrefix = "arXiv",
    primaryClass = "hep-lat",
    journal = "PoS",
    volume = "LATTICE2010",
    pages = "083",
    year = "2010"
}

@article{Borsanyi:2012zv,
    author = "Borsanyi, Szabolcs and Durr, Stephan and Fodor, Zoltan and Krieg, Stefan and Schafer, Andreas and Scholz, Enno E and Szabo, Kalman K",
    title = "{SU(2) chiral perturbation theory low-energy constants from 2+1 flavor staggered lattice simulations}",
    eprint = "1205.0788",
    archivePrefix = "arXiv",
    primaryClass = "hep-lat",
    doi = "10.1103/PhysRevD.88.014513",
    journal = "Phys. Rev. D",
    volume = "88",
    pages = "014513",
    year = "2013"
}

@article{BMW:2013fzj,
    author = {D{\"u}rr, Stephan and others},
    collaboration = "BMW",
    title = "{Lattice QCD at the physical point meets SU(2) chiral perturbation theory}",
    eprint = "1310.3626",
    archivePrefix = "arXiv",
    primaryClass = "hep-lat",
    reportNumber = "CPT-P005-2013, WUB-13-14",
    doi = "10.1103/PhysRevD.90.114504",
    journal = "Phys. Rev. D",
    volume = "90",
    number = "11",
    pages = "114504",
    year = "2014"
}

@article{Boyle:2015exm,
    author = "Boyle, P. A. and others",
    title = "{Low energy constants of SU(2) partially quenched chiral perturbation theory from N$_f$=2+1 domain wall QCD}",
    eprint = "1511.01950",
    archivePrefix = "arXiv",
    primaryClass = "hep-lat",
    doi = "10.1103/PhysRevD.93.054502",
    journal = "Phys. Rev. D",
    volume = "93",
    number = "5",
    pages = "054502",
    year = "2016"
}

@article{Cossu:2016eqs,
    author = "Cossu, Guido and Fukaya, Hidenori and Hashimoto, Shoji and Kaneko, Takashi and Noaki, Jun-Ichi",
    title = "{Stochastic calculation of the Dirac spectrum on the lattice and a determination of chiral condensate in 2+1-flavor QCD}",
    eprint = "1607.01099",
    archivePrefix = "arXiv",
    primaryClass = "hep-lat",
    reportNumber = "KEK-CP-346, OU-HET-898",
    doi = "10.1093/ptep/ptw129",
    journal = "PTEP",
    volume = "2016",
    number = "9",
    pages = "093B06",
    year = "2016"
}

@article{ETM:2009ztk,
    author = "Baron, Remi and others",
    collaboration = "ETM",
    title = "{Light Meson Physics from Maximally Twisted Mass Lattice QCD}",
    eprint = "0911.5061",
    archivePrefix = "arXiv",
    primaryClass = "hep-lat",
    reportNumber = "HISKP-TH-09-37, DESY-09-187, SFB-CPP-09-115, MS-TP-09-27, IFT-UAM-CSIC-09-58, ROM2F-2009-24",
    doi = "10.1007/JHEP08(2010)097",
    journal = "JHEP",
    volume = "08",
    pages = "097",
    year = "2010"
}

@article{Brandt:2013dua,
    author = {Brandt, Bastian B. and J{\"u}ttner, Andreas and Wittig, Hartmut},
    title = "{The pion vector form factor from lattice QCD and NNLO chiral perturbation theory}",
    eprint = "1306.2916",
    archivePrefix = "arXiv",
    primaryClass = "hep-lat",
    reportNumber = "HIM-2013-03, MITP-13-029, SHEP",
    doi = "10.1007/JHEP11(2013)034",
    journal = "JHEP",
    volume = "11",
    pages = "034",
    year = "2013"
}

@article{Gasser:1983yg,
    author = "Gasser, J. and Leutwyler, H.",
    title = "{Chiral Perturbation Theory to One Loop}",
    reportNumber = "CERN-TH-3689",
    doi = "10.1016/0003-4916(84)90242-2",
    journal = "Annals Phys.",
    volume = "158",
    pages = "142",
    year = "1984"
}

@article{ETM:2010cqp,
    author = "Baron, R. and others",
    editor = "Rossi, Giancarlo",
    collaboration = "ETM",
    title = "{Light hadrons from Nf=2+1+1 dynamical twisted mass fermions}",
    eprint = "1101.0518",
    archivePrefix = "arXiv",
    primaryClass = "hep-lat",
    reportNumber = "HU-EP-10-75, SFB-CPP-10-110, DESY-10-205, HU-EP-10/75, SFB/CPP-10-110, MS-TP-10-22, DESY 10-205",
    doi = "10.22323/1.105.0123",
    journal = "PoS",
    volume = "LATTICE2010",
    pages = "123",
    year = "2010"
}

@article{Amoros:1999dp,
    author = "Amoros, Gabriel and Bijnens, Johan and Talavera, P.",
    title = "{Two point functions at two loops in three flavor chiral perturbation theory}",
    eprint = "hep-ph/9907264",
    archivePrefix = "arXiv",
    reportNumber = "LU-TP-99-15",
    doi = "10.1016/S0550-3213(99)00674-4",
    journal = "Nucl. Phys. B",
    volume = "568",
    pages = "319--363",
    year = "2000"
}

@article{Clark:2009wm,
    author = "Clark, M. A. and Babich, R. and Barros, K. and Brower, R. C. and Rebbi, C.",
    collaboration = "QUDA",
    title = "{Solving Lattice QCD systems of equations using mixed precision solvers on GPUs}",
    eprint = "0911.3191",
    archivePrefix = "arXiv",
    primaryClass = "hep-lat",
    doi = "10.1016/j.cpc.2010.05.002",
    journal = "Comput. Phys. Commun.",
    volume = "181",
    pages = "1517--1528",
    year = "2010"
}

@inproceedings{Babich:2011np,
    author = "Babich, R. and Clark, M. A. and Joo, B. and Shi, G. and Brower, R. C. and Gottlieb, S.",
    collaboration = "QUDA",
    title = "{Scaling lattice QCD beyond 100 GPUs}",
    booktitle = "{International Conference for High Performance Computing, Networking, Storage and Analysis}",
    eprint = "1109.2935",
    archivePrefix = "arXiv",
    primaryClass = "hep-lat",
    doi = "10.1145/2063384.2063478",
    month = "9",
    year = "2011"
}

@inproceedings{Clark:2016rdz,
    author = "Clark, M. A. and Jo{\'o}, B{\'a}lint and Strelchenko, Alexei and Cheng, Michael and Gambhir, Arjun and Brower, Richard. C.",
    collaboration = "QUDA",
    title = "{Accelerating lattice QCD multigrid on GPUs using fine-grained parallelization}",
    booktitle = "{International Conference for High Performance Computing, Networking, Storage and Analysis}",
    eprint = "1612.07873",
    archivePrefix = "arXiv",
    primaryClass = "hep-lat",
    reportNumber = "FERMILAB-CONF-16-638-CD",
    doi = "10.5555/3014904.3014995",
    month = "12",
    year = "2016"
}

@misc{BijensPrivate,
  author       = {Johan Bijnens},
  title        = "{Private communication}",
  year         = {2026},
}

@misc{renoEtmc,
    author = {},
    collaboration = {Extended Twisted Mass},
    title         = "{Quark masses and non-perturbative renormalisation of quark bilinear operators with $\ensuremath{N_f=2+1+1}$ Wilson-clover twisted mass fermions}",
    year         = {In preparation},
}

@article{JUWELS,
author = {{J\"{u}lich Supercomputing Centre}},
title = {{JUWELS Cluster and Booster: Exascale Pathfinder with Modular Supercomputing Architecture at Juelich Supercomputing Centre}},
journal = {Journal of large-scale research facilities},
number = {A138},
volume = {7},
doi = {10.17815/jlsrf-7-183},
url = {http://dx.doi.org/10.17815/jlsrf-7-183},
year = {2021}
}
\bibliographystyle{JHEP}
\end{document}